\documentclass[a4paper]{article}

\usepackage[pages=all, color=black, position={current page.south}, placement=bottom, scale=1, opacity=1, vshift=5mm]{background}
\SetBgContents{
 
}      

\usepackage[margin=1in]{geometry} 

\usepackage{amsmath}
\usepackage{multirow}
\usepackage{amsthm}
\usepackage{amssymb}

\newcommand{\acapo}[1]{%
  \begin{tabular}{@{}l@{}}\strut#1\strut\end{tabular}%
}

\usepackage[utf8]{inputenc}
\usepackage{hyperref}
\hypersetup{
	unicode,
}

\usepackage[sort&compress,numbers,square]{natbib}
\theoremstyle{plain}

\theoremstyle{definition}

\usepackage{algorithm, algpseudocode} 
\usepackage{mathrsfs} 

\usepackage{lipsum}

\title{Non-pharmaceutical interventions during the COVID-19 pandemic: a rapid review}
\author{Nicola Perra$^{1}$\thanks{n.perra@greenwich.ac.uk}}

\date{
	$^1$Networks and Urban Systems Centre, University of Greenwich, London, UK \\ \today
}

\begin{document}
	\maketitle
	
\begin{abstract}
Infectious diseases and human behavior are intertwined. On one side, our movements and interactions are the engines of transmission. On the other, the unfolding of viruses might induce changes to our daily activities. While intuitive, our understanding of such feedback loop is still limited. Before COVID-19 the literature on the subject was mainly theoretical and largely missed validation. The main issue was the lack of empirical data capturing behavioral change induced by diseases. Things have dramatically changed in 2020. Non-pharmaceutical interventions (NPIs) have been the key weapon against the SARS-CoV-2 virus and affected virtually any societal process. Travels bans, events cancellation, social distancing, curfews, and lockdowns have become unfortunately very familiar. The scale of the emergency, the ease of survey as well as crowdsourcing deployment guaranteed by the latest technology, several Data for Good programs developed by tech giants, major mobile phone providers, and other companies have allowed unprecedented access to data describing behavioral changes induced by the pandemic.\\ 
Here, I aim to review some of the vast literature written on the subject of NPIs during the COVID-19 pandemic. In doing so, I analyze $347$ articles written by more than $2518$ of authors in the last $12$ months. While the large majority of the sample was obtained by querying PubMed, it includes also a hand-curated list. Considering the focus, and methodology I have classified the sample into seven main categories: epidemic models, surveys, comments/perspectives, papers aiming to quantify the effects of NPIs, reviews, articles using data proxies to measure NPIs, and publicly available datasets describing NPIs. I summarize the methodology, data used, findings of the articles in each category and provide an outlook highlighting future challenges as well as opportunities
\end{abstract}
\newpage
\tableofcontents
	
\section{Introduction}

Our interactions, movements, and behavior affect the spreading of infectious diseases. The unfolding of such illnesses, in turn, might drastically affect our actions. Although obvious, particularly during a pandemic, we still don't have a well understood and developed theory or even an accepted standardized approach to account for this observation. Capturing the feedback loop between human behavior and infectious diseases is one of the key challenges in epidemiology~\cite{vespignani2009predicting}. Arguably, it can be considered as the \emph{hard problem} of epidemiology.\\
Even before the COVID-19 pandemic, the literature tackling this issue was vast~\cite{verelst,funk}. One of the reviews on the subject noted a key challenge: only $15\%$ of the papers are based on empirical data, most models being ``purely theoretical and lack[ing] representative data and a validation process"~\cite{verelst}. In the small set of papers informed, at least partially, by empirical data we find interesting approaches. Researchers have designed games~\cite{maharaj2011participatory,chen2013behavioral}, surveys, and used datasets (e.g., television viewing) to infer social-distancing as well as attitudes, altruism and self-interest in the context of vaccinations~\cite{zhong2013modeling, gray2011will,rubin2009public,rubin2014design,shim2012influence,bayham2015measured}. Others have used surveys to measure risk perception during the H1N1 2009 pandemic~\cite{cohen2013vaccination,fierro2013lattice}, to estimate the perceived severity of the SARS outbreak~\cite{durham2012incorporating}, or characterize behavioral changes induced by the seasonal flu~\cite{gozzi2020towards}. Data from social media have been used to estimate the spread of awareness in the population during the H1N1 2009 pandemic~\cite{pawelek2014modeling,collinson2015effects,xiao2015media}. Another type of approach used epidemiological data (such as incidence) to calibrate diseases and behavioral models~\cite{poletti2011effect,funk2009spread}. A notable example of this method is Ref.~\cite{he2013inferring} where the authors used historical records from the 1918 influenza pandemic to fit three different models, one of which included behavioral responses, to determine the origins of the multiple epidemic waves observed. \\
Despite these efforts, the scarcity of direct observational data describing the feedback between human behavior and diseases represented a major obstacle. The reader might have noted the use of the past tense which is not, like many others in the text, a grammatical mistake. In fact, in the last year or so, things have dramatically changed. The COVID-19 pandemic took over the world and unfortunately, non-pharmaceutical interventions (NPIs) have been one of the only weapons against the disease. NPIs refers to a wide range of both top-down (i.e., governmental) and bottom-up (i.e., self-initiated) measures aimed at interrupting infection chains by altering key aspects of our behavior. Travels bans, curfews, social distancing, bans of social gathering, face masks, increased hygiene, remote working, school closures, and lockdowns are examples. The heterogeneity of such measures across space and time as well as the magnitude of the changes they induced offer an unprecedented opportunity to understand, measure, and model the link between human behavior and infectious diseases. Modern technology and many Data for Good programs created by tech giants such as Google, Apple, Facebook by major mobile phone operators such as Vodafone, Telefonica, Orange and by smaller companies such as Cuebiq, SafeGraph, Unacast have provided unparalleled lenses to capture the effects of both top-down and bottom-up NPIs. In this background, I aim to summarize some key observations, data produced, approaches, and knowledge developed in this turbulent year.

\section{Disclaimer}

The incredible work of the research community makes it extremely hard to review manually all papers related to COVID-19. Just to give an idea of the scale, a query on PubMed for ``COVID" returns more than $71,000$ results. On Google Scholar we can find instead more than $135,000$ matches. Within this large body of work, some lines of research are smaller than others. However, NPIs have virtually affected all aspects of human activity. Consequently, we can expect the research on the subject to be vast and diversified across many communities.  As I write we are in the middle of the second wave in Europe. A new, potentially more transmissible, variant has emerged in the UK. Hence, many more papers are being started, finalized, and published. Considering my research interests and expertise the focus of this work is directed mainly towards the epidemiological implications of NPIs. From the development of epidemic models to surveys aimed at quantifying awareness and their adoption. Hence, I have opted for PubMed to gather a big part of the papers in the sample. Inevitably, this introduced a bias towards interdisciplinary, biomedical, and life science journals. Surely a search on the ISI Web of Knowledge would provide a more exhaustive picture of the implementation and effect of non-pharmaceutical interventions during the pandemic. As described below, I have not done a systematic search on PubMed by, for example, carefully crafting a set of target keywords via an iterative or a consultation process. Hence, it is important to stress how this review is not systematic nor complete. However, it covers a large and diverse body of research providing a glimpse of the breadth in the area and highlights a few key results as well as the different approaches used. \\

\section{Data collection, inclusion principles, annotation, and classification}

\begin{figure}[t]
  \centering
   \includegraphics[scale=0.25]{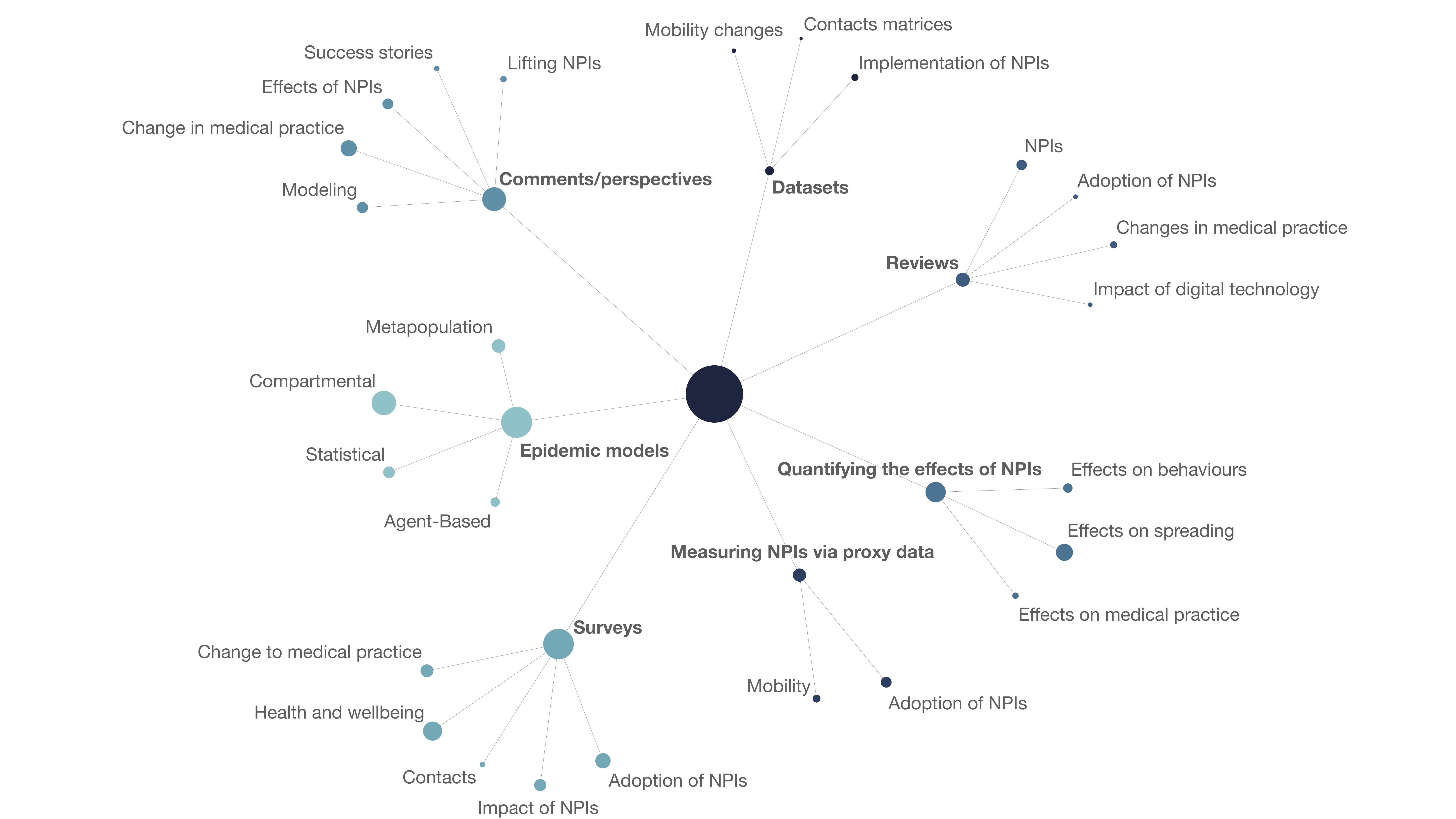}
  \caption{Schematic representation of the classification scheme. The central node describes all articles. The size of each node (i.e., category) is proportional to the number of articles.}
  \label{fig:fig1}
\end{figure}

Since the beginning of the pandemic, I have been following the research aimed at modeling the unfolding of the virus and at characterizing its societal effects. In doing so, I have compiled a list~\footnote{Until December $18^{th}$, $2020$} of $57$ influential papers that have shaped the research in the area and that, I believe, everyone would expect to see. Clearly, this list is biased by the research communities I am part of, those I follow, and my perception. Thus, I have thought to drastically extend the list using PubMed. To this end, on November $10^{th}$ I have submitted the following query: \emph{((non-pharmaceutical intervention) OR (non pharmaceutical intervention)) AND (COVID)}. The query targeted papers that mention explicitly NPIs and returned $387$ results. After some thinking on November $11^{th}$ $2020$ I have further extend the search querying PubMed for: \emph{(behavioral changes) AND (COVID)}. The idea was to expand even more the corpus by including papers that studied the impact of COVID-19 on our behaviors without mentioning explicitly NPIs. The second search returned $210$ results. In both cases the query was run without any particular other specification such as constraining the search to titles or abstracts. I have scanned titles and abstracts of these $597$ papers manually eliminating those referring to drugs trails, drugs development, clinical trials, clinical procedures and research protocols. I have removed duplications and papers outside the scope. Furthermore, I have kept only articles written in English. At the end of the process, and considering the initial list, I was left with $347$ papers which constitute the corpus of this review. \\
It is important to stress one more time how this is just a small subset of all related work. For example searchers in PudMed for \emph{restrictions AND COVID}, \emph{lockdown AND COVID}, \emph{(control measures) AND COVID} and \emph{(social distancing) AND COVID} return more than $2000$, $2900$, $3000$ and $8000$ results respectively. Furthermore, any systematic review should play with synonymous and similar words. This monumental work, which would end up with well more than fifteen thousand papers is left for the future and it would benefit from a large collaboration.\\ 
For each paper, I have manually extracted the following information: DOI, countries from author's affiliations, countries focus of the study (if any), journal and month of publication. Furthermore, I added annotations to describe which method was used (i.e., survey) and other relevant information (i.e., sample size, type of data used). After a first sweep, I have done several others to classify the corpus and create a personal, machine readable, taxonomy reported in Fig~\ref{fig:fig1}.\\

At the higher level, by considering the most representative methodology and/or aim of each paper, we find $7$ categories:

\begin{enumerate}
\item \textbf{Epidemic models ($29\%$)}: papers aimed at describing the unfolding of COVID-19 via epidemic models;
\item \textbf{Surveys ($28\%$)}: papers aimed at characterizing the impact of NPIs on several areas of human activity and/or their adoption via surveys;
\item \textbf{Comments and/or perspectives ($17\%$)}: papers that offer a reflection/perspective on NPIs in particular contexts;
\item \textbf{Quantifying the effects of NPIs ($12\%$)}: papers aimed at characterizing the effects of NPIs on epidemic indicators, behaviors and activities;
\item \textbf{Reviews ($6\%$)}: papers presenting a summary of the literature about NPIs in different contexts and areas;
\item \textbf{Measuring NPIs via proxy data ($5\%$)}: papers monitoring and measuring NPIs via data proxies;
\item \textbf{Datasets ($2\%$)}: papers describing and sharing data collections relevant for the study of NPIs.
\end{enumerate}

As in any classification scheme not all cases are clear cuts and there is some overlap. To simplify the analysis, I have assigned each paper to one and one only main category. Clearly, this implies same level of arbitrary in the classification. \\
By using \emph{Semantic Scholar}~\footnote{\url{https://pypi.org/project/semanticscholar/}} I have also extracted information about the authors and citations of each paper. Remarkably, these papers have been written by more than $2518$  authors (note that some were not indexed) and accumulated more than $9300$ citations~\footnote{As of December $19^{th}$ $2020$}. In Table~\ref{table:citations}, I provide the breakdown of number of authors, citations, median of citations and the top three articles (per number of citations) in each category. Epidemic models are the most cited category in the sample both for total number and median of citations. Surveys are the least cited considering the median of citations of the articles in the group. It is important to stress how this literature is very recent. Some papers have been published just one or two months ago hence they did not have enough time to accrue citations.

\begin{table}[]
\begin{tabular}{|l|l|l|l|l|}
\hline
\textbf{Category}               & \textbf{Authors} & \textbf{\acapo{Total \\Citations}} & \textbf{\acapo{Median \\citations}} & \textbf{\acapo{Top three \\articles for citations}} \\ \hline
Epidemic models                  & 774              & 4835                     & 5                                     &  $1^{st}$~\cite{Chinazzi_2020}, $2^{nd}$~\cite{Prem_2020}, $3^{rd}$~\cite{Flaxman_2020}                                        \\ \hline
Surveys                          & 751              & 775                      & 1                                     &       $1^{st}$~\cite{Zhang_2020_3}, $2^{nd}$~\cite{Reznik_2020}, $3^{rd}$~\cite{Jarvis_2020}                                      \\ \hline
\acapo{Comments and/or\\ perspectives}                        & 420              & 1049                     & 4                                     &       $1^{st}$~\cite{Eubank_2020_xx5}, $2^{nd}$~\cite{Oliver_2020_x3}, $3^{rd}$~\cite{Jiao_2020_x4}                                     \\ \hline
Quantifying the effects of NPIs & 405              & 2126                     & 2                                     &       $1^{st}$~\cite{Kraemer_2020_xx2}, $2^{nd}$~\cite{Pan_2020_xx1}, $3^{rd}$~\cite{Tian_2020_xx1}                                     \\ \hline
Reviews                          & 131              & 192                      & 3.5                                    &      $1^{st}$~\cite{De_Simone_2020_xx5}, $2^{nd}$~\cite{Romano_2020_xx5}, $3^{rd}$~\cite{Imai_2020_xx5}                                      \\ \hline
Measuring NPIs with proxy data  & 88              & 159                      & 2                                      &       $1^{st}$~\cite{Badr_2020_xx1}, $2^{nd}$~\cite{Weill_2020_xx1}, $3^{rd}$~\cite{Bonaccorsi_2020_xx1}                                     \\ \hline
Datasets                        & 105              & 37                       & 1.5                                      &    $1^{st}$~\cite{Pepe_2020_rf3},   $2^{nd}$~\cite{Willem_2020_xx1}, $3^{rd}$~\cite{Desvars_Larrive_2020_rf3}                                      \\ \hline
\end{tabular}
\label{table:citations}
\caption{Number of authors, total number of citations, median number of citations, and top three papers for citation for each of the categories. Citations and authors names have been extracted via Semantic Scholar on December $19^{th}$, $2020$.}
\end{table}

\section{Epidemic models}

\begin{figure}[t]
  \centering
   \includegraphics[scale=0.25]{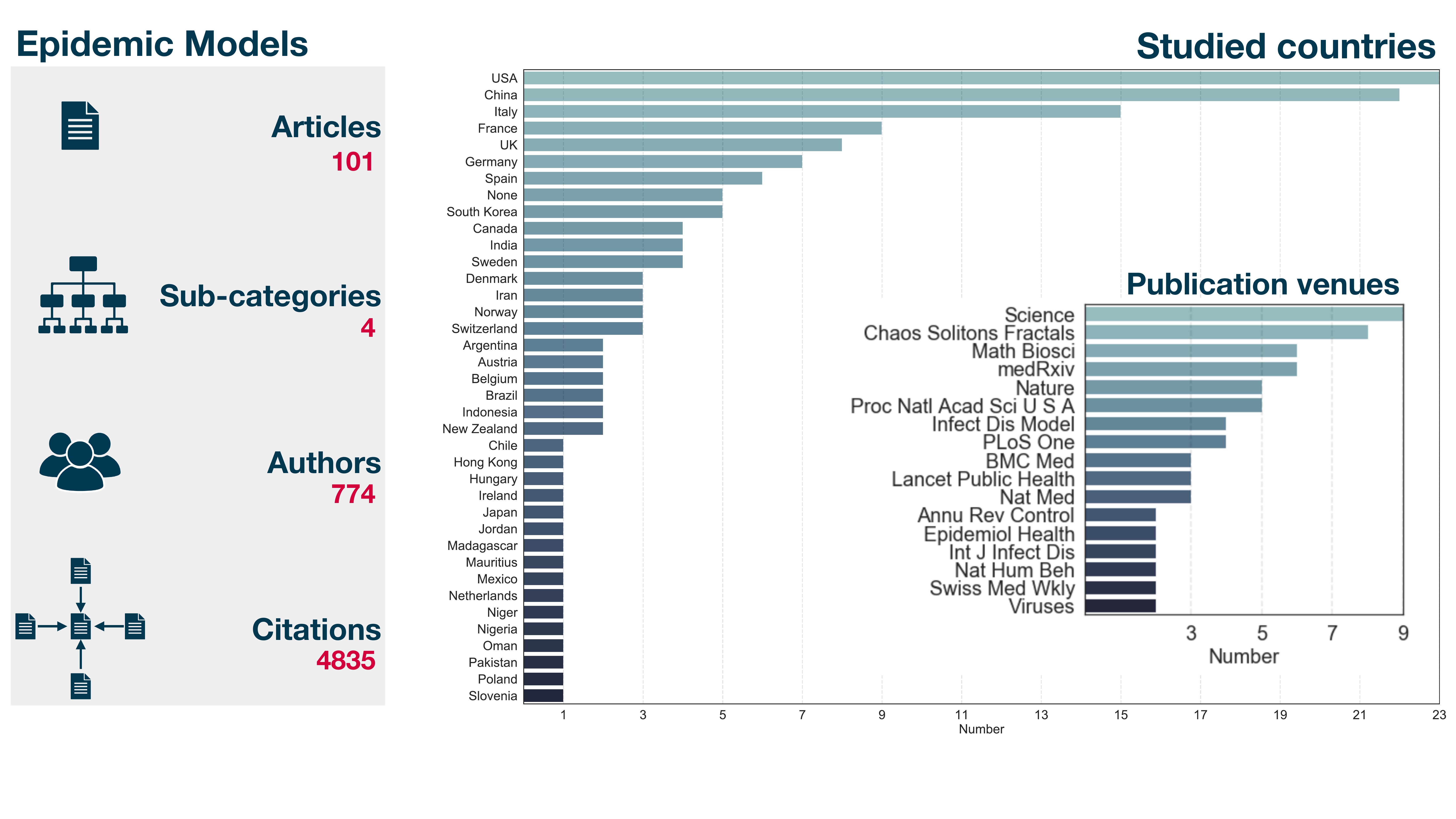}
  \caption{Total number of authors and citations of the papers in this category on the left. Note that these numbers are estimated from Semantic Scholar. On the right (top) histogram describing the number of countries subject of study in this category. Note how few papers that studied hundreds of countries are not counted in the histogram. On the right (bottom) most represented publication venues of the category. To improve visibility I am showing only journals featuring at least two articles}
  \label{fig:fig2}
\end{figure}

After the first suspicious cases of pneumonia in Wuhan were linked to a new virus, a set of key questions rapidly emerged. How infectious and fatal is it? How does it transmit? How many people are already sick? Can it be contained? What are the chances of importation in other countries? Is this the start of a new pandemic? When the virus was detected in Singapore, South Korea, France, UK, Italy, USA answers to many of these questions were still unclear. However, it did not take long to understand that widespread community transmission was probably taking place. Testing was initially targeted towards symptomatic individuals with travel history in China and the virus was spreading undetected via what is called \emph{cryptic transmission}. Cases, deaths and hospitalizations started to double at increasing pace in many countries. Asian states took draconian measures early. Europe reacted too late. Despite concerning news coming for Asia, the first European countries hit by waves of deaths and hospitalizations assumed, just days before, to be in perfect control. Emblematic is what the major of London declared the same days when Italy started a national lockdown: ``we should carry on doing what we’ve been doing"~\cite{mayor_of_london}. Things escalated quickly and in one week also the UK entered in a national lockdown. After the efforts and unprecedented NPIs put in place around the world, epidemic curves started to level off and also thanks to the good weather COVID hospitals started to empty, deaths rapidly decreased, and number of cases showed clear downwards trends. At this stage, the key questions revolved around how to relax the measures and reopen society. Should we keep the schools closed? Restaurants and pubs? What about international borders? Should face masks be mandatory? Summer $2020$, in the northern hemisphere,  has been extremely better than the dark months of spring. However, the infections started to ramp-up quite quickly in September and October. While countries have developed quite different strategies to manage winter, it became unfortunately clear that the virus was winning one more time. As I write these words, we are in the second national lockdown in England and from tomorrow some of the restrictions will be lifted. The questions now revolve around the Christmas break. What should people be allowed to do? How many people should meet during the holidays? What about Christmas shopping? One more time we are all facing a very challenging act balancing risks with hopes to go back to semi-normal life.\\
Although, the set of questions during the different phases of the pandemic changed many of the answers have been provided and/or informed by epidemic models. Not all suggestions coming from \emph{the Science}, as they say here in the UK, have been followed by our leaders, but it is clear that forecasts, predictions, scenario analyses coming from a variety of models were key to inform response strategies. In this section, I will revise some of the literature produced in the area. All models target the same virus, but the type of approach, data used, assumptions, how NPIs are modeled, limitations, strengths, predictive power or predictive ambitions are extremely diverse. \\
Before diving into the details, few words about the subset of research I will be reviewing. First of all, the large majority of models was tested or motivated by observations in one single country. USA, China, Italy, France and the UK are the most commonly used countries (see Fig~\ref{fig:fig2}). The list does not come as surprise considering the epicenter of the pandemic, data availability as well as the evolution of the pandemic in USA and Europe. However, few models in the sample consider hundreds of countries~\cite{Karnakov_2020,Loeffler_Wirth_2020,Chinazzi_2020,Bo_2021,Loeffler_Wirth_2020,Walker_2020}. Second, this body of research has been published by authors with affiliations in $40$ countries. USA, UK, China, and Italy are the top of the list. Third, there is a large variety also in terms of publication venues (see Fig~\ref{fig:fig2})). In fact, the sample of epidemic models has been published in $50$ venues (including $7$ papers in medRxiv). \emph{Science}, 
\emph{Chaos Solitons Fractals}, \emph{Mathematical Biosciences}, \emph{medRxiv}, \emph{Nature}, and \emph{PNAS} are at the top of the list. Finally, these papers have been written by more than $774$ authors and received more than $4835$ citations. A median of $5$ citations per article.

\subsection{Classification}

After analyzing the whole subset of papers and considering the typology of model adopted, I have identified $4$ categories. In particular: 

\begin{enumerate}
\item \textbf{Compartmental models}: consider a single population divided according to health statuses and in some cases age structure;
\item \textbf{Metapopulation models}: are based on a network of subpopulations (i.e. cities, regions, countries) connected by mobility;
\item  \textbf{Statistical models}: capture the evolution of the epidemic inferring key parameters and behaviors from data;
\item \textbf{Agent-based models}: capture the spreading patterns at the level of single individuals.
\end{enumerate}

In Table~\ref{table: epimodels} I present the full set of articles grouped according to the categories and in some cases sub-categories. I also report the country or countries used, if any. Here, same papers might be present in multiple categories. In fact, some of them adopted, often compare, different methods.

\begin{table}[]
\begin{tabular}{|l|l|l|}
\hline
\textbf{Category}              & \textbf{Subcategory}    & \textbf{Country/countries of study and reference} \\ \hline
\multirow{12}{*}{Compartmental} & SIR-like               & \acapo{Brazil, China~\cite{Cotta_2020}; China~\cite{Maier_2020,Lobato_2020,Din_2020}; China, Italy~\cite{Wangping_2020}; \\ China, France, Iran, Italy, South Korea, USA~\cite{Hsiang_2020}; \\
France, Iran, Italy~\cite{De_Visscher_2020}; Germany~\cite{Dehning_2020}; India~\cite{Debashree_2020}; \\ Italy~\cite{Calafiore_2020}; USA~\cite{Althouse_2020}; 187 countries~\cite{Loeffler_Wirth_2020}; \\ several European countries~\cite{Bhanot_2020,Karnakov_2020}}
                \\ \cline{2-3} 
                
                         & SIR-like age structured & 
China~\cite{Lei_2020}, Italy~\cite{Scala_2020}, 143 countries~\cite{Walker_2020}                   \\ \cline{2-3} 
                         & SEIR-like               & \acapo{Argentina, Japan, Indonesia, New Zealand, Spain, USA~\cite{L_pez_2020}; \\ Canada~\cite{Ogden_2020,Anderson_2020_fv2}; Canada, Germany, Italy~\cite{Alauddin_2020}; \\ China~\cite{Yang_2020, Hao_2020, Li_2020_3,Yan_2020}; China, Italy~\cite{Liu_2020_2}; China, UK~\cite{Overton_2020};\\Germany~\cite{Kantner_2020}; India~\cite{Mondal_2020,Mandal_2020}; Indonesia~\cite{Aldila_2020}; Ireland~\cite{N_raigh_2020}; \\ Mexico~\cite{Acu_a_Zegarra_2020}; Pakistan~\cite{Ullah_2020}; South Korea~\cite{Kim_2020,Kim_2020_2}; \\Switzerland~\cite{Lemaitre_2020}; \\ USA~\cite{Li_2020_2,Kain_2020,Childs_2020,Reiner_2020,Ngonghala_2020,Eikenberry_2020,Perkins_2020, Iboi_2020,Weitz_2020_fv2}; Theoretical~\cite{Stutt_2020,Zamir_2020}}                    \\ \cline{2-3} 
                         & SEIR-like age structure & \acapo{China~\cite{Prem_2020}; UK~\cite{Davies_2020,Brett_2020,Gosc__2020}; France~\cite{Di_Domenico_2020,Salje_2020}; Germany~\cite{Barbarossa_2020}; \\  Madagascar~\cite{Evans_2020}; Mauritius, Niger, Nigeria~\cite{van_Zandvoort_2020};\\ South Korea~\cite{Min_2020}; UK, USA~\cite{Burns_2020}; USA~\cite{Matrajt_2020}; Theoretical~\cite{Britton_2020}}                   \\ \cline{2-3} 
                         & SEIRS-like              & USA~\cite{Hoffman_2020}                    \\ 
                         \hline 
                         \hline
\multirow{2}{*}{Metapopulation} & Age unstructured     & \acapo{Canada~\cite{Karatayev_2020}; China~\cite{Rader_2020,Lai_2020,Ge_2020,Aleta_2020}; France, Poland, Spain~\cite{Medrek_2021};\\ Jordan~\cite{Kheirallah_2020}; Italy~\cite{Bertuzzo_2020,Carli_2020}; USA~\cite{pei2020differential,Chang_2020}; Europe~\cite{Ruktanonchai_2020}}
                  \\ \cline{2-3} 
                         & Age structured     & \acapo{Chile~\cite{Gozzi_2020}; Hungary~\cite{R_st_2020}; Spain~\cite{Arenas_2020_fv2}; Sweden~\cite{Sj_din_2020};\\ USA~\cite{Lau_2020}, 200+ countries~\cite{Chinazzi_2020}}                   \\ \hline
                         \hline 
Statistical              & NA                 & \acapo{Austria, Belgium, Denmark, France, \\ Germany, Italy, Norway, Spain, \\Sweden, Switzerland, UK~\cite{Bryant_2020}; Brazil~\cite{Candido_2020}\\China~\cite{Leung_2020,Quilty_2020}; China, Iran, Italy, South Korea~\cite{Zheng_2020};\\  Hong Kong~\cite{Cowling_2020}; India~\cite{Basu_2020}; Oman~\cite{Al_Wahaibi_2020}; Slovenia~\cite{Manevski_2020}; \\USA~\cite{Morley_2020_xx1}; 190~\cite{Bo_2021}; 11 European countries~\cite{Flaxman_2020}; 41 countries~\cite{Braunereabd9338}  }                 \\ \hline
\hline 
Agent-based              & NA                 & \acapo{China~\cite{Yang_2020}; China, Italy, USA~\cite{Wilder_2020}; Canada~\cite{Ogden_2020}; \\Italy~\cite{Bouchnita_2020}; 
USA~\cite{Aleta_2020_2,Aleta2020.12.15.20248273}; Theoretical~\cite{Kwon_2020,Silva_2020}}                  \\ \hline
\end{tabular}
\label{table: epimodels}
\caption{Summary of the epidemic models in the sample. The first column describes the main categories, the second the sub-categories, the last the country/countries focus of each study and the references}
\end{table}

\subsection{Compartmental models}

Compartmental models are the most common in the category. Probably this does not come as surprise since they require the least amount of data about the population under study. Nevertheless, they are a proven and powerful tool for describing the evolution of infectious diseases. It is important to note how compartmental models are featured also in other categories of models discussed below. In fact, it is extremely common to describe individuals' progression in the different phases of a disease (i.e., natural history of the disease) via compartments, each one representing health statuses. Susceptible, infectious, asymptomatic infectious, hospitalized, and recovered are classic examples. Other modeling approaches use a compartmentalization of the target population as one of the elements of the theoretical construct. Explicit social contacts structures, travel patterns and other features are added to model to describe key epidemiologically behaviors. In the category discussed here instead, these other details are simplified and expressed via effective functions, contact matrices, and rescaling of parameters.\\
As mentioned, compartmental models split the population according to health statuses. The number and type of compartments change as function of the disease under study, according to the hypotheses about the mechanisms driving the spreading of the virus and the behavior of the population. All compartmental models in the sample, and there are many, are a variation of two basic archetypes: SIR or SEIR. The sequence of letters, describing each compartments, captures the natural history of the disease. In SIR models susceptible individuals, in contact with infectious, develop the disease and move from $S$ to $I$.  This transition is often written as $S+I \xrightarrow{\beta} 2I$ where $\beta$ is the transmission rate. The $R$ compartment describes recovered individuals that are \emph{removed} from the disease dynamics. The transition from $I$ to $R$ is spontaneous and can be written as $I\xrightarrow{\mu} R$ where $\mu$ is the recovery rate. $\mu^{-1}$ is the infectious period. In SEIR models, we have another compartment $E$ describing individuals that were exposed to the virus but not yet infected. Thus the transmission of the disease can be written as  $S+I \xrightarrow{\beta} E+I$. As soon as the viral load reach certain levels exposed become infected $E\xrightarrow{\epsilon} I$. The sum of $\epsilon^{-1}$ and $\mu^{-1}$ is called generation time, defined as the average time from an infection to the next. In these archetypes, the per-capita rate at which susceptible contract the infection, the \emph{force of infection}, is often written as $\lambda=\beta I/N$. Hence, it is proportional to the transmission rate and to the probability of meeting an infectious individual in a homogeneously mixed population. These models are characterized by another key quantity called \emph{basic reproductive number}, $R_0$, defined as the number of secondary infections generated by an index case in an otherwise fully susceptible population. In case $R_0$ is above one the disease is able to spread and affect a finite fraction of the population. In SIR and SEIR models it is simple to show how $R_0=\beta/\mu$. In more complex models with several types of infectious compartments the expression is generally different but it can be calculated via the so called next generation matrix approach. See Ref.~\cite{blackwood2018introduction} for a very accessible and clear review.\\
The majority of models is the sample follow the SEIR archetype thus can be classified as SEIR-like models. If you ever had to quarantine after crossing a border, you have experienced first hand the reason why including the compartment of exposed is important: the incubation period peaks within few days but it has been reported to possibly extend for more than one week.  As we will see below, the general structure of these models is that of a simple SEIR model but typically several other compartments are added to account for individuals that are hospitalized, different types of manifestation of the disease (i.e., symptomatic, asymptomatic, paucy-symptomatic), as well as behavioral classes (i.e., individuals wearing face masks, reducing their contacts because concerned). COVID-19 is a quite complex disease with a wide range of manifestations that affects people's health and behaviors in different ways. \\
A subset of both SEIR and SIR like models include another important element: the age-structure of the population. In fact, COVID-19 does not affect age classes homogeneously. Older individuals have a much higher risk of hospitalization and mortality~\cite{verity2020estimates}. Furthermore, stratifying the population for age classes allows to account for the fact that contacts are not homogenous across age brackets. Including empirical contact matrices, which describe the rate of contacts within and across age classes, allows to move towards a more precise representation of the spreading of the virus and to account, explicitly, for different infection fatality rates (IFRs).  Furthermore, the ability to include contacts patterns across age groups allows to split contacts taking place at school, work, and in other contexts. This is particular important for models which aim to measure the impact of school closure, remote working, and other NPIs. After this general prelude, let's dive into the research that adopts compartmental models to study the spreading of COVID-19.\\

We start considering the body of modeling work based on SIR-like models. This subset of papers studies the effects of isolation and quarantine policies~\cite{Maier_2020}, models NPIs as effective changes of the parameters~\cite{Bhanot_2020,Debashree_2020,Wangping_2020,Lobato_2020,Cotta_2020,Dehning_2020,Karnakov_2020,Loeffler_Wirth_2020,Calafiore_2020,De_Visscher_2020}, investigates the impact of different policies~\cite{Hsiang_2020,Althouse_2020}, and defines the conditions for the optimal control of the spreading~\cite{Din_2020}. In the following, I will highlight some of these approaches.\\
Ref.~\cite{Maier_2020} presents a modification of the SIR scheme including a compartment $X$ to account for the portion of infectious individuals that are either quarantined or adopts other forms of NPIs, which effectively nullifying their ability to spread the virus. Furthermore, the model also considers that a portion of susceptible adopts NPIs and as result is removed from the epidemic dynamics. Interestingly, these two relatively simple modifications of the classic SIR model are enough to explain the sub-exponential growth observed in China. In fact, without any interventions a SIR model would imply an exponential growth in the early phases of the epidemic which is not what has been empirically observed in the Chinese context. A number of papers propose approaches to account for NPIs as effective variation or modulation of key SIR parameters such as the transmission rate $\beta$ and recovery rate $\mu$ which are either fitted from data or set a priory in scenario analyses. Two papers focusing on India~\cite{Debashree_2020}, Italy and China~\cite{Wangping_2020} extended the force of infection including a time-varying parameter $\pi(t)$ to modulate the \emph{bare} transmission rate $\beta$. The modulation is function of the NPIs put in place to curb the spreading of the virus. By fitting the model to epidemic indicators Ref.~\cite{Wangping_2020} estimated a $R_0$ for Italy above $4$ and in Hunan around $3$ in the early phases of the pandemic. A similar model has been proposed to infer the impact of NPIs on the epidemic evolution in Germany~\cite{Dehning_2020}. The author extended the SIR archetype considering that the transmission rate $\beta$ can change at certain points $t_i$, $\beta \rightarrow \beta_i$ with $i=1,\ldots,n$, from $\beta_{i-1}$ to $\beta_i$ over $\Delta t_i$ days. Hence, the index $i$ runs over the possible change points, each covering several days. For simplicity they assume a linear variation. The parameters $\beta_i$, $t_i$, $\Delta t_i$ among others are obtained via a Bayesian Markov Chain Monte Carlo (MCMC) comparing each potential model with real data. Interestingly, they observed that only the compound effects of NPIs, put in place at different moments, was enough to bring the reproductive number below one. A similar estimation framework has been used to model the pandemic in Brazil and China~\cite{Cotta_2020}. In this paper however, authors impose a precise functional form of the variation of the transmission rate: $\beta(t)=\beta_0 \exp^{-\gamma (t-t_x)}$ for $ t>t_x$ where $\beta_0$ is the value in absence of any interventions, $\gamma$ is the modulation introduced by NPIs and $t_x$ is the time in which the interventions are put in place. A Bayesian framework is also used in Ref.~\cite{Karnakov_2020} which provides a picture of key epidemic indicators such as $R_t$ for $51$ countries which can be explored via a dashboard~\footnote{\url{https://cse-lab.ethz.ch/coronavirus/}}. In this model the transmission rate is written as $\beta=R_t \mu$ (remember that $\mu$ is the recovery rate). The reproductive number is considered to go from $R_0$ to $k_{int}R_0$ with $k_{int}=(0,1)$ as function of the time due to NPIs implementation. The results indicate that the timing of NPIs is key for an effective suppression of the virus. Ref.~\cite{Loeffler_Wirth_2020} adopts an extension of the SIR archetype, considering time-varying recovery and infection rates, to model the spreading of COVID-19 in $187$ countries. The results, that can be explored via an interactive tool~\footnote{\url{https://apps.health-atlas.de/covid-19-grapher/}}, are obtained fitting the parameters to the real epidemic data.  The comparison between different countries highlights the heterogeneity in the response. For example Iceland was able to drastically reduce $R_t$ since early April thus suppressing the spread with efficacy. Sweden in contrast, opted for a mitigation strategy, and as result the decrease of $R_t$ was much slower. In Ref.~\cite{Calafiore_2020} authors extend the basic SIR model including, among other things, deaths and writing the transmission rate as $\beta(t)=\beta_1 b_1(t)+\beta_2 b_2(t)$ thus as a linear combination of two basic rates. The first weights more in the early phases of the spreading. The second kicks in when the NPIs are in full swing. Similarly they consider the recovery and mortality rates as linear combinations of so called basis functions. A key point of the methodology is defining the shape of such functions. Interestingly the authors adopt a sparse identification techniques applied to the real epidemic data finding a combination of exponentials and polynomials to be the best fit. The model, applied both to national and regional data in Italy, show the effectiveness of the first lockdown (that was implemented in the first half of March 2020) especially in the regions more affected. In Ref.~\cite{Hsiang_2020} authors use a mix of econometric and epidemiological methods to quantify the effect of a NPIs policies on the evolution of the pandemic. The problem formulation is similar to those linking economical policies to economic growth. The first are substitute by NPIs. The second, by the epidemic growth. In the early stage of an epidemic is easy to show that, according to an SIR model, $\log(I_t)-\log(I_{t'})=\exp^{g(t-t')}$ where $g$ is $\beta-\mu$. The authors link this observable to the set of NPIs policies implemented by the country $c$, in subnational unit $i$, in day $t$ as:
\begin{equation}
\log(I_{c,i,t})-\log(I_{c,i,t-1})=\theta_{0,c,i}+\delta_{c,i}+\mu_{c,i,t}+\sum_{p=1}^{p_c}(\theta_{p,c}policy_{p,c,i,t})+\epsilon_{c,i,t}
\end{equation} 
$\theta_{0,c,i}$ is the growth rate in absence of any policy, $\delta_{c,i}$ accounts for weekly patterns in data reporting, $\mu_{c,i,t}$ accounts for variations in reporting methods, which might induce jumps in the data, $p$ describes the policy adopted, $\theta_{p,c}$ the impact of each policy in the growth rate and $\epsilon_{c,i,t}$ is the classic error term. In other words, the growth rate is assumed to be a linear combination of each policy implemented. By fitting the model to data from China, South Korea, Iran, Italy, France and USA, the authors establish that the portfolio of measures implemented in these countries was enough to reduce the transmission and were beneficial in terms of health outcomes. The results also show the heterogeneous impact of the measures across countries. For example the impact of school closures on daily growth rate appear to be higher in Italy than France. Ref.~\cite{Althouse_2020} used online searches and mobility data to estimate the effect of heterogenous implementation of NPIs in the USA. Interestingly, they found that some specific locations (such as churches) saw an increase in attendance locally while, as result of stay home orders, they decrease nationally. This counterintuitive observation shows the effect of heterogeneity in NPIs. Local closure might induce some individuals to travel further for the same activity. The effect of this type of adaptive behavior might increase the number of cases. In fact, the changes in mobility might induce mixing of otherwise separated groups, thus accelerating the spreading. They confirm this intuition with a simple model that divides a population of $N$ individuals in $M$ gatherings. They assume that a fraction $X$ closes at time $t_c$ due to the implementation of NPIs. As result, a fraction of the population $Y$ does not complies with the measures and visit the $1-X$ gatherings still open. In these settings, the dynamics of the disease split at time $t_c$ in two. On one side, the open gatherings are still participating to the spreading of the virus. On the other, closed gatherings can just recover the individuals infected before. The results clearly show that the patchy adoption of NPIs induces worse health outcomes and highlights the negative effects of inconsistent geographical implementation of NPIs.\\
Three articles expanded the SIR modeling framework considering age structure and contact matrices~\cite{Walker_2020,Lei_2020,Scala_2020,Burns_2020}. In Ref.~\cite{Walker_2020} authors study the impact of COVID-19 and the effects of different NPIs strategies in low and middle income countries. Including age brackets and the share of contacts within/across them allows to account explicitly for the demographic distribution and mixing patterns of the population under consideration. This becomes very important when comparing the spreading of COVID-19 in different socio-economic contexts. In fact, low and middle income countries are typically younger populations, thus less at risk of serious forms of COVID-19. However, the social stratification is such that older individuals have higher contacts with other age brackets respect to what happens in higher income settings~\cite{Walker_2020} . Hence, there is smaller fraction of the population at high risk, but that fraction is more intensely connected with younger individuals thus more prone to be exposed to the virus. This interplay is fully explored by the authors which also consider a range of comorbidities, healthcare availability and quality. Furthermore, the authors consider two possible approaches in the management of the pandemic. The first is suppression of the first wave by strict NPIs which pushes $R_t$ below one. The second instead is mitigation, which reduces $R_t$ to values close to one. Suppression drastically limit the burden in the short run, but comes with high socio-economic costs and since it limits the spreading does not form the necessary immunity to avoid multiple peaks. Mitigation, allows for less stringent NPIs but is linked to a much higher number of cases/deaths and pressure to the healthcare system. While younger demographic structures might suggest that mitigation in low and middle income settings might be a more suitable strategy respect to high income contexts, the higher levels of intergenerational mixing, limited resources in terms of healthcare facilities, and comorbidities impose great deal of caution. It is interesting to notice how the response of these group of countries has been put in place early respect to the local circulation of the virus. The paper also point out the challenges and different angles governments have to consider when developing a strategy against COVID-19. Authors in Ref.~\cite{Burns_2020} consider two different strategies to curb the spreading of COVID-19 in schools. In doing so, they consider a stratification of the population (i.e., school) considering different cohorts (i.e., classes), day of infection (to account for different viral loads) and study the effects of NPIs that i) isolate students with fever ii) reduce the school week to four days. After fitting the model to the spreading of ILIs and COVID-19 in school settings, the authors found that, in the context of COVID-19, reducing the school week is more efficient measure.\\
The body of modeling work based on the SEIR scheme touches similar themes to those just mentioned, but due to the larger number of articles, there are also some new interesting angles. In particular, we find papers focusing, among other things, on the effects of face masks~\cite{Li_2020_2,Reiner_2020,Ngonghala_2020,Eikenberry_2020,Iboi_2020}, capturing different aspects of NPIs~\cite{Kain_2020,Childs_2020,Kim_2020,Kim_2020_2,Aldila_2020,Acu_a_Zegarra_2020,Liu_2020_2,Mandal_2020,Hao_2020,Yan_2020,Lemaitre_2020,Mondal_2020,Overton_2020,Weitz_2020_fv2,Anderson_2020_fv2}, studying the problem of optimal control of the pandemic~\cite{Perkins_2020,Ullah_2020,N_raigh_2020,Kantner_2020}, investigating approaches to lift measures put in place in the first wave~\cite{L_pez_2020,Alauddin_2020,Ogden_2020}, model the possible exposure to the virus via droplets or fomites~\cite{Stutt_2020,Zamir_2020}, and consider the impact on the health care infrastructure of COVID-19 and seasonal flu in the winter~\cite{Li_2020_3}. \\
In Ref.~\cite{Reiner_2020} authors present a detail scenario analysis of the winter season in each US state. Interestingly, in their model not only $\beta$ is considered function of time, but the force of infection is modulated by a mixing parameter $\alpha$ defined in such a way that $\lambda(t)=\beta(t)(I_1+I_2)^{\alpha}/N$ where $I_1$ and $I_2$ describe per-symptomatic and symptomatic individuals. Furthermore, they develop a pipeline modeling the link between the transmission rate and cases, deaths as well as a range of covariates such as mobility indicators, population density, smoking rates, masks use via a regression. In doing so, they build a linear regression model to link the implementation of different NPIs (e.g., school closures, stay home orders) on mobility considering data from Google mobility reports~\footnote{\url{https://www.google.com/covid19/mobility/}}, Facebook data for good~\footnote{\url{https://dataforgood.fb.com/docs/covid19/}}, SafeGraph~\footnote{\url{https://www.safegraph.com/data-examples/covid19-shelter-in-place?s=US&d=09-13-2020&t=counties&m=index}} and Descartes laboratories~\footnote{\url{https://www.descarteslabs.com/mobility/}}. Using the observations for all these variables in the first wave, they study the possible scenarios in the winter considering an adaptive strategy where NPIs are reinstated if a certain threshold of mortality rates is reached. The results confirm the effectiveness of NPIs and show that a $95\%$ adoption rate of mask covering could be crucial to mitigate the effects of disease resurgence. Furthermore, they estimate that universal face covering could help save more than hundred thousands of lives. The importance of face masks has been also show in  Refs.~\cite{Li_2020_2,Ngonghala_2020,Eikenberry_2020,Iboi_2020}. In more details, authors in Ref.~\cite{Ngonghala_2020} consider a model featuring susceptible, exposed, infectious, asymptomatic, critical (ICU), and recovered individuals. Furthermore, they split the population separating susceptible and exposed individuals under quarantine, as well as isolated infectious people. In doing so, they also assume that a fraction of the population adopts face masks that have an efficacy modeled as a free parameter. Interestingly, they apply the model to the US and New York state observing that, within the assumptions made, an adoption rate of $70\%$ in New York state and $80\%$ in the USA of mask with efficacy above $70\%$ could lead to an elimination of the virus. Ref.~\cite{Eikenberry_2020} proposes a different methodology by modeling the adoption of face masks splitting the population and compartments (susceptible, exposed, symptomatic infectious, asymptomatic, hospitalized, recovered and dead) in two: those wearing face masks and those not wearing them. Using data and observations from New York and Washington state their model suggests that high adoption rates ($80\%$) even of moderately effective masks ($50\%$) could seriously impact the spreading of COVID-19 by reducing projected deaths by $17\%-45\%$ over a two months period. Authors in Ref.~\cite{Kain_2020} manage to account super-spreading events and thus study the effects of NPIs aimed at \emph{chopping off the tails} of transmission rates. Typically, models that consider this aspect are based on explicit contact networks capturing the heterogeneity of social interactions. Compartmental  models are typically based on the assumption of homogenous mixing which by definition neglect the features of real contact patterns. The authors propose accounting for the heterogeneity in transmission rates by considering the effective (population-wide) effect of having individual transmission rates following a gamma distribution. In doing so, they are able to model the effect of NPIs targeting super-spreading events, by affecting the aforementioned distribution. Furthermore, they assume a time varying $\beta(t)=\beta_0 \beta_m^{\theta}$ where $\theta$ modulates the implementation of NPIs. Such measures induce variations in the transmission rate distribution. The model, fitted to data from several counties across USA, shows that targeted NPIs that cut the right tail of transmission rates distribution are more efficient than social distancing measures at the population level (affecting the average rather than the extreme values). This result is in line with classic results from epidemic spreading on heterogenous networks which show that targeted measures aimed at isolating few central individuals are much more efficient than average measures across the whole system~\cite{barrat2008dynamical,barabasi2016network}. Two papers based on data and observations in South Korea model NPIs considering that susceptible individuals might adopt protective behaviors as function of the number of infected individuals in the population~\cite{Kim_2020,Kim_2020_2}. 
In particular, they adapt the model proposed in Ref.~\cite{perra2011towards} where there are two classes of susceptible: $S$ and $S^F$. The latter describes scared individuals (F stands for feared) of infection that change behaviors to reduce their risks. The transmission rate for these individuals becomes $S^F+I \xrightarrow{\delta \beta} 2I$ where $\delta$ is the protection gained. People join the compartment as function of the number of individuals in self-isolation. This transition is not modeled as mass-action. Thus it is not driven by the fraction of individuals in self-isolation but by their absolute number. In fact, in a context such as South Korea, where the extinction efforts have been quite successful, the news of few cases might induce behavioral changes. Feared individuals might move back as function of the fraction susceptible (not scared) and recovered individuals. The authors fit the model to the epidemic curves at the country and/or regional level to estimate the extent to which behavioral changes, as modeled, have been adopted. They show that, as expected, regions with more cases are compatible with a lower adoption. A similar model has been proposed in Ref.~\cite{Aldila_2020} in the context of Indonesia. Here, susceptible individuals are split in two: unaware and aware of the risks of infection. People move in the aware compartment with a constant rate which models a media effect. Aware individuals are characterized by a lower transmission rate and might relax their behavior spontaneously. The model includes also undetected and detected infectious compartments. People in the first might be tested and transition in the second compartment. The results show how a combination of effective media campaigns and aggressive testing might reduce drastically the burden of the disease. Ref.~\cite{Acu_a_Zegarra_2020} proposes similar approach. In fact, the authors assume that a fraction $q$ of the population adopts NPIs while $1-q$ does not. The split takes place at the time when the measures are put in place. The transmission rate is modeled as linear decreasing function of time, which goes from $\beta_0$ to $\beta<\beta_0$ within a certain time window after the start of NPIs. The authors fit the model to the epidemic data from Mexico City finding that NPIs were critical to contain the first wave and that the data is compatible with high level of adoption rates. Ref.~\cite{Anderson_2020_fv2} proposes a SIER-like model splitting compartments in two accounting for those adopting physical distancing and those that do not. By using a Bayesian approach the model is fitted to the evolution of the first pandemic wave in British Columbia (Canada), few US States and New Zealand. The results for Canada suggest a widely adoption of physical distancing in the Canadian province and indicate that NPIs have reduced by about $80\%$ social contacts bringing the $R_t$ below its critical threshold. Similar results are found for New Zealand. In the USA the results suggest a reduction in New York, Florida, and Washington but not in California. In Ref.~\cite{Hao_2020} authors study the pandemic in Wuhan with the aim to reconstruct its progression as function of the interventions. In doing so, they propose an extension of the classic SEIR model considering individuals traveling in and out the population under study (note that this is modeled as an effective in-out flow) before the cordon sanitarie. The initial ascertainment rate is estimated considering the cases imported from Wuhan to Singapore which, due to aggressive surveillance, is assumed to have detected them all. Furthermore, the infectiousness of unascertained cases is considered to be $55\%$ respect to infectious individuals. The incubation period is set to be $5.2$ days and the pre-symptomatic infectious period $2.3$ days implying a latent period of $2.9$ days. Finally, they consider the transmission and ascertainment rates possibly different in four time periods describing the different phases of the pandemic. The values of these rates are estimated from data using Markov Chain Monte Carlo methods. The results highlight the features of the pandemic: high covertness ($87\%$ of infectious were unascertained) and transmissibility ($R_0=3.54$).\\ 
Several papers studied the implementation of NPIs as an optimal control problem~\cite{N_raigh_2020,Perkins_2020,Ullah_2020,Kantner_2020}. Authors in Ref.~\cite{N_raigh_2020} consider as optimal strategies that reduces the disease burden as well as the socio-economic cost of NPIs. It is interesting to notice how the structure and weight of the different components of the cost function is arbitrary and dependent on the ultimate goals. The authors propose one that puts high values to human lives and assumes that economic activity is proportional to the number of contacts, similarly to the transmission rates. This assumption allows to simply link NPIs to economic costs. Using data and observations from Ireland the authors found that  disease suppression is preferable to mitigation. Ref.~\cite{Kantner_2020} confirms that balancing economic costs of NPIs and avoid the overflow of health care facilities in an optimal mitigation strategy might be a ``tightrope walk close to the stability boundary of the system''. Furthermore,  authors of Ref.~\cite{Perkins_2020} point to the fact that the effectiveness of optimal mitigation strategies depends on the details of key epidemiological parameters and on the exact impact of NPIs. Hence confirming the challenges linked to the definition of optimal mitigation strategies. \\
A set of papers focus on the effects of relaxing NPIs after the first wave~\cite{L_pez_2020,Alauddin_2020,Ogden_2020}. Authors in Ref.~\cite{L_pez_2020} consider an extension of the SEIR model including confined susceptible, quarantined and deceased individuals. They model recovery and transmission rates as function of time fitted to the data from the first wave in different countries (Argentina, Indonesia, Japan, New Zealand, Spain, and USA). They then explore different reopening scenarios finding that lockdowns of two months would be ideal to prevent disease resurgence and that effective social distancing as well as other NPIs adopted by a larger fraction of the population could avoid the need for other lockdowns. Authors in Ref.~\cite{Alauddin_2020}  focus on data from Canada, Germany and Italy and used a neural network approach to fit an extended SEIR model that accounts for asymptomatic, symptomatic, hospitalized, and deceased individuals to real epidemic data. Their results suggest that reopening strategies should be gradual and stage-wise to avoid quick disease resurgence.\\
Two theoretical approaches consider explicitly the presence of the virus in the compartmental structure~\cite{Stutt_2020,Zamir_2020}. In particular, authors of Ref.~\cite{Stutt_2020} split a population in two: those wearing masks and those that don't. Then they consider that asymptomatic and symptomatic individuals create the inoculum in form of droplets. These are described by a compartment $D$ and move with a certain rate into the fomite compartment $F$ capturing the virus presence in contaminated surfaces. Finally, fomites decay. Interestingly, the transmission rates are driven by the interactions of susceptible with the $D$ and $F$ compartments. This allows to model more explicitly the role of face masks which offer some protection from droplets but not from fomites. Ref.~\cite{Li_2020_3} focuses on the possible impact of the seasonal flu in the management of the COVID-19 pandemic. In fact, flu symptoms might be confused with those of COVID-19. As result resources, key for the fights against the current pandemic, such as tests, might be diverted towards the flu. The authors propose to capture this possible negative interaction by extending the classic SEIR model to account for asymptomatic, symptomatic, as well as susceptible, exposed, and infected quarantined. Furthermore, they consider the possibility of having individuals with the flu in the pool of people to quarantine and subject to COVID-19 tests. They consider a transmission rate explicitly function of contacts that due to NPIs go from $c_0$ to a $c_b$ exponentially as function of time. Their results show how an uptake of the flu vaccine could help the depletion of resources necessary to contain COVID-19.\\
Several articles adopted SEIR-like models with age-structures to study the impact of different NPIs in countries~\cite{Salje_2020,Prem_2020,Davies_2020,Min_2020,Barbarossa_2020,Di_Domenico_2020}, reopening scenarios~\cite{Matrajt_2020,Gosc__2020}, investigate herd immunity~\cite{Brett_2020,Britton_2020}, and consider particular contexts such as African countries~\cite{Evans_2020,van_Zandvoort_2020}.\\
Two articles modeled the impact of the NPIs in the first wave in France and studied different reopening scenarios~\cite{Salje_2020,Di_Domenico_2020}. In particular, Ref.~\cite{Salje_2020} considers an extended SEIR model with contact matrices and age structure fitting the transmission rate $\beta$ before and during the lockdown from data. They studied different contact matrices during the lockdown (not estimating them from data). The baseline for the lockdown considers no contacts in school, a reduction of $80\%$ of work contacts, and $90\%$ in other locations. The results suggest that about $3\%$ of infected individuals was hospitalized (by May 11), and an average $0.5\%$ died across the board. The IFR for individual $80+$ was found to be $8.3\%$ while $0.001\%$ for those $20$ or younger. Furthermore, the results indicated that $R_t$ was reduced from $2.9$ to $0.67$ and $5.3\%$ of the population is considered to be infected by mid May. Ref.~\cite{Di_Domenico_2020} considers a similar model (though with a broader range of compartments to account for asymptomatic, paucy-symptomatic, mild, and severe infectious) for the I\^le de France, region where the capital is located. They consider a range of NPIs such as remote working, school closure, case isolation, banning of social events, and senior isolation that study them as single or as portfolio of interventions. Each NPIs affects the contact matrices. The lockdown is for example modeled considering a complete removal of contacts in school, remote working at $70\%$, $90\%$ of senior isolation, closure of all non essential activities but no case isolation. Fitting the model via a maximum likelihood approach to hospital admission data, the authors estimate an $R_0=3.18$ that was effectively reduced to $0.68$ by the lockdown. Furthermore, estimates for the IFR are in between $0.7-1.2\%$. These figures are a bit higher than those mentioned above. It is important to stress how this paper focuses on one of the most afflicted region of France while the first considered the entire country thus averaging with areas less affected. A similar approach, though with a simpler compartmental structure and no fit with real data but a scenario analysis informed by observations from Wuhan, has been proposed also in Ref.~\cite{Prem_2020}. The results, published at the end of March, indicate the impact of school closures and remote working on the mixing patterns, thus on the disease and the risks of lifting the restrictions too early.  Authors of Ref.~\cite{Davies_2020} consider instead the case of UK and study the impact of several NPIs such as school closure, social distancing, shielding, self-isolation and different portfolios of them on the burden of the disease. They use a simpler age-stratified SEIR model for each of the $187$ counties in the UK and model the variation of the contact matrices induced by NPIs. Interestingly, the results show that moderate levels of interventions (i.e., school closure, self-isolation, or shielding) even in combination would not be enough to avoid exceeding healthcare capabilities. The authors suggest that only more strict measures, such as lockdown, implemented for shorter periods in a background of general social distancing might be enough to avoid crushing the health system.\\
 Authors of Ref.~\cite{Gosc__2020} model different reopening scenarios after the first lockdown in London UK. In doing so, they extend the SEIR model to account for age-structure, contact matrices and other compartments such as asymptomatic, symptomatic and isolated infectious individuals. They study different scenarios: city wide extended lockdown, universal testing in a background of less stringent social distancing, shielding people $60+$, universal testing and face coverings, universal testing isolation of cases and their contacts and face coverings during the lockdown. The model confirms also in this context how strategies that combine a range of NPIs are the most efficient and offer alternatives to extended lockdowns. \\
Two articles tackle the issue of herd immunity. Ref.~\cite{Brett_2020} adopts an SEIR model with age structure and contact matrices fitted to the UK to investigate the feasibility of dynamic mitigation strategies that reduce the reproductive number in such a way to avoid overflowing the healthcare system. In other words, when the infections are decreasing some measures might be lifted until certain warning levels are reached. This implies the ability to modify the contact matrices as function of the disease progression. In doing so, hypothetically, the impact of the virus can be mitigated with less stringent measures. The authors show how, even in simple modeling settings, such dynamic mitigation strategies require an unpractical balancing of several poorly defined indicators. It is important to notice how mitigation and not suppression was the initial strategy proposed by the government in mid March, before a drastic u-turn towards disease suppression induced by a tsunami of hospitalizations. Authors in Ref.~\cite{Britton_2020} investigate the herd immunity threshold. This is the fraction of the population with immunity (either acquired through infection or vaccination) needed to avoid disease resurgence. In a homogeneously mixed population this fraction can be computed as $1-1/R_0$. Assuming a $R_0$ in the range of $2.5$ the herd immunity would be reached at $60\%$. The authors show that this threshold can be as low as $42\%$ considering heterogeneity in contact matrices and infection rates across age-brackets.\\
Finally, two articles studied NPIs in the context of African countries~\cite{Evans_2020,van_Zandvoort_2020}. Ref.~\cite{Evans_2020} investigates possible explanations behind the relative lack of COVID-19 cases in sub-Saharan Africa. Using data and observations in Madagascar, which had a moderate peak in mid July, they considered three competing explanations: i) low case detection ii) epidemiological differences due to climate, younger population, lower population density and transportation infrastructure in rural areas iii) effective NPIs. Interestingly, the results, obtained via an age-stratified SEIR model, suggest that the epidemiological markers of COVID-19 are consistent with those observed in other countries and that the observed pattern is compatible with a low case detection, late case importation, and effective NPIs. Authors in Ref.~\cite{van_Zandvoort_2020} study the evolution of COVID-19 and the effects of different types of NPIs in three African countries selected from youngest to oldest average age: Niger, Nigeria, Mauritius. In doing so, they adopt an age-stratified SEIR model modifying some of the parameters to better match the African context. In particular, they shifted case severity and IFR to younger ages. The results show how general physical distancing and self-isolation of symptomatic individuals, although important, might not be enough to contain the virus without lockdown levels. Furthermore, shielding of high risk people is key to avoid overflow of healthcare facilities.\\
Finally, one article investigates the possibility of waning immunity~\cite{Hoffman_2020}. The compartmental structure follows the SEIRS archetype where recovered might loose immunity and move back to the susceptible compartment. The authors consider also the effects of NPIs by using time-varying transmission and mortality rates. Furthermore they introduce a compartment $P$ to account for individuals that protect themselves from infection by adopting and complying with NPIs. They assume that susceptible move to this compartment with a time-varying rate which is fitted and move out with another time-varying parameter. The model is applied and fitted to the data from New York state. The results confirm the efficacy of the NPIs put in place and warn about the possibility of endemic infections in case of wining immunity.

\subsection{Metapopulation models}

Metapopulation models consider a number of sub-populations connected by means of human mobility. In doing so, the disease dynamics inside each patch (sub-population) follow a compartmental model like those described in the previous section, however sub-populations are now coupled by mobility flows. Thus, these types of models allow to capture the geo-spatial spreading of a disease which is defined by the dynamics within and across patches. In particular, metapopulation models can be represented as networks in which nodes describe sub-populations and links the mobility flows between them. The spreading of a virus is locally driven by the $R_0$ which is function of the reaction dynamics within each subpopulation. However, the global spreading is also dependent on the features of the mobility flows. This introduces another threshold typically called invasion threshold~\cite{colizza2008epidemic}. Considering explicitly the coupling between subpopulations allows to tackle broader set of questions respect to those amenable in single populations. What are the effects of national and international travel bans? When and from where the first cases have been imported? Local lockdowns are a good alternative respect to nation-wide measures? What is the chance of international importation of cases? What are the most affected sub-population and why? Furthermore, they allow to use importation and seeding events to estimate initially unknown epidemic parameters such as $R_0$ or the number of unascertained cases. These capabilities come with some costs. Access to mobility data and increase of computational complexity of the model are the two main aspects to consider. However, as we already saw data capturing historic as well as in near real time human mobility are readily available at different geographical scales and the use of compartmental models inside each subpopulation allows to scale up the models to hundreds of countries. On this point, it is important to mention how the definition of subpopulations is highly variable according to the data available or the goal of the research. They can describe neighborhoods of a city, cities, regions, countries etc. \\
The papers that used a metapopulation approach in the sample under investigation here cover several important angles such as the spatial heterogeneity of the pandemic due to local differences in the implementation of NPIs and/or features of population distributions~\cite{pei2020differential,Chang_2020,Rader_2020}, the first pandemic wave and the efficacy of the measures put in place~\cite{Medrek_2021,Kheirallah_2020,Lai_2020,Ge_2020,Aleta_2020,Arenas_2020_fv2}, and reopening scenarios~\cite{Ruktanonchai_2020,Bertuzzo_2020,Karatayev_2020,Carli_2020}. Let's dive in some details.\\
Ref.~\cite{Chang_2020} proposes a metapopulation model in which subpopulations are the neighborhoods of the ten largest metropolitan areas in the USA. The $98$ million people in these can interact within each node and/or visit a point of interest (POI) which might be a bar, hotel, gym etc.. The system is modeled as a bi-partite network in which the two types of nodes are neighborhoods and POI. The authors built the model leveraging hourly data extracted from mobile phone of users in the area (via SafeGraph). The data is used to construct the mobility flows and coupling in the system. The resolution of the data allows to factor in behavioral changes induced by the spreading of the virus. The model is fitted considering real data from the area under investigation. The model includes only three free parameters: i) transmission rates at POIs, ii) transmission rates in the neighborhoods and iii) the initial proportion of exposed individuals. Rhe authors assume that disease dynamic follows a SEIR compartmental model. Interestingly, the transition from S to E in each neighborhood $c_i$ is modeled as follows:
\begin{equation}
N\left ( S_{c_i}(t) \rightarrow E_{c_i}(t),t \right ) \sim Pois \left ( \frac{S_{c_i}(t)}{N_{c_i}}\sum_{j=1}^{n}\lambda_{p_j}(t)\omega_{ij}(t) \right )+Binom \left ( S_{c_i}(t),\lambda_{c_i}(t) \right )
\end{equation} 
where  $\lambda_{p_j}(t)$ is the rate of infection in POI $j$, $\omega_{ij}(t)$ the number of individuals from neighborhood $c_i$ visiting POI $j$ at time t, and  $\lambda_{c_i}(t)$ the infection rate in that neighborhood at time $t$. Note how the authors divided the infection dynamics explicitly considering events taking place in POIs and within each census area. Notably, the rate of infection in each POI is written as $\lambda_{p_j}(t)=\beta_{p_j}(t)\frac{I_{p_j}(t)}{V_{p_j}(t)}$ where the denominator is the total number of visitors in $j$ at each time and as usual $\beta$ is the transmission rate. The authors provide an expression for this rate considering it a function of the disease (which is one of the parameter fitted), and the contacts within each POIs. The latter is modeled accounting for the physical area and the average time people spend there. The rate of infection in each neighborhood is modeled similarly via the force of infection. The results indicate that a small set of POIs are responsible for the large majority of infections. Full served restaurants and hotels are on the top of the list. Furthermore, the model suggests that reducing the time spent in POIs is more efficient that reducing the mobility across the board. The authors find also heterogeneity in terms of infection rates in disadvantaged socio-economic groups which are not able to reduce their mobility as others. In Ref.~\cite{pei2020differential} authors built a metapopulation model for $3142$ US counties. The mobility flows up to mid March are taken from the official statistics while those after are estimated from inter county mobility towards POIs from SafeGraph. In fact, historical data is far from representative when top-down and bottom-up behavioral changes started to take off. Interestingly, the infection dynamics are split between day and night time. The model, fitted from the epidemic data of the counties more affected, shows a large heterogeneity in terms of the reduction of the reproductive number across counties. Among several others, the different extent of NPIs implementation is an explaining factor. Ref.~\cite{Rader_2020} uses detailed cases data from China to study the link between the shape of the epidemic curve and spatial features of cities. Interestingly, they found that the \emph{peakedness} of the epidemic is affected by population aggregation and heterogeneity in such as way that crowded cities are affected by higher attack rates and longer epidemics. The observations can be explained by considering the hierarchical structure of social contacts which are stratified in households, neighborhoods, cities, prefectures etc. Densely populated areas have higher rates of connections across households hence, given a level of NPIs aimed to cut such contacts, more dense areas will retain more connections across subpopulations which will result in higher attack rates and longer epidemics. Authors of Ref.~\cite{Lai_2020} develop a metapopulation model at the level of prefectures in China. In doing so, they use a range of datasets from Baidu to estimate mobility across subpopulations and contacts within them. The goal is to quantify the effects of NPIs in the country. Interestingly, the model suggests that without interventions the number of cases would have been 67-fold. Early detection and isolation of cases is more efficient than overall travel restrictions and social distancing. Similar results, data and models in the context of China have been obtained also in Refs.~\cite{Ge_2020,Aleta_2020}. Authors of Ref.~\cite{Kheirallah_2020} use the Global Epidemic and Mobility model~\footnote{\url{http://www.gleamviz.org}} to simulate the spreading of COVID-19 and quantify the impact of NPIs in Jordan. Details of the model are reported below. Ref.~\cite{Medrek_2021} studies the case of France, Poland and Spain dividing the countries in cells and adopting a cellular automata to model the transitions within and the movements between cells. By considering time-varying transmission and mobility rate the authors confirm the efficacy of the NPIs put in place in the countries, though they estimate a very large $R_0=(8,10.5)$. \\
Ref.~\cite{Ruktanonchai_2020} tackles the issue of reopening strategies after the first wave in Europe. In doing so, they augment the mobility data reports from Google via much higher resolution mobility flows obtained considering CDRs (call details records) from Vodafone in Spain and Italy. In particular, they develop a linear model to rescale the data from Google considering CDRs to better capture key epidemiological behaviors. Using such augmentation process across all countries in Europe they built a metapopulation model at the NUTS3 resolution. The results indicate that a coordinated relaxation strategy is more beneficial than an uncoordinated approach. Interestingly, coordinated intermittent lockdowns are found to reduce by half the periods of stricter restrictions needed in case of isolated implementations. Ref.~\cite{Bertuzzo_2020} investigates the reopening scenario in Italy. In doing so, they built a metapopulation model at the level of provinces. The mobility flows are taken from official statistics and are modified according to the estimate obtained and shared in Ref.~\cite{Pepe_2020_rf3}. The dataset will be described in more details below. Accounting for the NPIs put in place in Italy, they consider three values of $\beta$. The first two are constant across the country, the last is allowed to be region dependent. Using a hierarchical framework they fit the model to the epidemic data and then consider different reopening scenarios. The results show how an increase of $40\%$ in the effective rate, after the lockdown, would be enough to induce a disease resurgence and that detecting and isolating around $5\%$ of the daily exposed via aggressive testing would be enough to avoid a rebound of the disease even relaxing NPIs. Ref.~\cite{Karatayev_2020} uses a metapopulation approach considering all counties of Ontario Canada. Mobility across subpopulation is estimated via a survey that mapped commuters in region. They consider different transmission rates according to the health status (asymptomatic, pre-symptomatic, symptomatic) and allow a variability across subpopulations. The model is fitted to the data and used to investigate different reopening approaches after the first wave. Interestingly, the results show that local reopening are better. As cost function they use the person days of closure and number of cases. They also observe that local measures might be better in the early stages of the epidemic if testing rates are high and the threshold that triggers them is set to be low. Authors of Ref.~\cite{Carli_2020} propose a theoretical framework in which the transmission within, and reduction of mobility rates across, subpopulations of a metapopulation network are obtained solving an optimal control problem where a possible goal could be to avoid overflowing the healthcare system. \\
A subsets of papers use a metapopulation approach accounting also for the age-stratification and contact matrices within each subpopulation. In this group we find papers that evaluate the risks of importation of cases from China in the early phases of the pandemic~\cite{Chinazzi_2020}, model the first wave~\cite{Lau_2020,Sj_din_2020,Gozzi_2020,Tagliazucchi_2020,Arenas_2020_fv2} and investigate reopening scenarios~\cite{R_st_2020,Karatayev_2020}.\\
Ref.~\cite{Chinazzi_2020} is an emblematic example of the types of questions a metapopulation model can answer during the very uncertain initial phases of a pandemic. The authors used the Global Epidemic and Mobility model~\cite{gleam2010,balcan2009h1n1,y2018charting,balcan2009multiscale} to understand the risk of importations of cases from China, estimate key epidemiological parameters, and provide a more precise picture of the epidemic in the epicenter. The model is based on three layers. The first describes the population distribution. The world is divided into over $3200$ subpopulations constructed using a Voronoi tessellation of the Earth's surface. Subpopulations are centered around major transportation hubs (e.g., airports). They consist of cells with a resolution of $15x15$ arc minutes (approximately $25 x 25$ Km). The population is estimated via high resolution data~\cite{sedac}. The model accounts for other features of individual subpopulations, such as age specific contact patterns, health infrastructure if available~\cite{mistry2020inferring}. The second layer describes the mobility flows across subpopulations. These are split in two: short-range (commuting patterns) and long-range (air transportation). The final layer is the epidemic layer which is a compartmental model applied to each subpopulation. The authors used an age-stratified SEIR model and adopted an Approximated Bayesian Computation to estimate the posterior distribution of $R_0$ given the observations of international case importation from China. In other words, the authors used the cases that were popping out around the world in the very early phase of the pandemic to define $R_0$ compatible with such observations. After fitting the model, the authors were also able to estimate the effects of NPIs on the pandemic. The results indicate a $R_0=2.57$ (CI $2.37$-$2.78$) and suggest that the travel ban from/to Wuhan delayed the unfolding of the epidemic in China of only $3-5$ days. More significant were the estimated effects internationally, though the model suggests an intense seeding that took place from Wuhan and other Chinese cities after the travel ban. The results also indicate that features of the disease and the air transportation network are such that a strong travel ban from/to China is not effective without sustained social distancing. It is important to note how a first draft of this paper was posted on medrxiv in mid of February 2020. At the moment the discussion outside China revolved around the possibility of containment, and on the risk of a global pandemic. The metapopulation approach and the real data used to fed it allowed the authors to provide a picture of the spatio-temporal diffusion and diffusion potential of the virus which at that time was still highly debated. A similar model, without long-range mobility, has been proposed to describe the first wave and the impact of NPIs in the metropolitan area of Santiago de Chile~\cite{Gozzi_2020}. The model considers as subpopulation the municipalities of the capital of Chile and adopts data from mobile phones to estimate the mobility flows as well as their changes induced by NPIs. The model suggests that only the compound effects of subsequent measures was enough to suppress the epidemic. Furthermore, the results confirm how even in this country individuals from municipalities characterized by low Human Development Index (HDI) were not able to reduce their mobility as those living in high HDI locations. As result of these socio-economic factors poor areas registered higher cases and deaths. Authors of Ref.~\cite{Sj_din_2020} built an age-structured metapopulation model for Sweden at the municipality level. In doing so, they adopt the radiation model to estimate the mobility between subpopulations~\cite{simini2012universal} before and during the pandemic. They investigate different scenarios considering moderate to strong physical distancing. The results indicate that the scenario closer to the observation is the one with moderated physical distancing of people younger than 60, very strong distancing of those older than 60 as well as a general awareness and compliance to home isolation if symptomatic. Hence, this confirms how despite the lack of strict top-down NPIs respect to many other countries the Swedish population did change behavior. Unfortunately the health outcome was still much worse than other nordic countries which took a suppression rather than a mitigation approach. Ref.~\cite{Lau_2020} proposes another age-stratified metapopulation model for Georgia state (USA). The authors consider the state divided in small cells ($100$m $x$ $100$m) which are the nodes of the metapopulation system. Mobility is modeled considering a density function. The reduction of normal mobility due COVID-19 is estimated using the Facebook data. While typically metapopulation models consider mobility at the aggregated level of compartment (e.g., how many exposed from node $i$ travel to node $j$), authors here consider individual based travels and infections. In doing so, they are able to study the role of super-spreaders events. The model, fitted to the epidemic data via Bayesian inference, indicates that $2\%$ of cases are responsible for $20\%$ of infections and that the reproductive number was reduced below the critical value after two weeks of the shelter in place order. Also, infected non-elderly might be about three times more infectious than elderly thus driving the super-spreading events. Authors in Ref.~\cite{Arenas_2020_fv2} introduce an age-stratified metapopulation model for Spain. By using an Markov chain approach the authors are able to derive analytical expressions for the reproductive number as function of different types of NPIs. Furthermore, they show how the model can be calibrated to reproduce the observed evolution of the pandemic in the country.

\subsection{Statistical models}

The approaches described above model the spreading of infectious diseases mechanistically. Real epidemiological data is then used to constrain the phase space of parameters. Statistical models use assumptions about the progression of the disease to write down mathematical expression linking, for example, the number of cases at time $t$ with those a time $t-1$. Statistical inference approaches are then used to determine the parameters modulating such relationships. A classic example is the statistical estimation of $R_0$ and/or $R_t$. While in compartmental models $R_0$ is estimated looking at the difference between observed cases/deaths and those simulated by the model, statistical approach target directly the data estimating the parameters necessary to match the expected mathematical evolution of cases/deaths. \\
These approaches have been applied to the COVID-19 pandemic to gather information about the key epidemiological parameters and quantify the effects of NPIs by looking at deviations from expected and/or initial trends~\cite{Flaxman_2020,Manevski_2020,Al_Wahaibi_2020,Bo_2021,Basu_2020,Cowling_2020,Quilty_2020,Leung_2020,Candido_2020,Bryant_2020,Morley_2020_xx1,Braunereabd9338}. Let's dive into some details.\\
Ref.~\cite{Flaxman_2020} uses a discrete renewal process to estimate $R_t$ in $11$ countries in Europe (Austria, Belgium, Denmark, France, Germany, Italy, Norway, Spain, Sweden, Switzerland, the UK) and the effects of NPIs of its evolution. The authors assume that 
the number of infections in a country $m$ at time $t$ can be written as:
\begin{equation}
c_{t,m}= \left (1-\frac{\sum_{i=1}^{t-1}c_{i,m}}{N_m} \right)R_{t,m}\sum_{\tau=0}^{t-1}c_{\tau,m}g_{t-\tau}
\end{equation}
where the first term on the r.h.s. estimates the fraction of the population still available for infection (not already immune, or removed), $R_{t,m}$ is the reproductive number regulating the growth of infections, and $g$ is the generation distribution describing the time between being infected and infecting. Hence, the new number of infected individuals is function of how many have been infected in the past modulated by the generation distribution and the average number of infections each of these is able to produce. The authors consider six types of NPIs: school closures, self-isolation of symptomatic, banning of public events, partial or complete lockdown, social distancing, and any other government intervention. They then assume the effect of such interventions to be multiplicative and such that:
\begin{equation}
R_{t,m}=R_{0,m} \exp^{ \left (-\sum_{k}\alpha_k \delta_{k,t,m}-\epsilon_m \delta_{t,m}  \right)}
\end{equation}
where $delta_{k,t,m}$ indicates whether each intervention $k$ is in place, $\alpha_k$ is the effect of each intervention, $\epsilon_m$ is a country specific error term on the last intervention put in place by each country $\delta_{t,m}$. The parameters are fitted for all $11$ countries in a single hierarchical model. The results show that the portfolio of measures put in place was enough to suppress the epidemic. Lockdowns are found to be the most significant NPIs in terms of variations to the $R_t$. Furthermore, the models suggest that across all those countries a $3-4\%$ of the population has been infected up to early May. The same model, extended to consider more data such as hospitalizations and ICU patients, has been proposed for Slovenia~\cite{Manevski_2020}. The results suggest that the country has the smallest attack rate in Europe ($0.3\%$ of infected population)  and point to an IFR of $1.56\%$. A similar model has been also applied to the case of Brazil~\cite{Candido_2020}. The estimation of the $R_t$ point to the fact that the measures put in place were not enough to suppress the spreading, but were able to mitigate it. The authors consider also genomic data to shed light on the start of the local epidemic. The results point to the fact that the majority of importations in the genomic sample are linked to Europe and the local spreading was in place before the implementation of travels bans. A model with the same approach has been also proposed in Ref.~\cite{Braunereabd9338} to infer the effects of NPIs in $41$ countries in Europe as well as in other continents. Interestingly the results show how across the board reduction of gatherings to ten or less, school closure and remote working had the largest impact on the reproductive number. A model based on a renewal process has been also proposed in Ref.~\cite{Bryant_2020}. Authors define the reproductive number at time $t$ for country $m$ as:
\begin{equation}
R_{t,m}=R_{0,m}\exp^{-\sum_{k}\alpha_k \delta_{k,t,m}}
\end{equation}
where here the $\delta_{k,t,m}$ describes the mobility index linked to type of places $k$ (e.g., retail and recreation, grocery and pharmacy, public transport, workplaces, home) at time $t$ for country $m$ estimated from the Google mobility reports, and $\alpha_k$ is the effects of movements and interactions in each type of place on the disease spreading. The model is applied to $11$ European countries and confirms the importance of NPIs in the mitigation and suppression efforts. Interestingly, the model suggests that most significant variations in mobility are linked to groceries and pharmacies. However, variations of mobility to all categories of place are shown to be highly correlated with disease burden (measured in terms of deaths) with a delay of one month. Authors in Refs.~\cite{Al_Wahaibi_2020,Bo_2021,Cowling_2020,Leung_2020} adopt and/or extend the approach developed in~\cite{cori2013new,thompson2019improved} to estimate $R_t$. This method is readily available in Python and R and is called \emph{EpiEstim}. In short, the total number of infectious at time $t$ can be written as $\Lambda_t (g_s)=\sum_{s=1}^{t}I_{t-s}g_{s}$ where, as above, $g$ is the discrete serial distribution. The expected number of infected can be written as $E(I_t|I_0,I_1,\dots,I_{t-1},g_s,R_t)=R_t \Lambda_t(g_s)$. Without going to the details, the approach assumes that the number of case is extracted from a Poisson distribution providing an expression linking the probability of observing a number of cases in a time window given the $R_t$. This allows to apply Bayesian inference from data having as target $R_t$. Ref.~\cite{Bo_2021} applies the approach to data from $190$ countries and estimates the effects of different NPIs. Interestingly they found social distancing to be the most efficient in reducing the reproductive number followed by mandatory masks. Furthermore, they confirm that a portfolio of measures is required to suppress the spreading. Similar results have been reported in Ref.~\cite{Cowling_2020} in the context of Hong Kong. Beside estimating the effects of NPIs the authors report also the result of three surveys which show high level of awareness and adoption of NPIs such as face mask and avoid crowded places. Finally, EpiEstim has been used also to model the impact of NPIs in several cities in China~\cite{Leung_2020}. The results highlight the effectiveness of the aggressive NPIs strategy put in place and point to a very different confirmed case fatality ratio in Hubei and outside. In the epicenter of the pandemic it appears to be five times higher than outside. Authors in Ref.~\cite{Zheng_2020} adopt a state transfer matrix model, which similarly to compartmental models, to describe the different states of individuals that got in contact with the virus. Hence, this model could be classified as a compartmental model. However as it does not match directly any archetype I opted to add it to this category. The model is fitted to data from China, Iran, Italy and South Korea with the aim to characterize the impact of NPIs in those countries and project possible attack rates.

\subsection{Agent-based models}

Agent-based models push the mechanistic approach to its limit. In fact, they simulate the spreading of disease reproducing, to different extents, key aspects of the society where the disease is spreading. In particular, classic epidemiological agent-based models are based on a synthetic population that is built considering empirical socio-demographic features such as households sizes and compositions. A key aspect for the spreading of diseases are social contacts which are built accounting for socio-economic activities such as school, work places, and other types of activities. The nature of agent-based models requires a much wider range of data necessary to build the synthetic population and its interactions. The computational costs associated are also much more significative. A large body of literature has been devote to study the effects of considering real social contact networks on the spreading of infectious diseases~\cite{barrat2008dynamical,barabasi2016network}. These agent-based models do not typically account for all societal details such as households, schools and work places. As such they are not typically used to make accurate predictions but rather to develop a better understanding of how the features of real contact networks might affect the unfolding of a virus. They have been used to quantify the effects of contact tracing, isolation, vaccination campaigns among other things. \\
In the sample of papers under review here, we find full fledged agent-based models used to model the COVID-19 and the effects of NPIs at the levels of countries, regions, or cities~\cite{hoertel2020stochastic,Aleta_2020_2,Aleta2020.12.15.20248273,Wilder_2020,Yang_2020,Ogden_2020}. We also find more theoretical approaches modeling particular aspects of NPIs such as different strategies for isolation~\cite{Silva_2020,Kwon_2020,Bouchnita_2020}.\\
Authors in Ref.~\cite{hoertel2020stochastic} propose an agent-based model for France with the aim to understand the efficiency of different NPIs in the reopening phases after the first wave. This is a text book example of this category of models. The first important point to mention is the complexity of the approach which is exemplified by the number of parameters: $194$ in total. The majority ($140$) are to describe the socio-demographic features of the french population, $33$ are to model the contact networks, $21$ to describe the features of the virus. $192$ were obtained either from the literature or from sensible assumptions. The two remaining, which describe the contamination risk per minute per $m^2$ of contact and the fraction of undiagnosed cases, have been fitted to mortality data.  In the model households are assigned to a square grid. These are built considering a stratification which include age and gender to reproduce the french population. Furthermore, comorbidities (e.g., diabetes, obesity, different chronic diseases) are also taken into account. The contact network is built taking into consideration several types of interactions taking place at home, in school, work, as well as during other locations such as groceries. The construction of such synthetic population allows to model with high resolution several NPIs such as reopening of particular activities or isolation approaches. The results, which where published in July, paint quite a grim picture which unfortunately has been later confirmed: measures in place after the strict lockdown is lifted are not enough to avoid the overwhelming of the healthcare system. A similar approach has been proposed in Ref.~\cite{Aleta_2020_2}. The authors use high resolution data describing the movements and potential interactions of people in Boston. They build a multilayer synthetic population which model the socio-demographic features of Bostonians, their movements and possible interactions in different locations (i.e., households, workplaces, schools etc). The authors use the model to investigate the impact of different reopening scenarios. The results suggest that aggressive testing and household isolation could allow to reopen economic activities while keeping the health care facilities protected from overflow of cases. Ref.~\cite{Aleta2020.12.15.20248273} presents an agent-based model for the metropolitan areas of New York and Seattle. They model is informed with mobile phone data collected from about $700,000$ users by Cuebiq. The authors, similar to what done for Boston, built a weighted layered network capturing interactions in the community, schools, work places, and households. Furthermore, they use foursquare data to identify POIs. The results show the impact of NPIs in the two cities, provide information about the risk of infection according to the type of locations visited and estimate the size of transmission chains. In the same line with what reported above, the results suggest that the large majority of infections ($80\%$) are due to a small minority of individuals ($27\%$). Despite the big role played by large social event, the authors found that work places, restaurants, and grocery stores are the main drivers of the observed patterns. Authors of Ref.~\cite{Wilder_2020} propose an agent-based model to study the spreading of COVID-19 in Hubei, Lombardy (Italy), and New York City. The model nicely show how the details can be adapted to the goal and data available. The authors built synthetic populations form by $n$ individuals stratified for age, comorbidities and assigned to a household. Contacts are divided in two category: inside and outside households. Contacts in different contexts than home are modeled via contact matrices. The free parameters are fitted to data using a Bayesian framework. The results show how the efficacy of NPIs aimed at limiting contacts across age brackets vary according to the location under study. This suggests how a one fit all strategy might not be optimal and measures should be tailored to the specific socio-demographic features of each populations. Authors in Ref.~\cite{Yang_2020} use a network-based model where a representative sample of different cities interacts via a contact network. The authors model interactions within and across cities. These are modeled as clustered networks. Links inside each community are extracted from a configuration model. Links across communities assume an homogenous random graph. Connections between cities are built considering real data from Baidu. Since the units of the model are individuals, the number of travelers from city $i$ and $j$ are assigned randomly to nodes in $i$. These form a number of connections, which is the average in the system, with random nodes in $j$.  The population is scaled down to reduce the computational costs. This model does not consider explicitly the stratification of individuals in households, school or work locations. The results obtained confirm the importance of NPIs in controlling the pandemic in China. Authors in Ref.~\cite{Silva_2020} propose an agent-based framework where beside individuals, agents describe houses, business, government, and the health care system. Each type of agent is assigned with a set of features and enabled to take a set of actions. The framework allows for the implementation of a range of NPIs and measure the impact in terms of disease and economic burden.

\section{Surveys}

\begin{figure}[t]
  \centering
   \includegraphics[scale=0.25]{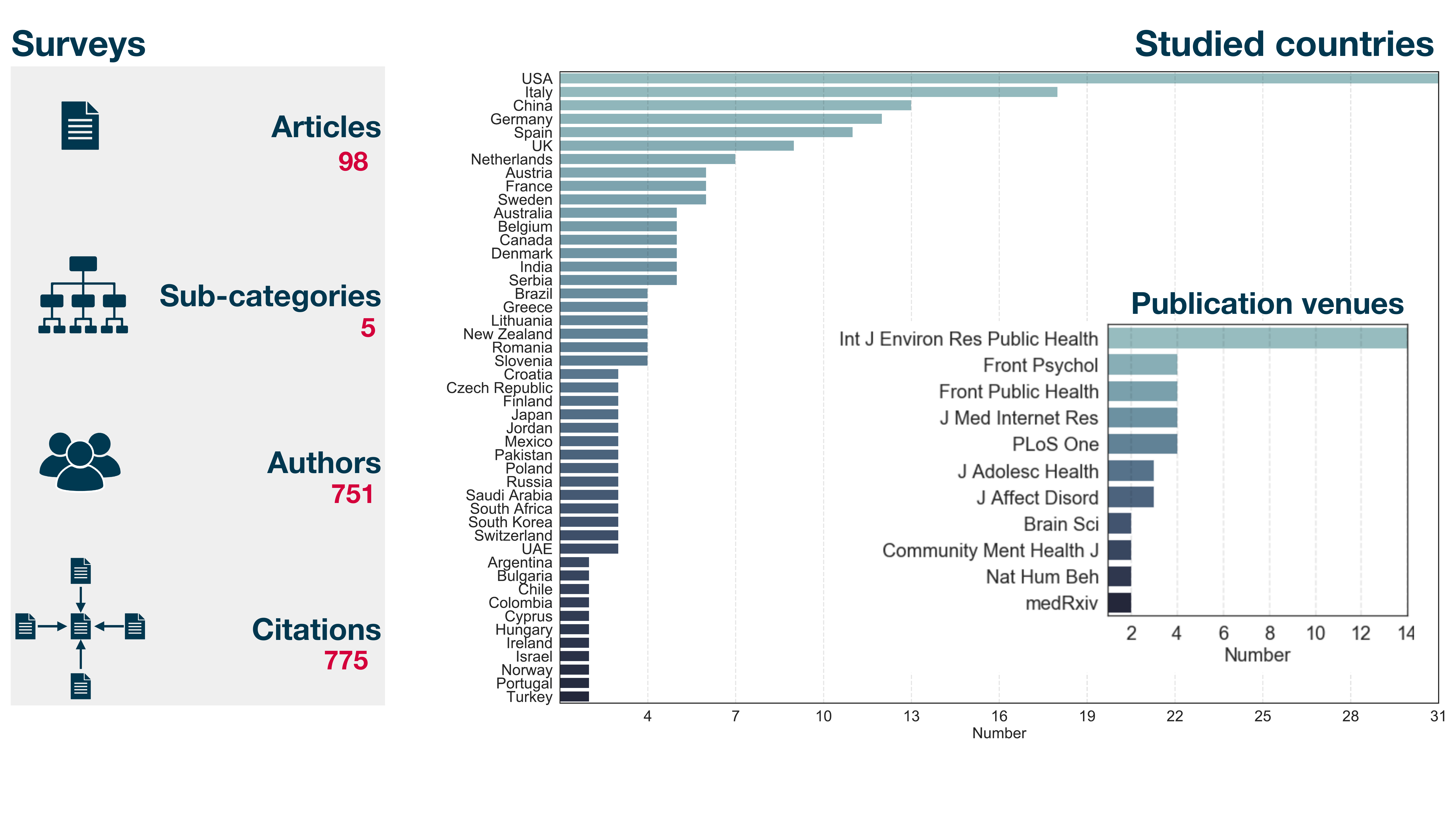}
  \caption{Total number of articles, sub-categories, authors, and citations of the papers in this category on the left. Note that these numbers are estimated from Semantic Scholar. On the right (main panel) histogram describing the number of countries subject of study in this category. On the right (in-set) most represented publication venues of the category. To improve visibility I am showing only countries featured at least in two studied and journals featuring at least two articles}
  \label{fig:fig3}
\end{figure}

The seismic impact of COVID-19 on our society has dramatically boosted the study of non-pharmaceutical interventions via explicit observations: surveys. This methodology allows to gather ground truth data by asking specific questions to a sample of the population rather than inferring answers from proxy data. Hence, they are a powerful tool to understand and measure causal links. In the era of big-data surveys might look old fashion and surpassed. However they are still an invaluable approach to characterize complex processes. The sample size is typically an issue castings doubts about the external validity of the findings. However, modern technology and having a larger part of the population stuck at home in front of a screen due to lockdowns offered the opportunity to reach large numbers of participants quite easily. \\
Before diving into the observations and results of this large body of research, let me spend few words to characterize the subset. First of all, it is interesting to notice how the surveys have been answered by participants in $85$ countries. While in the large majority of cases the survey targeted a specific country, four surveys were done in two countries, two in five, one was extended to participants in $59$ countries. Second, the most represented country is USA, followed by Italy and China. This top three is not surprising. China was the epicenter of the pandemic. Italy was the first country in Europe to be affected by the virus. USA is a powerhouse of scientific production and one of the most affected countries in the pandemic. In Fig.\ref{fig:fig3} I report a histogram considering, to improve the readability, just countries featured in more than one single study. All continents, except Antartica, are represented. Third, $41$ countries are found in the affiliations of this group of papers. USA, Italy, China and the UK are at the top of the list. Fourth, the research has been published in $65$ different journals (including $2$ preprint in the medRxiv not yet published elsewhere). The most represented journals are \emph{International Journal of Environmental Research and Public Health}, \emph{Frontiers in Psychology}, \emph{Frontiers in Public Health}, \emph{PLoS ONE} and the \emph{Journal of Medical Internet Research} (see Fig.~\ref{fig:fig3}). It is interesting to stress the diversity of the set of publication venues that ranges from \emph{Nature Human Behavior} and \emph{Science} to \emph{Journal of Medical Internet Research} and \emph{Seizure}. This, still limited sample, highlights the breadth of research on the subject. Finally, these papers have been written by more than $131$ authors and, as of December $19^{th}$ they received $192$ citations. See Fig.~\ref{fig:fig3} for more details.\\

\subsection{Classification}

After inspecting all the articles and considering their focus, I have assigned each survey to one of five categories. In particular: 

\begin{enumerate}
\item \textbf{Health and wellbeing indicators}: surveys aimed at monitoring, measuring or access the impact of NPIs on health and wellbeing;
\item \textbf{Adoption and awareness of NPI}: surveys aimed characterizing the adoption of NPIs identifying the key factors influencing changes in behaviors;
\item \textbf{Changes in medical practice}: describes studies focused on accessing the impact of NPIs on patients and health care;
\item \textbf{Effects of NPI}: surveys set out to understand the impact of NPIs on a range of human activities;
\item \textbf{Contacts}: surveys aimed at quantifying the effects of NPIs on human contacts.
\end{enumerate}

It is important to mention that similar topics will be found also in other types of study described in other sections. However, those don't rely on primary data collection methods but on data proxies. In other words, they use data that have been collected for other uses as proxy for behavioral markers.\\
In Table~\ref{table: survey} I report the full list of references organized according the taxonomy I have created. Furthermore, I have grouped the research considering the sample size and the country or countries from where the participants where recruited. Now, we are ready to dive into each subcategory. 

\begin{table}[]
\begin{tabular}{|l|l|l|l|}
\hline
\textbf{Category} & \textbf{Subcategory} & \textbf{Sample size} & \textbf{Country and Reference}\\
\hline
\multirow{15}{*}{\acapo{Health and\\ wellbeing indicators}} & \multirow{5}{*}{\acapo{Mental Health\\ and wellbeing}}       & $N>10^{5}$ & Italy~\cite{tintori2020adaptive}, 59 countries~\cite{Alzueta_2020} \\ \cline{3-4} 
                                                  &                                      & $N>10^3$   & \acapo{Bangladesh~\cite{Islam_2020}, China~\cite{Cai_2020,Liu_2021}, USA~\cite{Sutin_2020,Luchetti_2020,Patrick_2020}, \\Serbia~\cite{Sadikovi__2020} }  \\ \cline{3-4} 
                                                  &                                      & $N>5\times 10^2$    & \acapo{Canada~\cite{Trougakos_2020}, China~\cite{Song_2020}, Ghana~\cite{Asiamah_2020}, Jordan~\cite{Abuhammad_2020}, \\ Italy~\cite{Mazza_2020}, Russia and Belarus~\cite{Reznik_2020}, USA~\cite{Emerson_2020,Copeland_2020,Lee_2020}} \\ \cline{3-4} 
                                                  &                                      & $N>10^2$   & USA~\cite{Huckins_2020}  \\ \cline{3-4} 
                                                  &                                      & $N>10$     &  Netherlands~\cite{van_de_Groep_2020}\\ \cline{2-4} \cline{2-4}  \cline{2-4}  \cline{2-4} \cline{2-4} \cline{2-4}  \cline{2-4} 
                                             
                                                  &     \multirow{3}{*}{Health Behaviors}  & $N>10^{3}$    &  China~\cite{Liu_2020}, Spain~\cite{Balanz_Mart_nez_2020}, UK~\cite{Rogers_2020}, USA~\cite{Bourassa_2020,Knell_2020,Vidot_2020}\\ \cline{3-4} 
                                                  &                                      & $N>5 \times 10^{2}$     & Australia~\cite{Hammoud_2020}, USA~\cite{Barbosa_2020} \\ \cline{3-4} 
                                                  &                                      & $N>10^2$    &Spain~\cite{Medrano_2020,Romero_Blanco_2020}, USA~\cite{Nelson_2020}  \\ \cline{2-4} \cline{2-4} \cline{2-4}  \cline{2-4}  \cline{2-4} \cline{2-4} \cline{2-4}  \cline{2-4}  \cline{2-4} 
                                                  &        \multirow{3}{*}{Effects on patients}   & $N>10^3$   & Germany~\cite{Sch_fer_2020} \\ \cline{3-4} 
                                                  &                                      & $N>10^2$    &  Germany~\cite{Musche_2020}, Japan~\cite{Kishimoto_2020}, Italy~\cite{Passanisi_2020,Martinotti_2020}\\ \cline{3-4} 
                                                  &                                      & $N>5$     & Brazil~\cite{Browne_2020}, USA~\cite{Beukenhorst_2020} \\ \hline
\hline

\multirow{5}{*}{\acapo{Adoption and \\awareness of NPI}} & \multirow{4}{*}{NA}       & $N>10^{4}$ & \acapo{Australia, Austria, France, Germany, Italy,\\New Zealand, UK, USA~\cite{Galasso_2020,Allen_2020}, Japan~\cite{Muto_2020}}  \\ \cline{3-4} 
                                                  &                                      & $N>10^3$   & \acapo{China~\cite{Xu_2020,Wei_2020},Germany~\cite{L_decke_2020}, Jordan~\cite{Al_Dmour_2020}\\
India~\cite{Ashraf_2020, Kuang_2020}, Saudi Arabia~\cite{Tripathi_2020},\\ Slovenia~\cite{Lep_2020}, New Zealand~\cite{Gray_2020},\\ Taiwan~\cite{Wang_2020_xx5}; UK~\cite{Rogers_2020}, USA~\cite{Kasting_2020,Arp_2020,Clipman_2020,Fisher_2020}} \\ \cline{3-4} 
                                                  &                                      & $N>5\times 10^2$    & Serbia~\cite{Cvetkovi__2020}, Pakistan~\cite{Haque_2020} \\ \cline{3-4} 
                                                  &                                      & $N>10^2$    & \acapo{India~\cite{Haq_2020},Pakistan~\cite{Balkhi_2020}, Saudi Arabia~\cite{Siddiqui_2020},\\
 South Korea~\cite{Jang_2020}, USA~\cite{Barber_2020}}  \\ \hline
\hline

\multirow{15}{*}{\acapo{Changes in \\medical practice}} & \multirow{5}{*}{Mental Health}      & $N>10^3$ & Austria~\cite{Probst_2020} \\ \cline{3-4} 
                                                  &                                      & $N>5\times 10^2$   & Italy~\cite{Colizzi_2020}, 29 countries~\cite{Jeste_2020} \\ \cline{3-4} 
                                                  &                                      & $N>10^2$    & Austria and Germany~\cite{Korecka_2020} India~\cite{Panda_2020}, USA~\cite{Murphy_2020} \\ \cline{3-4} 
                                                  &                                      & $N>10$     &  \acapo{Australia,  Canada, Netherlands, UK, USA~\cite{Waller_2020}, \\ Germany~\cite{Kersebaum_2020}, Italy~\cite{Taddei_2020,Brondino_2020} }\\ \cline{2-4} \cline{2-4}  \cline{2-4}  \cline{2-4} \cline{2-4} \cline{2-4}  \cline{2-4} 
                                             
                                                  &     \multirow{2}{*}{Oncology}  & $N>10^{2}$    & Italy~\cite{Gebbia_2020}  \\ \cline{3-4} 
                                                  &                                      & $N>10$    &  \acapo{Austria, Belgium, France, Germany, Italy,\\ Luxembourg, Netherlands, Spain, Switzerland,\\ USA~\cite{Onesti_2020}} \\ \cline{2-4} \cline{2-4} \cline{2-4}  \cline{2-4}  \cline{2-4} \cline{2-4} \cline{2-4}  \cline{2-4}  \cline{2-4} 
                                                  &        \multirow{3}{*}{Other care}   & $N>10^3$   &USA~\cite{Palinkas_2020} \\ \cline{3-4} 
                                                  &                                      & $N>10^2$    & China~\cite{Li_2020}, Italy and Saudi Arabia~\cite{Monaco_2020} \\ \cline{3-4} 
                                                  &                                      & $N>20$     &  UAE~\cite{Ayas_2020}, USA~\cite{Auerbach_2020}\\ \hline
                \hline                                  
\multirow{15}{*}{Effect of NPIs} & \multirow{5}{*}{Leisure}      & $N>10^3$ & \acapo{USA~\cite{Applebaum_2020}, Sweden~\cite{H_kansson_2020_xx1},\\ 12 countries~\cite{Ugolini_2020}, 24 countries~\cite{Randler_2020}}\\ \cline{3-4} 
                                                  &                                      & $N>5\times 10^2$   & Sweden~\cite{H_kansson_2020} \\ \cline{3-4} 
                                                  &                                      & $N>10^2$    &South Korea~\cite{Choi_2020}, Sweden~\cite{H_kansson_2020_2}, USA~\cite{Graupensperger_2020}
  \\ \cline{2-4} \cline{2-4}  \cline{2-4}  \cline{2-4} \cline{2-4} \cline{2-4}  \cline{2-4} 
                                             
                                                  &     \multirow{1}{*}{Education}  & $N>10^{3}$    &  China~\cite{zhao2020effects}, Jordan~\cite{Elsalem_2020_xx1}, 32 countries~\cite{Myers_2020}  \\ \cline{2-4} \cline{2-4} \cline{2-4}  \cline{2-4}  \cline{2-4} \cline{2-4} \cline{2-4}  \cline{2-4}  \cline{2-4} 
                                                  &        \multirow{1}{*}{Mobility}   & $N>10^3$   & \acapo{Germany~\cite{B_nisch_2020}, Denmark, Netherlands, \\ Italy, Spain, UK~\cite{Sun_2020}} \\ \cline{2-4} \cline{2-4} \cline{2-4}  \cline{2-4}  \cline{2-4} \cline{2-4} \cline{2-4}  \cline{2-4}  \cline{2-4}         
                                                  &        \multirow{2}{*}{Consumption}   & $N>10^3$   & China~\cite{Song_2020_2} \\ \cline{3-4} 
                                                  &                                      & $N>10^2$    &Romania~\cite{Butu_2020}, Tunisia~\cite{Jribi_2020}  \\   \hline                            
\hline

\multirow{2}{*}{\acapo{Contacts}} & \multirow{2}{*}{NA}       & $N>10^{3}$ & China~\cite{Zhang_2020_2}, UK~\cite{Jarvis_2020} \\ \cline{3-4} 

                                                  &                                      & $N>5\times 10^2$    &  China~\cite{Zhang_2020_3}   \\ \hline
\end{tabular}
\label{table: survey}
\caption{Summary of the surveys in the sample of papers. The first column describes the main topic of each survey, the second the subcategory, if any. The third the sample size. Note that for simplicity I have group them in brackets. The last column describes the country or countries focus of the survey and/or the location of the participants.}
\end{table}

\subsection{Health and wellbeing indicators}

This body of research aims to characterize the effects of NPIs on i) mental health, ii) health behaviors, and iii) patients. These constitute three subcategories. Let's discuss them in order.\\
It interesting to notice the variety of angles and aspects covered in the surveys on mental health. In fact, they consider loneliness~\cite{Lee_2020,Luchetti_2020,Emerson_2020}, domestic abuse~\cite{Abuhammad_2020}, psychological mindset~\cite{Reznik_2020, Islam_2020,Sadikovi__2020, Sutin_2020, Alzueta_2020,Asiamah_2020,  tintori2020adaptive, Song_2020,Patrick_2020}, as well as particular groups of the population studying the toll on frontline workers~\cite{Cai_2020}, students~\cite{Copeland_2020, Huckins_2020}, parents and children~\cite{van_de_Groep_2020,Trougakos_2020,Mazza_2020,Liu_2021}. Several common themes as well as some conflicting results emerged from this research. When it comes to loneliness age, gender, and perceived concerns about the pandemic appear to be important factors. In particular, younger individuals, especially female, reported a larger increase in loneliness respect to adults~\cite{Lee_2020}. However, a direct comparison between participants in the $60-70$ and $70+$ age brackets show that the older group felt more lonely~\cite{Emerson_2020}. Hence, loneliness appears not to be a monotone function of age. Very young and older age classes are more affected by it. Furthermore, loneliness increases more for those reporting greater concerns about the pandemic. However, it is important to notice that while two surveys found an increase of loneliness during the first wave~\cite{Emerson_2020,Lee_2020}, another reported no significant variations from January to April~\cite{Luchetti_2020}. One survey focused on the effect of NPIs on domestic abuse~\cite{Abuhammad_2020}. The study considers a sample of $687$ women living in Jordan contacted via Whatsup. Tragically $40\%$ reported same level of domestic abuse during the health emergency. Being unemployed and married were the most important features linked to abuse. In terms of psychological mindset of participants living under some type of NPIs country of residence, age, gender relationship status, living arrangements, income, and education are important factors~\cite{Reznik_2020,Islam_2020,Sadikovi__2020,Sutin_2020,Asiamah_2020, Alzueta_2020,tintori2020adaptive,Song_2020}. In fact, a study conducted by recruiting participants in $59$ countries reports that socio-demographic features, having had symptoms, mobility restrictions, and having had issues transitioning to remote work or an increase of conflict at home explain around $20\%$ of the variance in depression and anxiety~\cite{Alzueta_2020}. Another survey estimates panic in $79\%$ of participants and found being older than $30$, educated, married, and living with a joint family to be predictors of it~\cite{Islam_2020}. In contrast, a survey conducted in Serbia found that worry, fear, boredom, anger, and annoyance decreased in time during the first wave~\cite{Sadikovi__2020}. Adherence to NPIs and exposure to media were linked to positive effects on worry and fear. Similarly, a survey aimed at capturing the effect of NPIs on five factor personality traits found that during the crisis neuroticism went down~\cite{Sutin_2020}. The authors speculate that this trend is due to the progression of the pandemic rather than to peoples' personality.\\
The second subcategory describes surveys aimed at capturing the impact of NPIs on a range of health behaviors such as sleep, physical activity, sex, and substance use. Two studies on children noted how COVID-19 restrictions led to less physical activity and sleep disturbances but to more screen and sleeping time~\cite{Medrano_2020,Liu_2020}. The findings hint to a complex picture with some positive and negative effects of home confinement. Physical activity and changes in sleeping patterns have been reported also in the adult population~\cite{Rogers_2020,Bourassa_2020,Knell_2020,Balanz_Mart_nez_2020,Romero_Blanco_2020}. A sample of more than $9000$ participants in the UK provides some hints about the magnitude of physical activity variation before and during the lockdown~\cite{Rogers_2020}. While $64\%$ reported to have kept similar levels of activity, $25\%$ did less and $11\%$ increased. Overall, the surveys on the subject highlight high BMI, depression, health conditions, gender (being female), and high risk perception as features connected to larger decrease of physical activity~\cite{Rogers_2020,Bourassa_2020}. Being an essential worker negatively impacted lifestyle~\cite{Romero_Blanco_2020}. Social distancing and stay home orders affected also other health behaviors such as sex and substance use. Two surveys done in Australia and USA show how NPIs have led to drastic reduction of sex with occasional partners and an increase of porn consumption~\cite{Hammoud_2020,Nelson_2020}. Alcohol consumption, above the drinking limits, has reported to increase during NPIs~\cite{Barbosa_2020}. Similarly, medicinal use of cannabis has increased among individuals with mental health conditions and $16\%$ affirmed to have changed to route of administration shifting away from smoking~\cite{Vidot_2020}. \\
The \emph{tsunami} of COVID-19 cases that flooded health care facilities had a drastic impact also on patients affected by other conditions. Several surveys aimed at provide a better understanding of this unfortunate phenomenon. They considered patients affected by diabetes, cancer, ALS, hypertension, mental or substance abuse disorders~\cite{Sch_fer_2020,Passanisi_2020,Kishimoto_2020,Martinotti_2020, Musche_2020,Browne_2020,Beukenhorst_2020}. Admittedly, the research focused on short term effects and on the shift towards telemedicine (i.e., remote care). A study conducted in Italy on diabetic patients revealed their good resilience and adaptation to technology~\cite{Passanisi_2020}. However, a similar research done in Japan hints to a negative impact of reduced physical activity on glycemic control~\cite{Kishimoto_2020}. A cohort of patients affected by cancer in Germany showed levels of stress and anxiety similar to a control~\cite{Musche_2020}. Drastic reduction of physical activity, with potentially detrimental effects, has been reported for adults with hypertension~\cite{Browne_2020}.

\subsection{Adoption of NPIs}

Several interesting themes emerge from the surveys aimed at understanding the factors driving the adoption of NPIs. First of all, across several countries and in different studies gender appears to be a key factor. In fact, female respondents are found more likely to perceive COVID-19 as a risk, agree and comply with NPIs~\cite{Galasso_2020,Cvetkovi__2020, Barber_2020,L_decke_2020, Muto_2020}. Similar trends are observed for individuals with high level of education~\cite{L_decke_2020,Balkhi_2020}. Age again does not appear to be a monotonic function: very young and old individuals share similar trends. In fact, young adults have reported a smaller decrease in mobility and old individuals less compliance to NPIs (especially if male)~\cite{Barber_2020,Cvetkovi__2020}. It is important to mention that age was not a relevant factor in a survey conducted in Germany~\cite{L_decke_2020}. The research points also to a big urban-rural divide and thus to the importance of socio-economic factors when it comes to compliance. In fact, several barriers to NPIs have been reported in peri-urban and rural areas in India~\cite{Ashraf_2020,Kuang_2020,Haq_2020}. Individuals living in urban settings report a better knowledge about COVID-19 and better opportunities to adopt hygienic behaviors.  The country of residence also seems to play a role when it comes to the trust level towards media during the emergency which might influence adoption.  A cohort in Jordan reported a positive influence, especially on the awareness levels~\cite{Al_Dmour_2020}. However, a survey in Slovenia, done with $48$ of the first confirmed case, points to low level of trust of the media and shows that perceived credibility of information linked to lower levels of negative emotional response and higher adherence to NPIs~\cite{Lep_2020}. A survey conducted in South Korea compared the adoption of NPIs during COVID-19 with results obtained during the spread of MERS-CoV~\cite{Jang_2020}. Interestingly social distancing and practicing other transmission reducing behaviors increased significantly during the COVID-19 pandemic. The subset of participants with a higher perception of risk was found to be more likely to adopt NPIs in both. Interestingly, several surveys confirm the expectation coming from the health belief model pointing to the fact that knowledge about the disease and perceived risks are predictors behavioral changes~\cite{Xu_2020,Gray_2020,Kasting_2020,Tripathi_2020,Siddiqui_2020,Haq_2020,Balkhi_2020}. The adoption of face masks has been found to be negatively affected by perception of shortage in New Zealand~\cite{Gray_2020}, and positively influenced by national guidelines in the USA~\cite{Fisher_2020}. Ref.~\cite{Allen_2020} analyzes the dataset collected from half millions of users that installed the platform for digital surveillance called \emph{How We Feel}. The app collects a range of demographic, symptoms, and behavioral data such as use of masks and social distancing practices. Though the aim of the paper is to investigate  wether the data collected can be used to estimate risk factors and exposure to the virus, it provides interesting insights about the behavior of the sample in terms of NPIs. The majority ($61\%$) of users reported to have left home despite the stay at home orders. Only $19\%$ of these attributed to work the reason to leave home. Of this subset of people that left home the majority reported to practice social distancing and wearing face masks. Though still a significant fraction did not. A minority of users reported that, despite testing positive, they went to work sometime without wearing a mask. Healthcare and essential workers were more likely to go to work after testing positive.
Though, people who tested positive, reported a clear reduction of contacts respect to those that were untested or tested positive.

\subsection{Changes on medical practice}

How did medical practice change to cope with NPIs and COVID-19? A very interesting subset of surveys provides some answers from both patients and doctors. I have identified three subcategories which refer to mental health, oncology and other medical specialties. \\
A survey among psychotherapists in Austria revels that the decrease of face-to-face appointments was not necessarily linked to an increase of remote sessions~\cite{Probst_2020}. Similarly, another in Italy reports the needs for more health care support~\cite{Colizzi_2020}. Even more concerning are the results of a survey that pooled participants in $29$ countries~\cite{Jeste_2020}. These are the caregivers of individuals with intellectual and developmental disabilities. The majority, $74\%$, reported that they lost access to at least one therapy. The majority nevertheless, $56\%$, was able to keep services though $36\%$ reported to have lost healthcare provider. Across several surveys, telemedicine has been found to be useful~\cite{Jeste_2020,Panda_2020,Taddei_2020}. Though, socio-economic disparities, linguistic disadvantage, and health issues have been reported as challenges of remote care. It is also interesting to notice the perception of telemedicine among practitioners~\cite{Murphy_2020,Korecka_2020}. Rapid transitions towards it, reduce revenue and modification of communication linked to it have been reported as challenges~\cite{Murphy_2020}. Furthermore, the judgment of the transition appears to be function of the number of patients supported~\cite{Korecka_2020}.\\
In the second subcategory we have two surveys dealing with changes in oncology care. The first used Whatsup to explore the topic covered in discussions between patients and doctors. The most common themes that emerged cover delays with care, queries about immunosuppression, and variation of lifestyle or activities~\cite{Gebbia_2020}. The second survey was conducted in $21$ oncological care institutions in $10$ countries. Remote triage before visits was implemented by $90\%$ of centers. Adoption of PPE was used across the board and only few clinical instrumental examination were performed. Telemedicine was adopted by $76\%$ and separated pathways for COVID-19 positive and negative patients were put in place. Return to work policies required a negative test in $76\%$ of centers.\\
In the third and final subcategory we find surveys targeting both patients and practitioners in a variety of specializations. From remote management of geriatric patients with heart conditions~\cite{Li_2020} and the effects of NPIs on transplant recipients/donors~\cite{Monaco_2020} to the changes adopted by several hospitals in academic centers~\cite{Auerbach_2020}, hearing and emergency care~\cite{Ayas_2020,Palinkas_2020}. Overall the observations on patients point to the interim efficacy of telemedicine and to high level of awareness about the risk of infection. The findings on medical practices highlight the drastic changes and adaption put in place to cope with the pandemic. Marked reduction of in person visits, stockpile of PPE and use of respiratory isolation units have been reported. Remarkably, in one study $46\%$ of practitioners reported  examples of non-COVID-19 diagnoses in patients admitted for COVID-19 evaluation and vice versa highlighting the challenges induced by the virus to the health care system~\cite{Auerbach_2020}.

\subsection{Effect of NPIs}

This group of surveys captures the impact of NPIs of a range of human activities such as i) leisure, ii) education, iii) mobility, and iv) consumption.\\
Due to social distancing, travel limitation, and stay home orders the possibility for leisure has been dramatically reduced. A survey done in a cohort of around $4500$ participants in $25$ countries investigate the impact on bird watchers~\cite{Randler_2020}. A large majority, $85\%$, changed behavior reducing traveling and targeting local spots. They reported also a significant change in social interactions. The increase of time spent inside has induced variations in the locations selected when going outside~\cite{Ugolini_2020}. Walking paths have been changed to include urban gardens, tree lined streets and traveling by car in green areas. In others words, the confinement has boosted the interest and value of nature. The stop of many sport events and leagues has impacted also gambling~\cite{H_kansson_2020}. While sport gambling has dropped, the attention has been shifted towards other forms. The gambling habits of the majority of $2000$ respondent to another survey reported did not report a change in gambling habit~\cite{H_kansson_2020_xx1}. However, the minority that did was affected by gambling problems and reported an increase in alcohol consumption. The toll of NPIs on sport activities has been quite significant for elite athletes~\cite{H_kansson_2020_2,Graupensperger_2020}. Level of stress and anxiety have been observed, especially among women, and linked to worries about the future of the sport. A survey conducted in South Korea links the tendencies towards keeping practicing sport with the perception of risk and worry about getting infected~\cite{Choi_2020}. Age and the type of sport were found to play a modulating role.\\
Educational activities have been also altered. Like in many other sectors the need for physical distancing required drastic shifts in the delivery of classes as well as on research production. A survey conducted in China extended to parents, students and teachers point to the fact that while online teaching was positively valued, parents and teachers were found to be more concerned about the possible side effects~\cite{zhao2020effects}. A survey delivered in a cohort of students in Jordan aimed to understand the impact of remote learning and in particular on online exams~\cite{Elsalem_2020_xx1}. A sizable fraction of students ($32\%$) reported more stress linked to this exam mode. Navigation of questions and technical issues were reported to be main drivers of stress. Furthermore, online exams appeared to affect negatively eating habits, sleep, physical activity and led to increase in smoking. A study conducted on principal investigators in USA and Europe, point how  NPIs, despite the reduction of activities outside home, have reduced the time spent doing research especially for female scientists, those in experimental fields and with young children~\cite{Myers_2020}.\\
As we will see in more details later, one of the most explored aspect of non-pharmaceutical interventions is mobility. Beside the interest in the way population adapted movements/travels, characterizing variations in mobility is key to model the spreading of the virus. As result, we will speak about NPIs and mobility in several sections. Though the aim will be the same, what changes is the methodology used to characterize changes in mobility. Here, we report two studies done with cohort of $2000$ individuals in Germany~\cite{B_nisch_2020} and more than $1000$ people in five European countries~\cite{Sun_2020}. In both cases, smartphone apps were used to collected information about mobility (distance travelled, steps count) as well as other indicators (heart rate, sleep duration, bedtime, social app use). The results point to a drastic reduction of distance travelled (more than half in the lockdown period)~\cite{B_nisch_2020}. The reduction did not cease immediately after the end of lockdown. Age and gender groups behaved similarly, reporting a consistent relative variation respect to pre lockdown period. Reduction of steps and increase homestay have been reported also outside Germany~\cite{Sun_2020}. Furthermore, use of phone and social media apps have increased. Consistently with what mentioned above, sleeping patterns have been altered. More sleep and later bed time have been reported~\cite{Sun_2020}.

\subsection{Contacts}

NPIs are designed to reduce transmission chains. Hence, several targeted directly or indirectly social contacts. Different surveys have been designed and deployed to capture the effect of NPIs on contacts and thus estimating the impact on the disease spreading. A survey conducted in Wuhan and Shanghai during the outbreak revealed a reduction of factor $7-8$ of social contacts~\cite{Zhang_2020_3}. Using this data together with information about contact tracing the authors estimated, via an epidemic model, this reduction was enough to contain the spreading. Another survey conducted, considering also data after the end of the lockdown in China, shows that the relaxation of NPIs brought an $5-17\%$ increase in social contacts which are anyway $3-7$ times lower than pre-pandemic values and still below the levels associated to disease resurgence. A similar study conducted during the lockdown in the UK report a consistent, though smaller, reduction contacts (a factor 4)~\cite{Jarvis_2020}. Nevertheless the change was enough to push the $R_0$ below the critical threshold of one.

\section{Comments and/or perspectives}

\begin{figure}[t]
  \centering
   \includegraphics[scale=0.25]{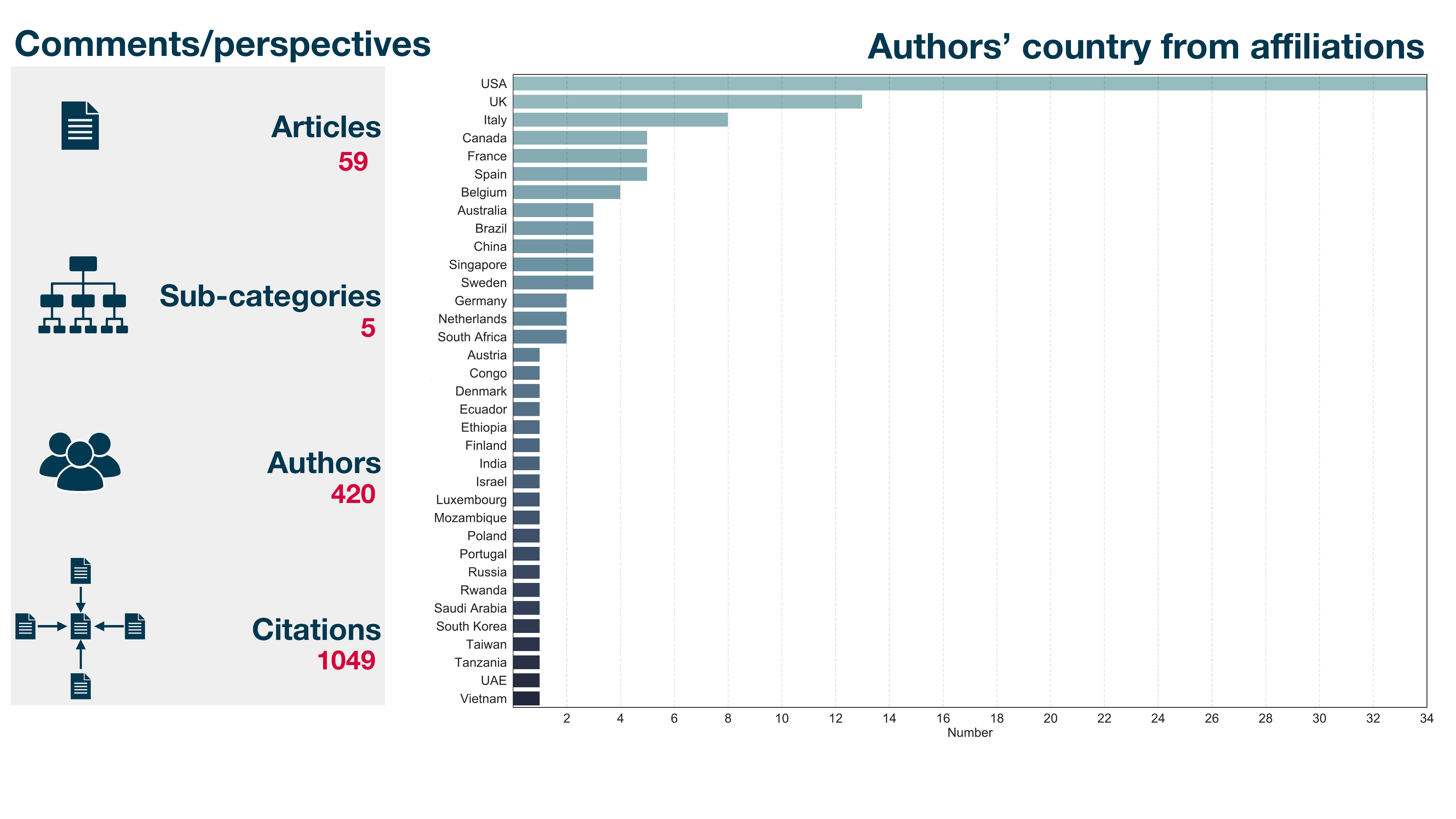}
  \caption{Total number of authors and citations of the papers in this category on the left. Note that these numbers are estimated from Semantic Scholar. On the right histogram describing the number of countries considering authors' affiliations.}
  \label{fig:fig4}
\end{figure}

A significative fraction of the sample can be classified as comments and/or perspectives. Here researchers reflect on future challenges, point out opportunities, summarize findings and approaches developed in the fight against COVID-19. In doing so, they do not necessarily present  new analysis or findings. Rather, the goal is to highlight key aspects, draw the attention to specific points, critically discuss evidence, questions assumptions as experts in the field.\\
Before summarizing this body of work, let me spend few words to provide a general description of it. While the large majority of articles provide a general outlook to a specific topic, some draw from a specific context. In fact, we find $6$ comments focused from observations in USA~\cite{Samuels_2020_x,Goldman_2020_x,Lau_2020_x,Connolly_2020_x,Sweeney_2020_x,Gostic_2020_x2}, $3$ in Italy~\cite{Leonardi_2020_x,Furfaro_2020_x,Gualano_2020_x}, $2$ in Canada~\cite{Elbeddini_2020_x,Zastepa_2020_x} and the UK~\cite{Jones_2020_x,Thompson_2020_x3}. We have then articles that focus on Africa~\cite{Nachega_2020_x4}, China~\cite{Sun_2020_x3}, Ethiopia~\cite{Mohammed_2020_x} , France~\cite{Furfaro_2020_x}, 
Poland~\cite{Torres_2020_x4}, Taiwan~\cite{Chien_2020_x4}, Vietnam~\cite{Nong_2020_x4},South Korea~\cite{Oh_2020_x4} and
Singapore~\cite{Lateef_2020_x4}. Comments and perspectives have been written by researchers with affiliations in $36$ (see Fig.~\ref{fig:fig4}). The most represented countries are USA, UK, and Italy. Finally, this research has been published in $50$ journals (including one preprint in medrxiv). Really a wide range of publication avenues ranging from journals focusing on pediatric psychology to interdisciplinary journals such as \emph{Science Advances}. The most representative journals, each with two comments/perspectives, are \emph{BMC Medicine}, \emph{Journal of Addiction Medicine}, \emph{Journal of Travel Medicine} and  \emph{Science}. These papers have been written by more than $420$ authors and received more than $1049$ citations. A median of $4$ citations per article.

\subsection{Classification}

Considering their topics, comments and/or perspectives can be classified in five categories (see Table~\ref{table: comments}): 

\begin{enumerate}
\item \textbf{Change to medical practice}: describe articles that analyze challenges and changes to medical procedures made necessary by NPIs;
\item \textbf{modeling}: refers to comments and/or perspectives discussing different aspects, challenges, approaches, and opportunities to model the unfolding of COVID-19;
\item \textbf{Effect of NPIs}: captures articles reflecting on the changes induced by NPIs on a range of activities;
\item \textbf{Lifting NPIs}: describes the set of articles about different strategies and challenges to tackle when relaxing NPIs;
\item \textbf{Success stories}: describes comments and/or perspectives about lessons learnt in countries that manage to successfully suppress COVID-19;
\end{enumerate}

\begin{table}[]
\begin{tabular}{|l|l|}
\hline
\textbf{Category}       & \textbf{Topic and references} \\ \hline
Change to medical practice &  \acapo{Mental health care~\cite{Goldman_2020_x,Sweeney_2020_x,Anmella_2020_x,Kopelovich_2020_x,Gualano_2020_x,Lateef_2020_x4}; Emergency care~\cite{Wong_2020_x,Sangal_2020_x};  \\Substance abuse~\cite{Hsu_2020_x,L_pez_Pelayo_2020_x,Samuels_2020_x,Ornell_2020_x,Zastepa_2020_x}; Oncological care~\cite{Jones_2020_x,Pothuri_2020_x}\\ Health care practices~\cite{Baweja_2020_x, Leonardi_2020_x,Lau_2020_x,Plevinsky_2020_x,Elbeddini_2020_x,Correia_2020_x}; Pediatric care~\cite{Plevinsky_2020_x,Pritchard_2020_x}\\Other care~\cite{Bartolo_2020_x,Mohammed_2020_x,Connolly_2020_x,Furfaro_2020_x,Lapid_2020_x,halliwell2020restructuring}}
                            \\ \hline
                            \hline
Modeling                   &      \acapo{modeling approaches~\cite{Thompson_2020_x3,Cobey_2020_x2,Thompson_2020_x2,Vespignani_2020_x2,Poletto_2020_x2,Panovska_Griffiths_2020_xx5,Eubank_2020_xx5}; 
Transmission dynamics~\cite{Lee_2020_x4,Colebunders_2020_x3,Althouse_2020_x4,Bouayed_2020_x4,Feng_2020_xx5}\\
Mobile phone data~\cite{Oliver_2020_x3};
$R_t$ estimation~\cite{Gostic_2020_x2}}                       \\ \hline
\hline
Effect of NPIs             &    \acapo{Effects on: children~\cite{Lambrese_2020_x4,Jiao_2020_x4};  education~\cite{Torres_2020_x4,Pacheco_2020_x4};
gambling~\cite{H_kansson_2020_x3}\\ other diseases~\cite{Sun_2020_x3}; the environment~\cite{El_Zowalaty_2020_x4,Vanapalli_2021_x4}; food consumption~\cite{Khan_2020_x3}; \\corporate world~\cite{Yeo_2020_x3}; porn comsuption~\cite{Mestre_Bach_2020_xx1}}                           \\ \hline
\hline
Lifting NPIs               &      \acapo{African context~\cite{Nachega_2020_x4}; 
Vietnamese context ~\cite{Nong_2020_x4};
 Preventing second wave~\cite{Wilder_Smith_2020_x4}\\ Effects on children back to school~\cite{Pelaez_2020_x4}}                       \\ \hline
 \hline
Success story              &      South Korea~\cite{Oh_2020_x4}, Taiwan~\cite{Chien_2020_x4}                         \\ \hline
\end{tabular}
\label{table: comments}
\caption{Summary of the comments and/or perspective in the sample. The first column describes the category of each article in this group. The second, the topics and references}
\end{table}

\subsection{Change to medical practice}

This is the largest group of comments and/or perspectives. We find articles focused on a range of medical specialities: mental health care~\cite{Goldman_2020_x,Sweeney_2020_x,Anmella_2020_x,Kopelovich_2020_x,Gualano_2020_x,Lateef_2020_x4}, emergency care~\cite{Wong_2020_x,Sangal_2020_x}, substance abuse~\cite{Hsu_2020_x,L_pez_Pelayo_2020_x,Samuels_2020_x,Ornell_2020_x,Zastepa_2020_x}, oncological care~\cite{Jones_2020_x,Pothuri_2020_x}, health care practices~\cite{Baweja_2020_x, Leonardi_2020_x,Lau_2020_x,Plevinsky_2020_x,Elbeddini_2020_x,Correia_2020_x}, pediatric care~\cite{Plevinsky_2020_x,Pritchard_2020_x}, and other care~\cite{Bartolo_2020_x,Mohammed_2020_x,Connolly_2020_x,Furfaro_2020_x,Lapid_2020_x,halliwell2020restructuring}. \\
Ref.~\cite{Sweeney_2020_x} draws on the experience developed when counseling patients suffering from anxiety disorders in one of the worst hit region in the USA during the first wave: New York City. The observations highlight how the changes in behaviors induced by NPIs shifted the sense of familiarity, predictability or daily activities towards the unknown and drastically limited social support. The variation of ordinary routine is linked to increase distress. Social support, which is one of the key pillars of well-being, has been also challenged by the closure of activities, and social distancing. The article calls for establishing new routines and to make efforts to keep high level of social engagements. Authors in Ref.~\cite{Kopelovich_2020_x} reflect on the challenges linked to mental health care delivery in the context of social distancing. While the needs to move a portion of the services remotely are clear, the authors call a for diversification of the offer that includes in-person visits for patients that might lack the technology required for tele-health, digital interventions, community outreach, family support and national warm lines. Authors in Ref.~\cite{Sangal_2020_x} tackle the practical issues of social distancing in the challenging context of emergency departments. A wide range of standard procedures and workflows have been drastically changed to limit the changes of transmission between patients and health workers.  Authors in Ref.~\cite{Hsu_2020_x} highlight the opportunities of digital phenotyping via smartphones to reduce the limitations induced by NPIs during the pandemic in substance abuse care. The authors point to the use of such technology, which can be used at scale, for early detection of possible issues thus targeting the limited resources and reducing risks. In the context of oncologic care (in particular pancreatic cancer) Ref.~\cite{Jones_2020_x} presents a discussion of the variations put in place by UK clinicians. The article, recommends for prioritization of specific treatments, strategies to postpone surgical procedures, and for approaches that limit hospital visits. Ref.~\cite{Leonardi_2020_x} proposes a set of measures  in hospitals to reduce the risks of infection in patients and health workers. Authors call for aggressive and periodic  testing of practitioners, use of appropriate PPE and air filtering systems, suspensions of surgical training and experimental treatments among other things. Authors in Ref.~\cite{Plevinsky_2020_x} discuss the changes to care in pediatric settings. They acknowledge the key role of telemedicine, discuss the opportunities, barriers, and call for future research aimed at characterize the possible short and long term effects of this change in care. Finally, Ref.~\cite{Mohammed_2020_x} describes the effects of NPIs on veteran care in the USA. Thanks to the efforts already in place, the transition to telemedicine in this context has been reported to be quick and efficient. Nevertheless, also in this case, the authors call for future studies aimed at measuring the short and long term effects of such variation in care especially in mental health. 

\subsection{Modeling}

Several comments and/or perspectives focus on different aspects of modeling COVID-19 dynamics and NPIs. In particular, we find articles discussing epidemic modeling efforts~\cite{Thompson_2020_x3,Cobey_2020_x2,Thompson_2020_x2,Vespignani_2020_x2,Poletto_2020_x2,Panovska_Griffiths_2020_xx5,Eubank_2020_xx5}, analyzing known and unknowns of  transmission dynamics~\cite{Lee_2020_x4,Colebunders_2020_x3,Althouse_2020_x4,Bouayed_2020_x4,Feng_2020_xx5}, presenting opportunities to model and control the spreading of COVID-19 enabled by novel data streams such as mobile phone data~\cite{Oliver_2020_x3}, and discussing the challenges in $R_t$ estimation~\cite{Gostic_2020_x2}.\\
Ref.~\cite{Vespignani_2020_x2} discusses the main challenges in developing epidemic models for COVID-19. From highlighting the differences between ``peace'' and ``war'' modeling due to lack of data and a fluid landscape of knowledge about the virus under study in the middle of a pandemic, to the importance and challenges of considering super-spreading events. Authors in Ref.~\cite{Poletto_2020_x2} reflect on the key role played by predictive modeling in the emergency stressing the wide range of methodologies that have been put in place as well as the key challenges ahead to develop standard metrics to measure their performance.  Authors in Ref.~\cite{Lee_2020_x4} provide an overview of the \emph{engines} responsible for the spreading of SARS-CoV-2. Contacts in households, presence of asymptomatic and pre-symptomatic transmission as well as super-spreading events are considered as key factors. The importance of understanding and capturing super-spreading events to accurately model and suppress COVID-19 is also matter of discussion in Ref.~\cite{Althouse_2020_x4}. Ref.~\cite{Oliver_2020_x3} discusses the opportunities and challenges offered by mobile phone data to inform, guide, and help designing NPIs policies during the different phases of a pandemic. In the early phases, mobile phone data can be fundamental to help standard surveillance efforts aimed at fast detection of cases and possible exposed contacts. During the exponential growth phase, when the shields of NPIs are lifted as response, mobile phone data can be crucial to monitor mobility reduction at the population level thus quantifying the effectiveness of the measures as well as to inform realistic epidemic models (as we saw in previous sections). After the peak, this type of data can be crucial to help contact tracing, identify with the help of epidemic models risks of hotspots of local disease resurgence. While these opportunities are real and they come with serious challenges. In fact, the authors note how the lack of a widespread use of this data as a go-to tool to face epidemic spreading is linked to several problems. Lack of a ``digital mindset'' of key institutions, reluctance of corporations collecting the data to share it, privacy concerns, the ``academic mindset'' which often does not allow to formulate solutions in relevant actionable terms,  remain challenges for to tackle in the future. Finally, as we saw in the section about statistical epidemic model, the estimation of $R_t$ is a classic approach to characterize the unfolding of infectious diseases. The authors of Ref.~\cite{Gostic_2020_x2} discuss several methods and, by using synthetic data, present possible practical challenges that might bias the results. Interestingly, they suggest the adoption of EpiEstim (which we discussed above) and recommend against other methods which either use data after the time $t$ (so can be used only retrospectively) or are prone to biased estimates if the underlying assumptions do not hold. 

\subsection{Effect of NPIs}

This category captures comments and/or perspectives discussing the effects of non-pharmaceutical interventions on specific groups such as children~\cite{Lambrese_2020_x4,Jiao_2020_x4}, on education~\cite{Torres_2020_x4,Pacheco_2020_x4} or other activities such as gambling~\cite{H_kansson_2020_x3} and food~\cite{Khan_2020_x3}, on other diseases~\cite{Sun_2020_x3}, the environment~\cite{El_Zowalaty_2020_x4,Vanapalli_2021_x4}, the corporate world~\cite{Yeo_2020_x3} and porn consumption~\cite{Mestre_Bach_2020_xx1}. \\
Ref.~\cite{Lambrese_2020_x4} discusses the issues children face while confined at home during lockdowns. The authors suggest parenting strategies that structure the day in a set of activities with assigned roles and adopt technology to keep them socially connected. Authors in Ref.~\cite{Jiao_2020_x4} highlight the behavioral and emotional disorders in children as result of the pandemic. Ref.~\cite{Torres_2020_x4} addresses the effectiveness of e-learning for medical students. The observations from a cohort of students in Poland showcase the effectiveness of the technology that allows to meet several learning outcomes but point out the limitations to reach others linked to practical skill development which is key for doctors. Authors of Ref.~\cite{Khan_2020_x3} reflect on the effects of NPIs, in particular those that increase the time spent at home and limit physical activities, on food consumption. The authors introduce the concept of \emph{covibesity} to highlight the possible negative consequences of home confinement: increase of take away orders, food and alcohol consumption. They also reflect on the marketing shifts of the food industry which has leveraged the increased time spent online boosting online advertising in particular for children. Authors in Ref.~\cite{Sun_2020_x3} discuss how the strict NPIs put in place in China have also reduced the burden of influenza. In fact, social distancing and increase hygiene are key factors which might interrupt transmission chains of a range of viruses beside COVID-19. Ref.~\cite{El_Zowalaty_2020_x4} highlights one of the silverlining of NPIs put in place to curb the spreading of COVID-19: global reduction of mobility and thus a much smaller fingerprint on the environment. The authors point to the fact that positive change to the environment are proven to be possible and hope for behavioral changes in this direction even after the end of the pandemic. Ref.~\cite{Yeo_2020_x3} discusses the reaction of leadership in the corporate world to systemic crises such as COVID-19. The results from interviews and observations highlight how in crisis times leaders tend to be guided by compassion and how the scale of the crisis might help leaders to question what really matters which might help them to better capture the ethos of the company. Finally Ref.~\cite{Mestre_Bach_2020_xx1} reflects on the increase in porn consumption due to NPIs, the possible negative effects associated with it and call for future studies investigating this phenomenon.

\subsection{Lifting NPIs}  

Four articles provide a perspective about the relaxation of NPIs in different countries and contexts~\cite{Nachega_2020_x4,Nong_2020_x4}, reflecting on the risks of a second wave~\cite{Wilder_Smith_2020_x4}, and on the challenges children going back to school after a long period of isolation face~\cite{Pelaez_2020_x4}. \\
Ref.~\cite{Nachega_2020_x4} details the challenges faced by African countries when relaxing NPIs. In fact, the limited resources and the specific contexts of each country in the continent might impede adequate testing, contact tracing, and adoption of NPIs to avoid a rebound of infections. The authors warn about the grim consequences of other COVID-19 waves and call for efforts similar to those in place to flight HIV, TB, and malaria. Authors in Ref.~\cite{Wilder_Smith_2020_x4} discuss the differences between mitigation and suppression which we have presented above. They argue that suppression approaches based on lockdowns are to be preferred to mitigation strategies and how they might provide the time necessary to build up the testing and contact tracing infrastructure to avoid disease resurgence. The article published in July warned about the lack of such infrastructure despite the relaxation of measures in most countries and how this might lead to a second wave. The warning unfortunately manifested to be a well poised concern.

\subsection{Success stories}

While most of the countries in the world have not met the challenges posed by COVID-19 and saw multiple significant waves of infection as well as lockdowns, few managed to suppress the virus effectively. South Korea and Taiwan are examples. Their experience and lessons learnt have been focus in two perspectives~\cite{Oh_2020_x4,Chien_2020_x4}. In both cases the strategy revolved on early actions, rapid development of testing capabilities, aggressive contact tracing, social distancing and high adoption of face masks. 

\section{Quantifying the effects on NPIs}

\begin{figure}[t]
  \centering
   \includegraphics[scale=0.25]{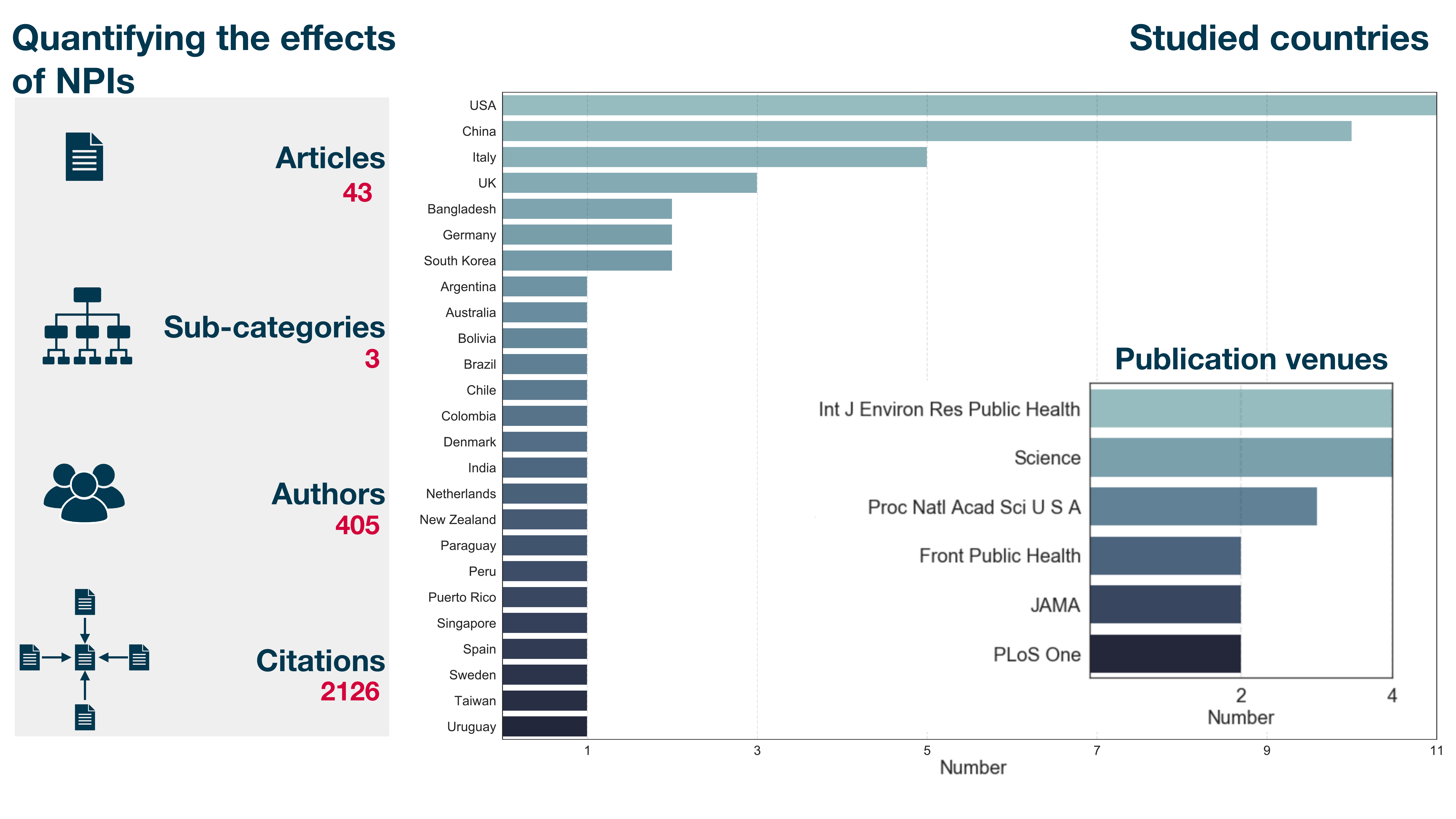}
  \caption{Total number of authors and citations of the papers in this category on the left. Note that these numbers are estimated from Semantic Scholar. On the right (top) histogram describing the number of countries subject of study in this category. Note how I have removed from the count, few papers that studied hundreds of countries as I could not find machine readable lists. On the right (bottom) most represented publication venues of the category. To improve visibility I am showing only journals featuring at least two articles. }
  \label{fig:fig5}
\end{figure}

Non-pharmaceutical interventions are not a monolith. By now it should be clear how they refer to a wide range of measures. A first binary classification can be used to split them according to whether they emerge bottom-up as voluntary reactions of the population or whether they are pushed top-down by authorities. In the first class, we find spontaneous use of face covering, increased hygiene, changes of diet, avoidance of transportation, social gatherings and more in general social distancing. In the second we find border closures, travel constraints, school closures, curfews, lockdowns, stay at home orders, the mandate use of face covering among others. While we were quite familiar with the first type, for example during a particularly aggressive seasonal flu, the second, often draconian, measures came as a surprise for many since they are real social and economical \emph{nuclear options}. It is important to mention how in reality the difference between the two is not a clear cut and, in a complex landscape as the COVID-19 emergency, they might be concurrent. Furthermore, the spatio-temporal unfolding of the virus, the sequence of events, economic and social pressure, political ideologies, and past experiences with similar health emergencies induced high level of heterogeneities in the response, visible not only across but within countries. As mentioned above, while Italy was shutting down in an unprecedented national lockdown, the UK was opting for a mitigation strategy (to develop herd immunity) rather that acting to suppress the virus. The subsequent wave of hospitalizations and deaths forced a u-turn. New Zealand, Singapore, Taiwan, Vietnam, and South Korea are all countries that acted early and aggressively. Many of them did not even had the luxury of being affected later respect to others. \\   
The articles in this section aim to understand, measure and quantify the efficacy of different NPIs in the mitigation/suppression of COVID-19. Some investigate their impact on the unfolding of COVID-19. Others focus on a range of behaviors beside those directly targeted by the interventions, variations induced on medical practice, and financial markets (see Table~\ref{table: effects}). It is important to mention how a large part of the first subgroup could be also classified as epidemic models. In fact, some measure the evolution of $R_t$ and others use compartmental models. These papers however aim at singling out the impact on NPIs on the spreading,  thus I thought to highlight them here.\\
As usual, before diving in the research, let me say few words about the sample of papers. They report analysis and observations from $149$ countries. In fact, while many of them focus on a particular country, others consider a wide range of them. USA, China and Italy are again at the top of the list (see Fig.~\ref{fig:fig5}). The $405$ authors of the articles in this category are affiliated with institutions from $27$ countries. USA, China, and the UK are the most represented. Furthermore, the research has been published in $32$ journals ranging from \emph{PNAS} and \emph{Nature Human behavior} to \emph{Lancet Oncology} and \emph{JAMA}. The most represented journals are the \emph{International Journal of Environmental Research and Public Health} and \emph{Science}. As of December $19^{th}$ these paper received  $2126$ citations.

\subsection{Classification}
The sample of papers studying the impact of NPIs can be further classified in three main categories: 

\begin{enumerate}
\item \textbf{Effects on spreading}: describes papers that quantify the impact of NPIs on COVID-19 spreading;
\item \textbf{Effects on behaviors}: describes articles aiming to characterize how NPIs have affected a range of behaviors and activities;
\item \textbf{Effects on medical practice}: describes research efforts quantifying the impact of NPIs on health care
\end{enumerate}

Let's dive into the details. 

\begin{table}[]
\begin{tabular}{|l|l|}
\hline
\textbf{Subcategory}       & \textbf{Country and references} \\ \hline
Effects on spreading & \acapo{ Argentina, Bolivia, Brazil, Chile, Colombia, Paraguay, \\Peru, and Uruguay~\cite{Gonz_lez_Bustamante_2021_xx1}; Australia~\cite{Costantino_2020_ff}
Bangladesh~\cite{Rahman_2020_ff}; \\
China~\cite{Ali_2020_ff,Pan_2020_xx1,Kraemer_2020_xx2,Tian_2020_xx1,Jing_2020_ff,Lei_2020_ff,Sun_2020_fv2};
China, Germany, India, Italy, Singapore, \\South Korea, Spain, UK, USA~\cite{Kaur_2020_ff}; 
Italy~\cite{Guzzetta_2021_ff,Lavezzo_2020_ff};\\
New Zealand~\cite{Jefferies_2020_ff};
Puerto Rico~\cite{Valencia_2020_ff};\\
South Korea~\cite{Huh_2020_ff};
Taiwan~\cite{Hsieh_2020_ff};\\
USA~\cite{White_2020_ff,Chernozhukov_2020_xx1,Auger_2020_xx1,Kissler_2020_ff,Jalali_2020_ff,Baker_2020_fv2}; 44 countries~\cite{Jardine_2020_wq2}; 53 countries~\cite{Castex_2020_xx1};\\ 56 countries~\cite{Haug_2020_ff}; 142 countries~\cite{Koh_2020_ff};
149 countries~\cite{Islam_2020_xx1}}
                            \\ \hline
                            \hline
Effects on behaviors     &       \acapo{China~\cite{Li_2020_xx1};
China, Italy~\cite{Su_2020_xx1};
Denmark, Sweden~\cite{Sheridan_2020_xx1}; Germany~\cite{Schlosser_2020_fv2}\\
USA~\cite{Slavova_2020_xx1,Jacobson_2020_xx1,Ward_2020_xx1,Nguyen_2020_xx1};
20 countries~\cite{Rousseau_2020_xx1};\\ 67 countries~\cite{Zaremba_2020_xx1}}                    \\ \hline
\hline
Effects on medical practice             &         Italy~\cite{Quiros_Roldan_2020_re22}; Netherlands~\cite{Schuengel_2020_re22}; UK~\cite{Maringe_2020_re22,Wood_2020_re22}                     \\ \hline
\end{tabular}
\label{table: effects}
\caption{Summary of the articles aimed at measuring the effects of NPIs. The left column describes the main categories in this group of articles. }
\end{table}

\subsection{Effects on spreading}

These articles study the impact on the unfolding of the disease focusing on the measures adopted in a  specific country~\cite{Gonz_lez_Bustamante_2021_xx1,Costantino_2020_ff,Rahman_2020_ff,Ali_2020_ff,Pan_2020_xx1,Kraemer_2020_xx2,Tian_2020_xx1,Jing_2020_ff,Guzzetta_2021_ff,Lavezzo_2020_ff,Jefferies_2020_ff,Valencia_2020_ff,Hsieh_2020_ff,White_2020_ff,Chernozhukov_2020_xx1,Auger_2020_xx1,Kissler_2020_ff,Jalali_2020_ff,Sun_2020_fv2}, proposing multi country analyses~\cite{Kaur_2020_ff,Jardine_2020_wq2,Castex_2020_xx1,Haug_2020_ff,Koh_2020_ff,Islam_2020_xx1}, and characterizing the effects of NPIs on other diseases~\cite{Lei_2020_ff,Huh_2020_ff,Baker_2020_fv2}.\\
Ref.~\cite{Costantino_2020_ff} studies the impact of travel bans in Australia from China. In doing so, authors use historical data describing the mobility flows from China to Australia, the evolution of the epidemic in China and developed an age-structured SIER-like model to describe the unfolding of the local pandemic given importations from the epicenter of the outbreak. The analysis of the initial phases of the spread suggests that the travel ban has been effective in reducing the number of infected seeds coming in the country which might have resulted in faster local growth of cases. Authors of Ref.~\cite{Ali_2020_ff} use publicly available data to create a database capturing $1407$ transmission pairs outside Hubei province. These describe the onset of symptoms and, for about half, the social relationships, infector and infectee pairs in the first month of the emergency. The analysis of the data suggests how the implementation of NPIs was key to shorten the serial interval from $7.8$ days, in the early phases of the spreading, to $2.2$ days.  As more cases were promptly identified and isolated the opportunities for the virus to  start new infection chains decreased drastically. From a methodological stand point the authors suggest the use of time-varying estimates of serial interval in the computation of key epidemic indicators such as the reproductive number to avoid biased results. In Ref.~\cite{Pan_2020_xx1} authors analyzed the features of $32,583$ cases in Wuhan. They considered as set of features such as socio-demographic, occupation and severity of illness. Their analysis confirms the strong impact of NPIs which in about two months were able to bring the reproductive from values above $3$ to values well below one. Health workers have been affected at higher rates respect to the general population (more than 3 times), and the portion of the severe cases drastically reduced from $~50\%$ the initial phases to $~10\%$ over the intervention period. Authors in Ref.~\cite{Sun_2020_fv2} study more than $1000$ infected individuals and their close contacts (more than $15,000$). The dataset captures almost $20,000$ exposure events in Hunan (China) from mid January to early April 2020. The primary infections have been identified by the surveillance network, their closed contacts have been followed for at least two weeks and some of them were tested if developed symptoms before February 7. After this data all contacts were required to be tested even if asymptomatic. The authors studied the transmission risks according to the type (i.e., household, social, community, family, healthcare), duration, timing of exposure and as function of the NPIs put in place. By using a logistic regression model they find that household contacts are the most risky type of exposures. Extended family, social and community contacts followed. Interestingly, contacts in healthcare settings have been found to have the lowest risk, hinting to the efficacy of protective equipment and measures put place in those settings. The results show that the effects of NPIs varies as function of the type of contact. In fact, the lockdown increased household contacts increasing the risk for further transmission at home. The same measure instead drastically reduced community and social contacts. Longer exposures increase the risk of infection and transmission risk was found to be higher around the time of first symptoms. Children were confirmed to be less susceptible that individual with age between $26$ and $64$. People older than $65$ was found significantly more susceptible. The data allowed the authors to study serial intervals and the impact of NPIs on their distribution. Consistently with other observations, the serial interval decreased from $7.3$ days for individuals isolated more than $6$ days after symptoms to $1.7$ days for those isolated $2$ days after showing signs of the disease. Interestingly, the authors are able to describe the type of case detection (passive or active surveillance) as the pandemic unfolded and NPIs strategies adapted. While in the early phases (before Jan 27) about $80\%$ of cases were detected and isolated only after showing symptoms, thus via passive surveillance, with a median time from onset to isolation of more than 5 days, after the $4^{th}$ of February this time difference become negative ($-0.1$ days) thanks to active contact tracing. This clearly shows the key role of aggressive testing and tracing strategies to reduce the chances of further transmission. Ref.~\cite{Kraemer_2020_xx2} uses real-time mobility data, case and travel history across China to access the impact of NPIs in the country. The authors show how in the early phases the spatial distribution of cases is captured by and highly correlated with mobility data. The correlation drops significantly soon as after the implementation of NPIs (in particular travel restrictions). The results point also to clear decrease in the serial distribution showing the impact of strict surveillance and highlighting the effectiveness of the implemented measures to suppress the spreading of COVID-19. Ref.~\cite{Tian_2020_xx1} analyze epidemic and mobility data across China to describe the impact and effect of NPIs during the first $50$ days of the pandemic. The mobility data was constructed aggregating information from location-based services (mobile apps), migration and Baidu datasets. The authors use a linear regression model to link the arrival of cases in other cities with the travel bans in Wuhan. The results show a very high correlation ($0.98$) and that cities with larger populations and more connected with Wuhan had earlier importations. The authors build also a linear regression model the link between the timing and type of NPIs implemented in $296$ cities and the number of cases in the first week of local outbreak. The results suggest how 
cities that acted early had a lower burden and that stopping intracity transportation, entertainment venues, and banning public gatherings are the NPIs more associated with reductions in cases. Interestingly, the results show also that the cordon sanitaire in Wuhan, though important for the suppression of the disease, induced only a delay of about $3$ days in the arrival of the disease in other cities. Ref.~\cite{Jing_2020_ff} uses detailed data from contact tracing in China to estimate the transmission patterns in households settings. The results suggest higher infectivity during the pre-symptomatic phase, higher susceptibility for people $60+$, and highlight the importance of case isolation and aggressive contact tracing of close contacts. Similar conclusions have been drawn in the context V\'o, a small municipality in Veneto (Italy)~\cite{Lavezzo_2020_ff}. The city has been the first to report a confirmed COVID-19 death in Italian soil and was one of the first hotspots of the pandemic in the country. The authors conduced two surveys collecting demographic, clinical, and contact network information as well as testing the presence of infection in a large fraction of the population ($85.9\%$ and $71.5\%$ respectively). The results show how when the lockdown was put in place (first survey) the prevalence was about $1\%$. After the lockdown (second survey) the prevalence was slightly smaller than $3\%$. Remarkably, about $42\%$ of cases were asymptomatic. The results informed the local policy makers and, due to the high number of asymptomatic, testing has been conducted to all contacts of positive individuals in the region of Veneto. This approach has been critical to suppress effectively the virus during the first wave. Authors in Ref.~\cite{Jefferies_2020_ff} present an analysis of all confirmed cases in New Zealand. The country is often considered as one of the models for COVID-19 suppression. In fact, the early and strong implementation of NPIs has resulted in (almost) disease elimination. The authors study a range of features for each subject: demographic, health outcomes, transmission patterns, serial intervals, and testing coverage across the different phases of the response. The results show how the implementation of NPIs reduced the serial intervals and isolation times. Testing increased across time and the response has shown to limit disparities across socio-economic groups. Ref.~\cite{Valencia_2020_ff} presents the analysis of NPIs in the context of Puerto Rico. The results, which focused on the initial phases of the outbreak, suggest an high efficacy of the measures. The authors estimate that without interventions the death toll could have been three times higher. Authors of Ref.~\cite{White_2020_ff} study the differences of disease burden across US States linking them with the heterogenous implementation and adoption of NPIs. In doing so, they included a wide range of correlates such as population density, fraction of population above $65+$, life expectancy, years of schooling, income, vaccination rate and testing rates. Furthermore, they adopted two indexes proxies of the tendency of each subpopulation to adopt measures which might benefit the community and the tendency to follow rules. Though in the USA NPIs widely varied across states the authors identified five which are the most adopted: declaration of state-emergency, limit size of social gatherings, school closure, strict restaurant operations, and stay at home orders. In the early phases of the pandemic lower density and higher vaccination rates were significantly associated  with slower growth rates. Restricting restaurant was the only variable predictive of doubling time. However, the authors note how states that implemented more actions earlier have been more effective in reducing the growth. In Ref.~\cite{Chernozhukov_2020_xx1} authors use an econometric approach to correlate policies, informations about the spreading, behavioral changes measured as variation in mobility from the Google mobility reports, with disease progression in the USA. The findings confirm both top-down and bottom-up NPIs had an impact in the mitigation effort. Ref.~\cite{Auger_2020_xx1} presents a statistical analysis of the effect of school closures on the progression of the disease in the USA. In doing so, they consider a range of covariates such as other NPIs, demographic and socio-economic informations. The results show that closures of school were linked, temporally, with a reduced disease burden measured both in terms of cases and deaths. Interestingly, they observe that the effectiveness of school closures has been higher when they were implemented earlier. The authors note how measuring the effect of a single NPIs when several are in place is particular hard. Finally, authors in Ref.~\cite{Kissler_2020_ff} study a population of $1746$ pregnant women in New York City that shared  information about their residence (borough), age and SARS-CoV-2 test outcomes. Authors adopt a Bayesian approach to quantify the prevalence in each borough from the sample. Furthermore they consider general mobility data captured from Facebook to study the link between prevalence and adoption of NPIs (estimated from the mobility data). The results suggest a city-wide prevalence of $15.6\%$ up to early May which however varied from $11\%$ in Manhattan to $26\%$ in South Queens. Interestingly, prevalence was low in borough with higher reduction in movements out and evening into them. \\
Authors in Ref.~\cite{Islam_2020_xx1} analyze the impact of NPIs in $149$ countries. In doing so, they built a linear regression model where the dependent variable is the logarithm of the number of cases at time $t$ and the independent variables describe the days since the first confirmed case, a dummy variable describing the present or absence of interventions, the days since the start of the interventions, and the population size. They assume a week delay from the implementation of NPIs to their effect. They also consider a range of other covariates such as different demographic and socio-economic indicators. Furthermore, they consider five NPIs: school closure, remote working, limits on social gatherings, closure of public transportation, and lockdown. Interestingly, in the period between $1^{st}$ of January and $30^{th}$ of May, the authors observe that at least one of those NPIs was put in place in all $149$ countries. All five have been put in place in $118$ countries, at least four in $25$ and just three in $4$. Tanzania put in place two (school closure and limits on social gatherings) and Belarus only one (school closure).  These observations clearly show the global impact of the COVID-19 pandemic. The results suggest how the implementation of any NPIs is linked to a decrease of $13\%$ in incidence. Higher GDP per capita, higher percentage of population about $65$ and country security index were associated with stronger reductions. The effects of implementing all five rather than four measures are found to be similar. A smaller reduction was found in countries implementing only three of them. Combinations of school closures, remote working and limits on mass gatherings with or without closure of public transport was linked to decreases in disease burden. Early implementation of lockdowns has been found associated with large reduction of incidence. Authors of Ref.~\cite{Haug_2020_ff} aim to estimate the impact of NPIs in $56$ countries by looking at the variations of $R_t$.  In particular, they studied the dataset discussed in details below, which provides a taxonomy and temporal information about the range of NPIs implemented across the globe. Interestingly, they compute the ranking of effectiveness considering four different methods: linear regression where each NPIs together with other covariates is used to estimate the variation of $R_t$, lasso regression done assuming that without intervention $R_t$ would be constant and that variations are linked to the implementation of NPIs, random forest considering the variation of $R_t$ and  neural networks that approximate $R_{t+1}$ taking as input the set of NPIs and $R_t$. The estimation of $R_t$ is done using EpiEstim which we have discussed above.  The results suggest how lockdown, limits on social gathering, remote working and school closures are the most effective NPIs. Borders restrictions are also found to have a significant impact. Other less costly measures such as risk communication strategies that promote healthy behaviors are found to be important. It is interesting to note how food programs and financial support are also significant as they help people to adopt NPIs without the risk of harsh economic consequences. Ref.~\cite{Koh_2020_ff} investigates the impact of NPIs in $142$ countries. In doing so, they link, via a regression model, the variations of $R_t$ with NPIs considering also several country specific covariates. The results confirm the effectiveness of lockdowns and travel bans. Furthermore they suggest that partial lockdown might be as effective as full lockdowns if implemented earlier. Authors in Ref.~\cite{Castex_2020_xx1} combine data from the Oxford COVID-19 Government Response Tracker (which is discussed below)~\cite{hale2020variation}, epidemiological indicators, Google mobility reports, and socio-economic data from the World Bank to study the impact of NPIs in $53$ countries. Differently than the other approaches mentioned above, here the authors use a compartmental SIR model with time-varying parameters and link the logarithm of the transmission rate for each country to the NPIs and other covariates via a regression model. Their findings suggest that the impact of school closures and remote working decreases with employment rate, portion of elderly, country extension, population density and increase for GDP per capita as well as expenditure in health care. The authors reflect on the fact that adoption rates of NPIs could be less pronounced in case of low risk perception, younger population, and a stronger health care system. Authors in Ref.~\cite{Kaur_2020_ff} study the time series of the cumulative number of cases in a range of countries such as India, USA, and Singapore. They study variance, autocorrelation, skewness and other indicators to capture signals of slow down linked to implementation of NPIs. The results confirm how the timing of intervention is a key variable that might explain successful suppression strategies.\\
Authors in Ref.~\cite{Lei_2020_ff} investigate the effects of NPIs induced by COVID-19 on the seasonal flu in China. The drastic measures put in place to hamper the spreading of SARS-CoV-2 were extremely efficient to stop transmission chains also of the flu. The findings suggest a reduction of $64\%$ respect compared with previous years. In the context of South Korea and with a wider range of diseases (i.e., chickenpox, mumps, invasive pneumococcal disease, scarlet fever and pertussis) Ref.~\cite{Huh_2020_ff} estimates the impact of NPIs put in place in response of COVID-19 on these diseases. They use an ARIMA (autoregressive integrated moving average) model built considering historical data to compare estimated incidence of each disease with observations. The results suggest a large reduction ($36\%$) for chickenpox. For mumps they observe a marginal reduction only in the young population ($<18$ years old). Reductions of about $80\%$ were reported for respiratory viruses. Ref.~\cite{Baker_2020_fv2} investigates the effects of NPIs on two endemic infections in the USA: respiratory syncytial virus and seasonal influenza. By studying detailed surveillance data and using an epidemic model, they authors note how behavioral changes induced by COVID-19 have reduced their transmission by at least $20\%$. As result of the increased fraction of susceptible population the results indicate the chances of large future outbreaks.

\subsection{Effects on behaviors}

Within this category we find articles exploring the impact of NPIs on online activities~\cite{Su_2020_xx1,Rousseau_2020_xx1,Li_2020_xx1,Nguyen_2020_xx1}, health~\cite{Slavova_2020_xx1,Jacobson_2020_xx1,Ward_2020_xx1}, mobility~\cite{Schlosser_2020_fv2},  spending~\cite{Sheridan_2020_xx1}, and performance of financial markets~\cite{Zaremba_2020_xx1}. \\
Authors in Ref.~\cite{Su_2020_xx1} study the effect of the lockdown in Twitter and Weibo users in Lombardy and Wuhan. The first is the most affected Italian region in the first wave. The second is the epicenter of the pandemic. In doing so, the mined content posted two weeks before and after the lockdowns. They extracted psycholinguistic features from the posts. The results suggest that in Wuhan the attention shifted towards religion, emotions and social groups. In posts written in Lombardy the authors noted a decrease of stress and an increase to topics related to leisure. Ref.~\cite{Rousseau_2020_xx1} studies data from Google Trends in $20$ European countries to investigate possible shifts in attention towards nature and environmental topics during the pandemic. Their findings suggest that, since the beginning of the COVID-19 crisis and the implementation of strict NPIs, interest towards nature increased significantly. The effect is not so visible for environmental topics instead. Ref.~\cite{Li_2020_xx1} studies the link between posts and mobility from Sina Weibo in Sichuan, China. In more details, authors investigate variations in the overall content posted and mobility (proxy of behavioral changes) as function of the different NPIs and COVID-19 phases. Initially the increase of cases is linked to increase of negative emotions and decrease of mobility. As the number of cases kept increasing the authors observe a decrease in negative sentiments posted and a stabilization of mobility. After the peak decrease of negative posts and increase of mobility is reported. Overall the results show how the peak of negative emotion proceeded the epidemic peak as well as the variation of mobility patterns thus hinting to a significant bottom-up reaction in the population. Ref.~\cite{Nguyen_2020_xx1} investigates the increase of anti-asian sentiments expressed on Twitter as result of the pandemic. Unfortunately, even political leaders publicly referred to COVID-19 as the \emph{chinese virus}. The authors mined about three million tweets from USA and used SVM (support vector machine) to estimate the sentiment associated to them. The collection considered a dictionary of more than $500$ keywords defined in previous race-related studies and a database of racial slurs. Unfortunately, the authors detect an increase of anti-asian posts by $67\%$ comparing pre and post-COVID-19 posts. \\
Ref.~\cite{Slavova_2020_xx1} studies the runs of emergency medical services for opioid overdose in Kentucky (USA) before and after the state of emergency declaration in the state. In a background of decrease of runs for other conditions the authors observed an increase of calls for opioid overdose by $17\%$, and an increase of $50\%$ in calls for suspected overdose with deaths at the scene. Overall the results are tragic, alarming and highlight the possible consequence of health emergencies that drastically modify societal activities. Authors in Ref.~\cite{Jacobson_2020_xx1} study the variations in Google searches related with mental health before and after the stay at home order declaration in several US states. The results indicate that searches linked to negative thoughts, anxiety, suicidal ideation among others increased before the stay at home order and level off at the time of the announcements. Despite the research is based on a very large number of searches, the period of analysis covers only one week. Hence, more work is needed to investigate these trends, potentially considering second/third waves and subsequent NPIs. \\
Ref.~\cite{Schlosser_2020_fv2} uses mobile phone data to study the changes in mobility patterns in Germany. Interestingly, they investigate a time period covering the first lockdown and the relaxation of NPIs put in place. The results suggest a long lasting variation in the mobility patterns that remained in place despite the lifting of top-down constraints. In fact, the authors show how long range movements across the country have been considerably reduced and as consequence the mobility network in the \emph{new normal} is considerably more local and clustered.\\
Authors in Ref.~\cite{Sheridan_2020_xx1} use real-time transactions data in Sweden and Denmark to understand the effects of COVID-19 and different NPIs strategies on consumer spending. The results suggest how during the emergency spending dropped on average by $25\%$ in Sweden and by $29\%$ in Denmark. The small differences is particularly interesting considering that Denmark adopted a suppression strategy with a lockdown rather than a mitigation strategies with general social distancing and other restrictions without the social nuclear option of a shutdown. The findings suggest how the reduction in spending is largely caused by the virus. Considering the much higher disease burden in Sweden these results highlight how suppression strategies are a better control method when considering also economic factors. Finally, authors in Ref.~\cite{Zaremba_2020_xx1} study the link between performance of financial markets and NPIs in $67$ countries. They consider stock returns volatility and study their association with NPIs. The results suggest higher volatility (which might induce large-scale sale of risky assets and higher cost of capital) in countries that implemented stricter NPIs policies. Interestingly, information campaigns and events cancellations are found to be the major contributors of volatility. 

\subsection{Effects on medical practice}

Here we find four articles investigating and estimating the effects of NPIs on medical practice. Authors in Ref.~\cite{Maringe_2020_re22} use historical data to estimate the effects of diagnostic delays for breast, colorectal, esophageal and lung cancer in the UK. In fact, NPIs limited drastically visits, specialistic follow-up and routine screenings. The results paint a very grim picture with a projected increase of avoidable cancer deaths. These go from $8-10\%$ for breast cancer to $15-17\%$ for lung cancer. The authors call for urgent policy interventions to mitigate these tragic losses. In the context of a cohort of patients living with HIV in Italy, Ref.~\cite{Quiros_Roldan_2020_re22} highlights the negative impact of the pandemic and NPIs. Visits missed, decrease of new diagnosis respect of historical baselines, and a drop of antiretroviral therapy dispensation paint a grim picture even in this case. Ref.~\cite{Wood_2020_re22} aims to characterize the increase in ICU deaths linked to limited capacity. In the context of the UK, the modeling scenarios developed by the authors suggest that a significant impact could be obtained doubling beds and reducing length of stay by $25\%$. Finally, Ref.~\cite{Schuengel_2020_re22} studies the effect of NPIs in a long term care institution in the Netherlands specialized in intellectual disabilities. The authors focused on incidents reported during the emergency respect to historical data. The lockdown has caused a reduction across all types of incidents however, those linked with aggression spiked up quickly soon after highlighting the need for developing specific protocols which are better suited for patients with intellectual disabilities.  

\section{Reviews}

\begin{figure}[t]
  \centering
   \includegraphics[scale=0.25]{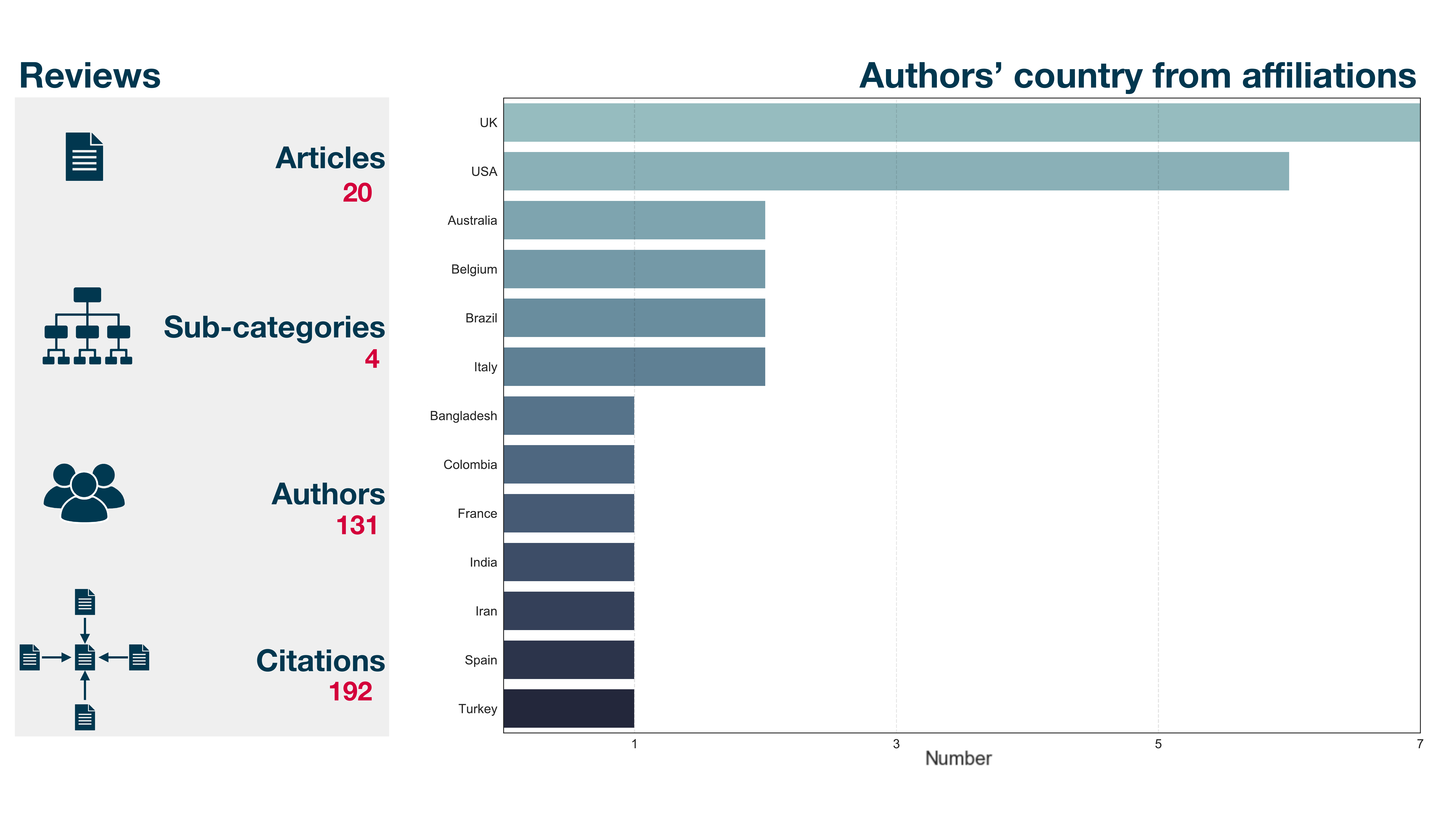}
  \caption{Total number of authors and citations of the papers in this category on the left. Note that these numbers are estimated from Semantic Scholar. On the right histogram describing the number of countries representing authors' affiliations.}
  \label{fig:fig6}
\end{figure}

Not surprisingly, I am not the first to think about writing a review about NPIs during the COVID-19 pandemic. In fact, the sample of papers I am analyzing here contains $21$ articles which can be classified as reviews. Some of them could be also considered as perspectives and/or comments, but I opted to include them here as they provide a broad overview of a particular topic or context. This section will be much shorter than the others. In fact, considering the nature of these articles, providing a summary of them is probably more appropriate for meta-analyses.\\
The efforts behind the compilation of these articles come from more than $131$ authors with affiliations in $13$ countries (see Fig.~\ref{fig:fig6}). The UK and USA are at the top of the list. Furthermore, in terms of publication venues, we find $20$ journals that range from interdisciplinary domains such as \emph{Nature Communications} to more specialized such as \emph{BMC Oral Health}. These articles already received $192$ citations.

\subsection{Classification}

By looking at the angle and focus, the sample of reviews can be further grouped in $4$ categories (see Table~\ref{table: reviews}):

\begin{enumerate}
\item \textbf{Non-pharmaceutical interventions}: describes articles which focus on different aspects of NPIs such as types, effectiveness, and their modeling;
\item \textbf{Change in medical practice}: captures articles summarizing approaches, results, and success stories about the adaptation of medical practices to NPIs;
\item \textbf{Adoption of NPIs}: considers articles discussing the issues and challenges linked to adoption of NPIs in different contexts;
\item \textbf{Impact of digital technologies}: describes reviews summarizing the opportunities and challenges that new technologies (such as mobile phones) offer in the fight against COVID-19
\end{enumerate}

\begin{table}[]
\begin{tabular}{|l|l|}
\hline
\textbf{Category}       & \textbf{Topic and references} \\ \hline
Non-pharmaceutical interventions &\acapo{Effectiveness of NPIs~\cite{Ray_2020_xx5,Pati_o_Lugo_2020_xx5,Imai_2020_xx5,Hens_2020_xx5,Alvi_2020_xx5}; \\Effects of NPIs~\cite{Pedrosa_2020_xx5,Kneale_2020_xx5,Cooper_2020_xx5,Brooks_2020_xx5,Lemke_2020_xx5}; Epidemic modeling~\cite{Estrada_2020_xx5}} \\ \hline
\hline
Change in medical practice     &   
\acapo{Surgery~\cite{Wall_2020_xx5,De_Simone_2020_xx5}; Ophthalmology~\cite{Romano_2020_xx5}; \\Disease control~\cite{Dinleyici_2020_xx5}; Dentistry~\cite{Banakar_2020_xx5}}    \\ \hline
\hline
Adoption of NPIs            &    \cite{Jiwani_2020_xx5,Seale_2020_xx5}                 \\ \hline
Impact of digital technology           & \cite{Budd_2020_xx5,Grantz_2020_xx5}                    \\ \hline
\end{tabular}
\label{table: reviews}
\caption{Summary of reviews articles in the sample. The first column describe the categories. The second instead, the topic and references}
\end{table}

\subsection{Non-pharmaceutical interventions}

In this category we find reviews that discuss the effectiveness of NPIs~\cite{Ray_2020_xx5,Pati_o_Lugo_2020_xx5,Imai_2020_xx5,Hens_2020_xx5,Alvi_2020_xx5,Feng_2020_xx5}, their effects~\cite{Pedrosa_2020_xx5,Kneale_2020_xx5,Cooper_2020_xx5,Brooks_2020_xx5,Lemke_2020_xx5} and modeling~\cite{Estrada_2020_xx5}.\\ 
In particular, Ref.~\cite{Ray_2020_xx5} reviews the modeling efforts measuring the effects of NPIs in the Indian context. Authors in Ref.~\cite{Pati_o_Lugo_2020_xx5} summarize the evidences about NPIs efficacy and approaches adopted across the globe. They highlight how portfolios of measures appear to be more effective than single interventions. Similar conclusions are drawn in Ref.~\cite{Imai_2020_xx5} where the authors also stress the key importance of right timing in the implementation of NPIs. Ref.~\cite{Alvi_2020_xx5} reviews the evidence about the efficacy and challenges of both pharmaceutical and non-pharmaceutical interventions.\\
Ref.~\cite{Pedrosa_2020_xx5} reviews the literature focused on the psychological impacts of NPIs with particular attention to minorities and disadvantaged groups, which unfortunately have been more affected by both disease and NPIs. Authors in Ref.~\cite{Kneale_2020_xx5} review the evidences about the impact of NPIs on education. They highlight the importance of considering the ripple effects that school closures have on students, parents, economy, and mental health. Ref.~\cite{Cooper_2020_xx5} analyzes the literature focused on the effects of NPIs on eating disorders. In doing so, it provides a range of suggestions and approaches to tackle the issue. Authors in Ref.~\cite{Brooks_2020_xx5} study the literature about the psychological impact of the pandemic on pregnant women highlighting the particular needs of this high-risk group. In the same spirit of considering a particular group of individuals, Ref~\cite{Lemke_2020_xx5} investigates the effects of NPIs and COVID-19 on professional drivers pointing out how the emergency has introduced a wide range of new stressors that sum up to those already in place. Finally, Ref.~\cite{Estrada_2020_xx5} provides a summary of the modeling approaches put forward to model COVID-19 and capture NPIs. 

\subsection{Change in medical practice}

In this category we find articles that summarize the evidence, approaches and suggestions on how to adapt medical practice in the middle of the COVID-19 pandemic. We find articles focused on surgery~\cite{Wall_2020_xx5,De_Simone_2020_xx5}, ophthalmology~\cite{Romano_2020_xx5}, disease control (routine vaccination campaigns)~\cite{Dinleyici_2020_xx5} and  dentistry~\cite{Banakar_2020_xx5}. 

\subsection{Adoption of NPIs}

Here we find two articles that provide a summary about the factors influencing the adoption of NPIs. Ref.~\cite{Seale_2020_xx5} confirms how adoption of health behaviors is linked to a range of factors, form socio-economic to psychological. Authors in Ref.~\cite{Jiwani_2020_xx5} summarize the practical issues linked  to NPIs adoption in several African countries where lack of water and soap are unfortunately widespread.   

\subsection{Impact of digital technology}

Here, we find two reviews that highlight the incredible opportunities offered by digital technologies to capture behaviors, their changes and inform modeling as well as policies~\cite{Budd_2020_xx5,Grantz_2020_xx5}. Interestingly, they also discuss the issues linked to privacy, legal, ethical concerns and biases in the data.

\section{Measuring NPIs via proxy data}

\begin{figure}[t]
  \centering
   \includegraphics[scale=0.25]{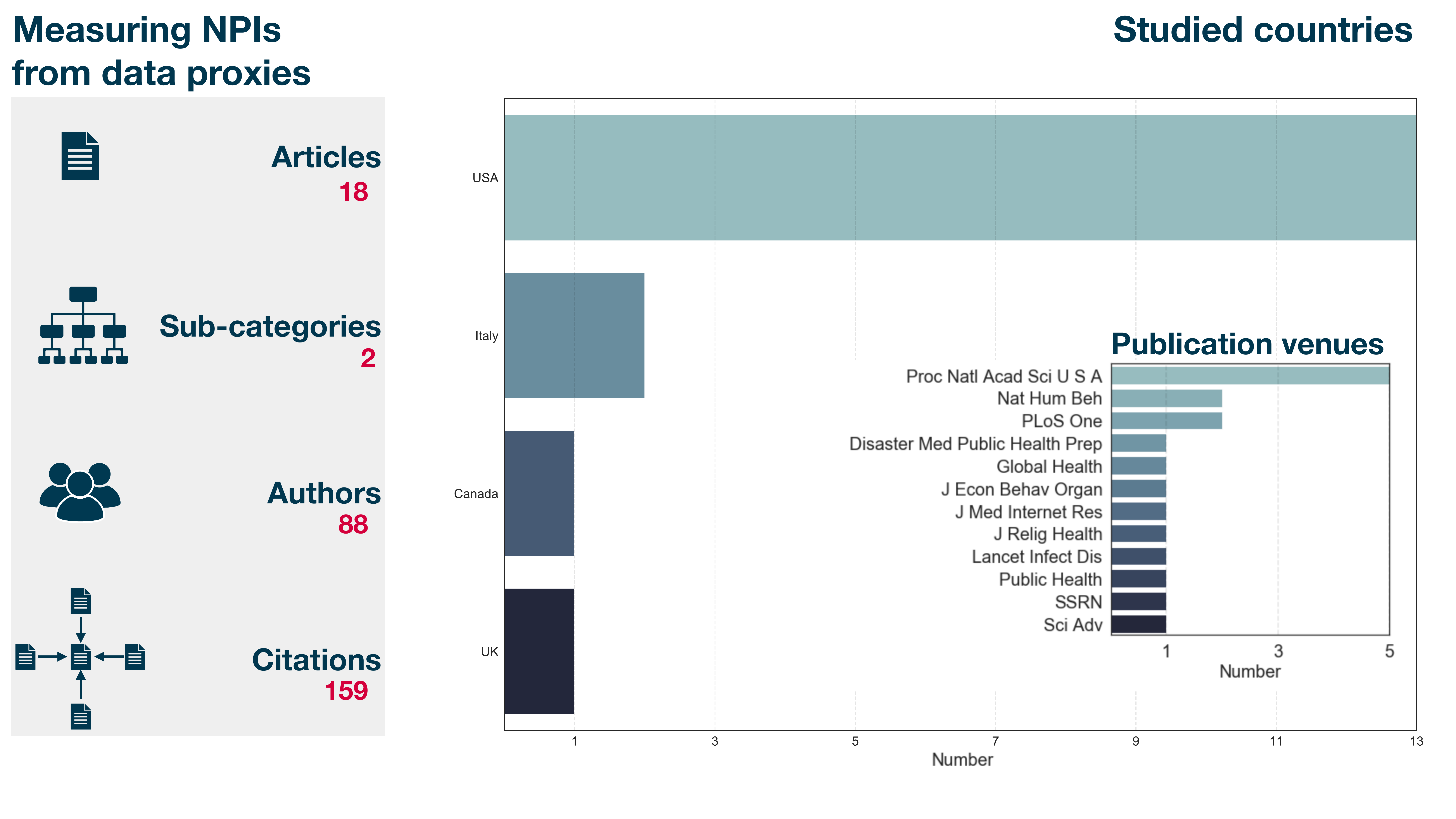}
  \caption{Total number of authors and citations of the papers in this category on the left. Note that these numbers are estimated from Semantic Scholar. On the right (top) histogram describing the number of countries subject of study in this category. Note how few papers using data from hundreds of countries are not counted in the histrogram. On the right (bottom) most represented publication venues of the category.}
  \label{fig:fig7}
\end{figure}

Every call we make or receive, every time we connect to internet via our mobile phones, all posts we leave on social media platforms, all queries we submit in search engines and all the times we swipe our credit cards are just few examples of our daily activities that leave digital traces. The data is recorded and stored by a wide range of companies to enable services, meet legal obligations, and billing purposes. However, such data offers unprecedented opportunities to measure and understand a variety of human behaviors which are of epidemiological relevance~\cite{salathe2012digital}. Information about which cell towers we are connected to when making a call or surfing the web can be used to infer our location and movements. The content, topic, and other meta data linked to our posts on Twitter, Facebook or online searches might be studied to estimate our health status, our concerns about an health emergency, judge our level of exposition to low quality information or our mobility. Our digital payments might be used to detect life changing moments (i.e., pregnancies), infer our mobility patterns, or gathering socio-economic  information about people similar to us, people living in our neighborhoods, city or country. In general, this data offers high-resolution proxies for many societal activity and behaviors. Due to serious privacy risks and concerns such data is stored in what we can call digital vaults. Luckily, over the last few years corporations, academia, and institutions at various levels have been working to unleash the potential hidden in these proxies while taking all the necessary steps to preserve privacy and to respect stringent regulations. In fact, a range of so called \emph{Data for Good} initiatives have been put in place with the aim of leveraging high-resolution datasets collected by companies in non-for-profit applications. I said luckily because these ecosystems, their success, and great potential were in places before the COVID-19 pandemic. This, together with the scale and magnitude of the emergency, has allow tech giants such as Google, Apple, Facebook and large mobile phone operators such as Telefonica, Vodafone, Orange as well as smaller companies such as SafeGraph, Cuebiq among others to promptly step-up and provide invaluable data to the scientific community. It is hard to imagine these efforts put in place so quickly, safely, and extensively just five-ten years ago. COVID-19 is probably the largest data for good effort ever assembled. \\ 
In this category, we find articles aiming to measure, quantify, monitor and capture NPIs via data proxies. It is interesting to stress the range of data types and sources used. From information collected via mobile phone apps, services, GPS, and wifi connections to posts on Twitter, Reddit, views of Wikipedia pages or credit card records. \\
Though the large majority of studies in the sample are based on data collected in the USA, the penetration of mobile phones allowed researchers to mine data from nearly all countries in the world (see Fig.~\ref{fig:fig7}). These research efforts have been produced by more than $88$ authors, affiliated with institutions from $7$ countries. Again USA, Italy and UK are on the top of the list. In terms of publication venues, this sample has been published in $12$ journals. \emph{PNAS}, \emph{Nature Human behavior} and \emph{PLoS One} are in the top of the list. It is not surprising to see these interdisciplinary journals in this category. In fact, the research efforts are based on a fusion of different approaches and techniques which are not bounded by specific scientific domains. In terms of citations, this sample collected (as of Dec 19) $159$ citations.

\subsection{Classification}

Considering their topics and aims these articles can be classified in two categories (see Table~\ref{table: measuring}):

\begin{enumerate}
\item \textbf{Adoption of NPI}: describes articles that tackle the issue of NPIs adoption;
\item \textbf{Mobility changes}: are papers that study the effects of NPIs on human mobility
\end{enumerate} 
 
Let's dive in some details.

\begin{table}[]
\begin{tabular}{|l|l|l|}
\hline
\textbf{Category}             & \textbf{Proxy data type} & \textbf{Country and reference} \\ \hline
\multirow{5}{*}{Adoption of NPI}  & \acapo{Mobile phones and/or vehicles\\ through location intelligence\\ and measurement platforms}              &             USA~\cite{Charoenwong_2020_xx1,Hill_2020_wq2,Hsiehchen_2020_xx1,Grossman_2020_xx1,Gollwitzer_2020_xx1,Wright_2020_xx1}                   \\ \cline{2-3} 
                                  & Twitter                  &         Twitter population~\cite{Gallotti_2020_xx1,Sutton_2020_xx1}                       \\ \cline{2-3} 
                                  & Wikipedia and Reddit     &   Canada, Italy, USA, UK~\cite{Gozzi_2020_xx1}                            \\ \cline{2-3} 
                                                    & Google                 &   USA~\cite{hartwell2020association}; 130 countries~\cite{M_Herren_2020_xx1} 
\\ \cline{2-3}                      
                                  & TV                       &    USA~\cite{Kim_2020_xx1}                            \\ \hline
                                  \hline
                                  \multirow{3}{*}{Mobility changes} & \acapo{Mobile phones and/or vehicles\\ through location intelligence\\ and measurement platforms}            &       USA~\cite{lee2020human,Xiong_2020_xx1, Weill_2020_xx1,Badr_2020_xx1}                       \\ \cline{2-3} 
                                  & Facebook                &     Italy~\cite{Bonaccorsi_2020_xx1}                           \\ \cline{2-3} 
                          & Citymapper app               &        22 countries~\cite{Vannoni_2020_xx1}                                              \\ \hline
\end{tabular}
\label{table: measuring}
\caption{Summary of the articles measuring NPIs via proxy data. The first column describes the main categories. The second the type of data used. The third, the country of study and the references}
\end{table}

\subsection{Adoption of NPIs}

Here, we find articles aimed at quantifying the adoption of NPIs via proxy data such as mobile phones collected by location intelligence companies~\cite{Charoenwong_2020_xx1,Hill_2020_wq2,Hsiehchen_2020_xx1,Grossman_2020_xx1,Gollwitzer_2020_xx1,Wright_2020_xx1}, Twitter~\cite{Gallotti_2020_xx1,Sutton_2020_xx1}, 
COVID-19 ad-hoc apps~\cite{Allen_2020}, Wikipedia and Reddit~\cite{Gozzi_2020_xx1}, Google~\cite{M_Herren_2020_xx1,hartwell2020association}, and TV viewings~\cite{Kim_2020_xx1}.\\
Authors in Ref.~\cite{Charoenwong_2020_xx1} use data from SafeGraph and Facebook to quantify social distancing at the county level in the USA. In particular, SafeGraph provides information about access to locations and POIs. Home locations are obtained considering the typical position of each phone in the night.  Work locations are inferred considering where phones spend three (i.e., part-time) or six (i.e., full-time) hours outside home. These locations are identified using data from January, hence pre-NPIs. Aggregating all the observations at the county level the authors identified a social distancing index for each county $i$ in day $t$  defined as:
\begin{equation}
SD_{i,t}=\frac{H_{i,t}}{N_{i,t}-W_{i,t}}
\end{equation}
where $H_{i,t}$ describes individuals that appear not to have left home in day $t$, $W_{i,t}$  those that appear to have gone at work, and $N_{i,t}$ the total number of devices in county $i$. The authors measured also the (online) social connectedness of each county considering Facebook friendship data as:
\begin{equation}
SC_{i,j}=\frac{FB_{i,j}}{U_{i}U_{j}}
\end{equation}                 
where $FB_{ij}$ is the number of social connections in Facebook between county $i$ and county or country (if outside USA) $j$, $U_{x}$ is the number of Facebook users in each considered location. They also study the timeline of NPIs put in place, epidemic indicators and built a regression model to find the association between social distancing with social connectedness, NPIs, and cases. The results show that counties with higher online social ties with China and Italy where more likely to comply with NPIs. This point to the fact that being exposed to informations from highly affected areas might influence risk perception and thus behavior change. Furthermore, the results show differences between political leanings, education levels, and stands about climate changes. The link between political affiliations and NPIs adoption is studied also in Ref.~ \cite{Hsiehchen_2020_xx1}. Here, authors use mobile phone data to test wether aggregated variation in mobility are associated with political leanings at the state level. By using a regression model the authors find a significant negative correlation between fraction of republicans and NPIs adherence (measured in terms of mobility reduction). This link was found to be significant adjusting for population, poverty rates, Gini index, urbanization levels, and portion of essential workers. Similarly authors in Ref.~\cite{Grossman_2020_xx1} study mobile data collected by SafeGraph to measure the association between political views and mobility reductions. In doing so, they considered behaviors, aggregated at the level of counties, after governors issued recommendations to physical distance. These typically preceded stay at home orders or other top-down restrictions. By using a regression model, built considering a wide range of covariates, the author confirm the role of political leanings in response to such recommendation thus in self-initiated behavioral changes. The recommendations had a significant effect on overall mobility, comparable to the stay at home orders. The effects are found to be stronger in democratic counties and even stronger in case of republican governors. The link between mobility reduction and political affiliation is studied also in Ref.~\cite{Gollwitzer_2020_xx1}. Here, the authors used mobile phone data collected by Unacast from about $15$M users in the USA as well as consumption levels of conservative traditional media (i.e., TV). Their results confirm a correlation between adherence to NPIs and political leanings at the level of counties. In fact, counties that in 2016 voted for Trump show, on average, $14\%$ less mobility reduction and those that went to Hillary Clinton. Also in this case, political affiliation was found to be a stronger predictor of mobility variation than other variables such as density, income, and demographic informations. The authors point out how counties that voted for Trump have been hit harder by the virus and show higher infections and fatality rates. Authors of Ref.~\cite{Hill_2020_wq2} use mobile phone data collected by Cuebiq to investigate the possible association between adoption of NPIs (proxied as mobility reduction) and religiosity of US states (measured via six indexes). The results from a regression analysis suggest that states classified as more religious showed a lower and slower reduction of mobility. Furthermore, stay at home orders appear to have weaker effects in such states. The results highlight once again how adoption of behavioral changes is a multifaceted problem linked to a wide range of perceptions and beliefs.  Authors in Ref.~\cite{Wright_2020_xx1} use mobile phone data collected by Unacast  to study the socio-economic factors influencing adherence to NPIs, measured as variation in mobility following stay at home orders, in the USA. By using a regression approach the authors confirm how low income counties complied less with the orders than high income ones. These effects are visible considering confounding factors as political leanings, population density, unemployment rates among others.\\
Authors in Ref.~\cite{Gallotti_2020_xx1} tackle a key issue which might have dire consequence in adoption of protective behaviors: the spreading of mis-information on social media. Focusing on Twitter, they collected and mined $100$ Millions of tweets and develop a \emph{Infodemic Risk Index} to monitor the exposure to low quality information of users across countries. Their results show that the spread of unreliable information typically precedes the local rise of cases potentially affecting adoption to NPIs. Interestingly however, after the initial phases reliable information seems to take over. Ref.~\cite{Sutton_2020_xx1} studies the communication strategies of official accounts (i.e., public health, emergency services, officials) on Twitter during the evolution of the emergency. In fact, the ability to reach effectively a large user base to promote safe and verified recommendation as well as explaining measures and regulations in place is key. The results show how the inclusion of engaging content such as videos/photos and hashtags increases sharing. Posts with content linked to epidemic indicators, technical information, content related to measures to contain the spreading and flatten the curve were re-shared more widely. Authors in Ref.~\cite{Gozzi_2020_xx1} study users attention to the COVID-19 emergency considering as proxy Wikipedia page views and discussion on Reddit. The results show how these were initially driven by media coverage (traditional media such as news articles and YouTube videos published by news organization). However, while media coverage remained high, the authors found evidence of decline in public attention raising questions about risk perception and COVID-19 fatigue on NPIs  compliance. Authors of Ref.~\cite{hartwell2020association} study the association between NPIs related keywords on Google Trends and disease burden in each US state. Using a regression model they find that attention towards NPIs related keywords, measured using proxy queries in Google, is associated ($0.42$) with higher deaths per capita and case-fatality ($0.60$). They also confirm the association between political leanings. Ref.~\cite{M_Herren_2020_xx1} studies the association between mobility changes measured via the Google mobility reports and economic, social and governmental features in $130$ countries. The results indicate a reduction of mobility across the board with patterns which are specific to the evolution of the epidemic in each location. Countries with more authoritarian governments (where Google is however less likely to have access) were found to be more responsive to changes of mobility as cases increased. This point to another dimension of compliance to NPIs. Finally, Ref.~\cite{Kim_2020_xx1} studies local television coverage and survey data in rural settings in the USA. The results suggest that despite local news tend to be \emph{urban-centric} they can still affect adoption of NPIs. In fact, the authors find that rural residents are more likely to adopt social distancing if they live in a media market more affected by the disease, which is driven by close by cities. The survey data allowed the authors to confirm, at least for a sample, that these effects are linked to local news viewings. The authors note how though significant these effects are about half respect to those driven by partisanship.

\subsection{Mobility changes}

About half of the papers that measured NPIs via proxy data focused on the changes in mobility that travel bans, school closure, remote working, local and national lockdowns have induced in the population. While there is rich variety of data sources used, this data is collected via mobile phones by location intelligence and measurement platforms~\cite{lee2020human,Xiong_2020_xx1, Weill_2020_xx1,Badr_2020_xx1,Hill_2020_wq2,Hsiehchen_2020_xx1}, tech giants such as Facebook~\cite{Bonaccorsi_2020_xx1,Charoenwong_2020_xx1} and Google~\cite{M_Herren_2020_xx1,Kraemer_2020_xx1}, as well as other mobile apps~\cite{Vannoni_2020_xx1}. Indeed the high penetration of smart phones, their portability, the vital role they took in our daily lives, the popularity of some apps and services make them the perfect tool to infer high-resolution data about our movements and their possible changes. \\
Ref.~\cite{lee2020human} study data collected via mobile phone as well as vehicles to characterize the variations of aggregated mobility in the USA during the early phases of the pandemic (January-April). The results indicate how the overall mobility decreased markably following the national emergency declaration. Interestingly, the results confirm how the population reacted before the stay at home mandates by reducing their mobility voluntarily maybe nudged by news and increase risk perception. The observed patterns however shows high levels of spatial heterogeneity linked to income and population density. Authors in Ref.~\cite{Xiong_2020_xx1} study the data collected from more than $100$M users in the USA from third party data providers. They focused on county level mobility patterns. The data can be explored via an interactive dashboard \url{https://data.covid.umd.edu/}. The observations show how travels outside counties decreased by $35\%$ when the national emergency was declared. The reduction however did not last long. By using a simultaneous equations model, they investigated the link between the inflow in each county and the number of infections. The model is based on two expressions. One for the number of cases in county $i$ in a day $t$, $y_{i,t}$. To limit fluctuations this number is computed considering $t \in [T,T+7]$, thus a seven day window. The second is an expression for the inflow $x_{i,t}$ in the same county. In this case $t \in [t-l,l]$ where $l$ is a lag period. In both expressions they include autoregressive terms, as well as county and temporal covariates. Using this approach, they observed a particularly strong association between the mobility and infections in states that relaxed the NPIs. Ref.~\cite{Weill_2020_xx1} adopts a range of datasets: exposure indexes (which effectively estimate the number of devices in contact in 14-days periods, as done by many contact tracing apps) computed by PlaceIQ~\cite{couture2020exposure}, Google mobility reports and data from SafeGraph. Their observations indicated how poorer areas decrease mobility less than wealthier. Even more, they observed a reversal: wealthier areas were more mobile than poorer before the emergency. This findings confirm observations from many others articles and, considering that low income neighborhoods are affected by low access to health care and higher level of preexisting conditions, point to the fact that NPIs and the pandemic are affecting society heterogeneously, targeting disproportionally already challenged groups. Authors of Ref.~\cite{Badr_2020_xx1} use mobile phone data collected by Teralytics, a location intelligence company, describing the mobility across and within counties in the USA. They defined a mobility ratio (MR) metric to measure the variations of mobility patterns in each county $i$ at time $t$ as:
\begin{equation}
MR_{j,t}=\frac{V_{j,j,t}+\sum_{k\neq j} V_{k,j,t}+\sum_{k\neq j} V_{j,k,t}}{V_{j,j,t_0}+\sum_{k\neq j} V_{k,j,t_0}+\sum_{k\neq j} V_{j,k,t_0}}
\end{equation} 
where $V_{k,j,t}$ describes the number of trips from county $k$ to $i$ at time $t$ and $t_0$ is the baseline values before the emergency thus in absence of NPIs. Note how in both numerator and denominator the second and third terms are not the same quantity repeated by consider in and out flow each county. They then computed the growth rate ratio for epidemiological data. They compare the moving average of the number of cases in each $3$ and $7$ days periods. They studied the association between the two metrics in $25$ counties. The results show a correlation between variation of mobility (induced by NPIs) and disease growth ration above $0.7$ in $20$ out of $25$ counties. The changes in mobility however are not visible in the progress of the disease before $9-12$ days and up to three weeks, which is compatible to generation times. Furthermore, they confirm how changes in mobility preceded official restrictions. Authors in Ref.~\cite{Bonaccorsi_2020_xx1} use Facebook mobility data to study the impact of strict NPIs put in place in the first wave in Italy. They also consider socio-economic data. Their results confirm, also in the Italian context, how in cities with higher fiscal capacity where able to reduce more mobility. Furthermore, they observe stronger reductions in municipalities with higher levels of inequalities. Authors in Ref.~\cite{Vannoni_2020_xx1} use data from $41$ cities in $22$ countries collected via the transportation app Citymapper. They study the association between mobility and NPIs policies. Across the board the find a strong correlation between the decline in mobility and the COVID-19 Government Response Stringency Index~\cite{hale2020variation}. The variation of mobility linked to a specific NPIs is stronger for closing public transport ($18\%$) probably due to the nature of the data, workplace closure ($15\%$), restriction of internal movement ($13\%$), school closure ($10\%$) and borders restrictions ($5\%$).

\section{Datasets}

\begin{figure}[t]
  \centering
   \includegraphics[scale=0.25]{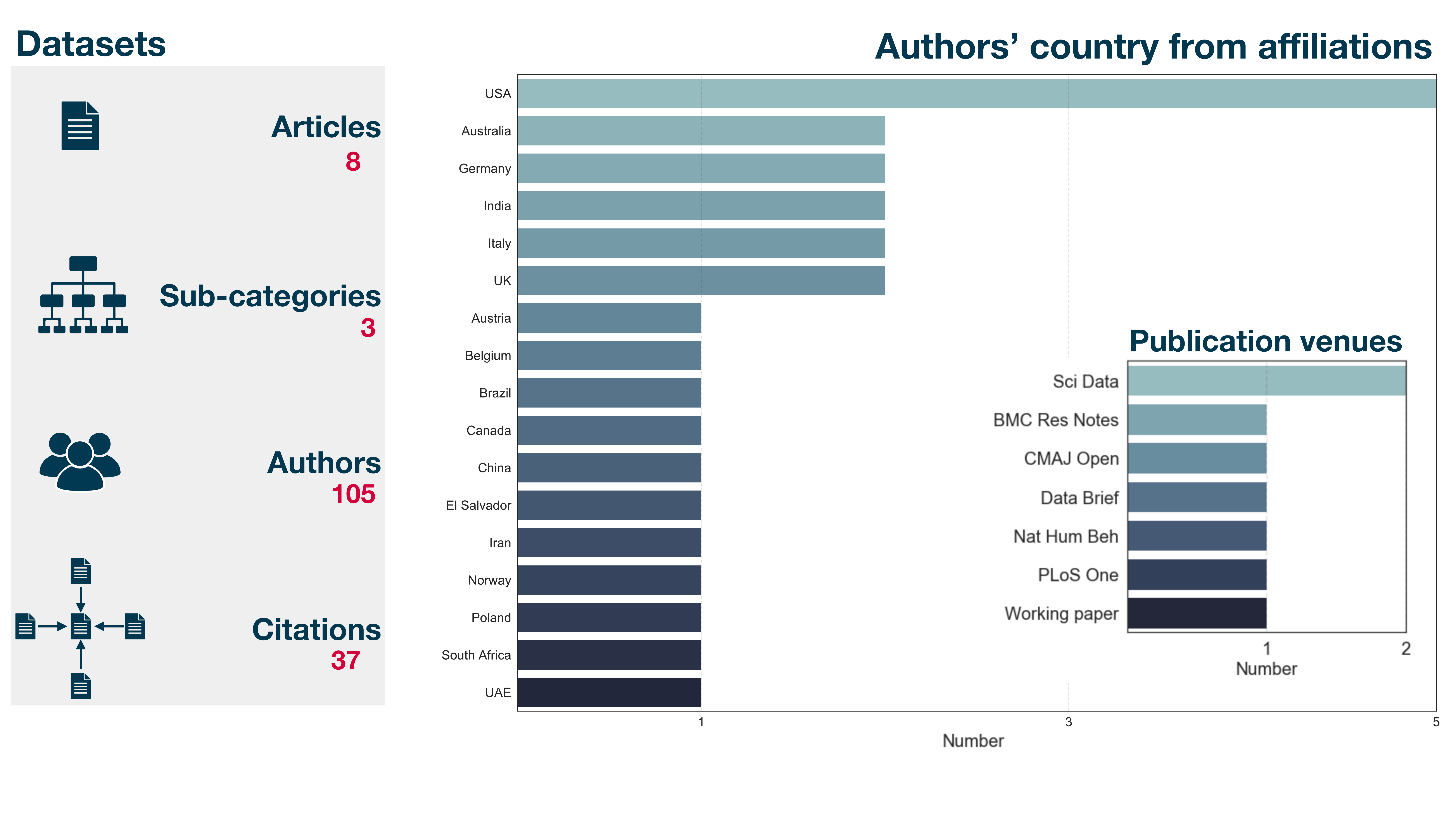}
  \caption{Total number of authors and citations of the papers in this category on the left. Note that these numbers are estimated from Semantic Scholar. On the right (top) histogram describing the number of countries describing authors' affiliations. On the right (bottom) most represented publication venues of the category.}
  \label{fig:fig8}
\end{figure}

The last category describes publicly available datasets describing different aspects of NPIs. Here, I am considering only datasets that have been featured in publications and shared with the community. There are many popular others that we mentioned several times above such as Google mobility reports, Apple mobility reports, and Facebook Data for Good datasets which have been put forward by the corporate world through their data for good initiatives. \\
The datasets, especially those describing the timeline of NPIs, cover large part of the countries in the world. They have been shared thanks to the work of more than $105$ authors affiliated in institutions from $17$ countries (see Fig.~\ref{fig:fig8}). In terms of publication venues, we find six journals. The most represented, with two publications, is \emph{Scientific Data} which is a journal from the Springer Nature family dedicated to description and sharing of datasets. Others journals we find are \emph{Nature Human behavior} and \emph{PloS One}. As of December $19^{th}$ these articles received $37$ citations, with a median of $1.5$.

\subsection{Classification}

The datasets in the sample can be divided in three categories (see Table~\ref{table: dataset}): 
\begin{enumerate}
\item \textbf{Implementation of NPIs}: describes datasets providing detailed information about the timelines and typology of NPIs;
\item \textbf{Mobility}: captures variations to human mobility induced by NPIs;
\item \textbf{Contacts}: describes datasets providing information about physical contacts which can be used to feed many of the age-structured epidemic models discussed above
\end{enumerate}

Let's dive into some details.

\begin{table}[]
\begin{tabular}{|l|l|}
\hline
\textbf{Category}       & \textbf{Country and references} \\ \hline
Implementation of NPIs &  \acapo{Government response trackers: Canada~\cite{McCoy_2020_rf3}, 195 countries~\cite{Cheng_2020_rf3},\\ 180 countries~\cite{hale2020variation}, 56 countries~\cite{Desvars_Larrive_2020_rf3}; Universities response in USA~\cite{Cevasco_2020_rf3}}\\ \hline
\hline
Mobility     &   \acapo{Australia, Brazil, China, Ghana, India, Iran, Italy, Norway, South Africa, \\USA~\cite{Barbieri_2020}; Italy~\cite{Pepe_2020_rf3}}
    \\ \hline
    \hline
Contacts            &    \acapo{Belgium, China, Finland, France, Germany, Italy, Hong Kong, Luxembourg,\\ Netherlands, Poland, Peru, Russia, South Africa, UK, Vietnam, Zambia, \\Zimbabwe~\cite{Willem_2020_xx1}}                \\ \hline
\end{tabular}
\label{table: dataset}
\caption{Summary of the articles presenting a dataset relevant for NPIs. The first column is the main category. The second provides information about country/countries of study and references}
\end{table}

\subsection{Implementation of NPIs}

Here we find datasets describing institutional interventions. The majority of them focused on governmental responses. One considered the specific context of Universities. \\
Ref.~\cite{hale2020variation} describes the \emph{Oxford COVID-19 Government Response Tracker}. The data can be downloaded and explored via this url: \url{https://ourworldindata.org/policy-responses-covid}. The dataset provides information about $17$ indicators capturing governmental responses. The taxonomy is organized in four main categories. The first is \emph{containment and closure} which is further divided in school closures, workplace closures, public event bans, restrictions of social gathering size, closure of public transport, stay at home requirements, national and international travel restrictions. The second type is \emph{economic response} which is divided in income support, debt/contract relief for households, fiscal measures, giving international support. The third category is \emph{health systems} which captures public information campaign, testing policies, contact tracing, emergency investments in health care as well as vaccines, facial coverings and vaccination policies. The fourth describes other responses. Some of the variables are ordinal, others are numerical. Furthermore, $11$ variables provide information at a geographical level. The resolution can go from country to city. Beside the data, the authors provide \emph{ready to use} indexes that can be readily adopted in modeling efforts. These are: 1) the overall governmental response index, 2) the stringency index, 3) the containment and health index, 4) the economic support index.\\   
Ref.~\cite{Desvars_Larrive_2020_rf3} provides data describing the implementation of NPIs in $56$ countries that can be downloaded form this url: \url{https://github.com/amel-github/covid19-interventionmeasures}. The taxonomy created is organized in four levels. The first is divided in 1) case identification, contact tracing and related measures, 2) environmental measures, 3) healthcare and public health capacity, 4) resource allocation, 5) risk communication, 6) social distancing, 7) travel restriction, and 8) returning to normal life. The other levels provide finer and finer description of each measure. The data comes as a timeline of events. Each one with a specific ID, country, state, region, date, classification in the four levels, status (which describe whether the measure is an extension of an old measure), comments, and source of information. \\ 
Authors in Ref.~\cite{Cheng_2020_rf3} present the \emph{CoronaNet} dataset with the timeline of NPIs implemented in $195$ countries. The dataset is available here \url{https://www.coronanet-project.org/}. The data is organized in a taxonomy that provides information about i) type of NPIs implemented (16 in total) ii) the type of government implementing the intervention (i.e. national or provincial) iii) geographical target of the measure (i.e. national, provincial) iv) the human or material target of the NPI (i.e. face masks, travelers) v) the directionality of the policy (i.e. outbound, inbound) vi) the mechanism of travel targeted by the measure (i.e. trains, flights), vii) whether the measure is mandatory or voluntary, viii) the nature of the enforcer (i.e. government, military) ix) the timing of the policy (i.e. date of announcement and/or implementation). Besides the description of the dataset, Ref.~\cite{Cheng_2020_rf3} applies a Bayesian approach to study the evolution of NPIs strategies across countries. They observe a clear acceleration of more costly measures in mid-March 2020, when indeed we observed a rapid escalation as mentioned above.\\
Authors in Ref~\cite{McCoy_2020_rf3} apply the taxonomy defined by the Oxford COVID-19 Government Response Tracker to the case of Canada. The data, together with many other relevant datasets from Canada and other countries, can be found here: \url{https://howsmyflattening.ca/#/data}. \\
Finally, authors in Ref.~\cite{Cevasco_2020_rf3} describe a dataset describing the strategies adopted by $575$ US universities. The data can be explored here: \url{https://covidtracking.com/}. By looking at official COVID-19 pages created by Universities, which describe the portfolio of measures put in place, the authors studied five (binary) variables: 1) remote learning 2) discourage campus housing 3) travel bans 4) close campus 5) remote working. It is interesting to note how the analysis of the data, points out very high heterogeneity in terms of NPIs in the country.  

\subsection{Mobility}

In this category we find datasets describing the effects of NPIs on human mobility. \\
Ref.~\cite{Pepe_2020_rf3} provide information about the mobility of more than $80,000$ users in Italy, estimated via mobile phones by Cuebiq, a location intelligence and measurement platform. The data covers the period from January 18 to April 17 and is available here: \url{https://data.humdata.org/dataset/covid-19-mobility-italy}. The dataset provides $3$ different indicators. The first is the daily origin-destination matrices between Italian provinces. These are normalized directed flows computed by tracing the movements across provinces of the users. The authors count as a stop only the locations where users stay for more than one hour. In doing so, transits are discounted. The second measure is the radius of gyration of users which provides information about their range of mobility. This metric is computed weekly. The third metric is the proximity network which provides information about the possible social contacts between users. These are relevant for modeling the transmission dynamics of the virus. To build this network, authors consider time windows of one hour and connect all users within a radium of $50m$. \\
Ref.~\cite{Barbieri_2020} describes a survey conducted among more than $9,000$ individuals in Australia, Brazil, China, Ghana, India, Iran, Italy, Norway, South Africa, USA. The answers are made available here: \url{https://doi.org/10.7910/dvn/eiquga}. The data provides demographic information and answers to several questions about the transportation modes before and after the start of the pandemic. Furthermore, the authors queried participants about their perception about COVID-19 related risks in transportation system.

\subsection{Contacts}

In this category we find a research project called \emph{SOcial Contact RATES} (Socrates)~\cite{Willem_2020_xx1}. The authors assembled a range of contact matrices, across different countries (Belgium, China, Finland, France, Germany, Italy, Hong Kong, Luxembourg, Netherlands, Poland, Peru, Russia, South Africa, UK, Vietnam, Zambia, Zimbabwe), both in pre and post COVID-19 periods. The data and interactive tool to explore and extract it in ready to use format is available here: \url{http://www.socialcontactdata.org/}.

\section{Conclusions and outlook}

What are the main take-home messages from the sample of the literature considered?\\
Many articles helped to define the features of the virus and its transmission. SARS-CoV-2 is characterized by large $R_0$. The majority of models estimate values in the $3-4$ range, depending on the region/country. Due to lack of adequate testing (especially in the early phases) and a large fraction of asymptomatic infections the virus spread largely undetected. Several data-driven modeling efforts show heterogeneity in disease transmission. A minority of cases is responsible for a large fraction of infections. Furthermore, some locations such as restaurants, hotels, workplaces are risker locations for infections. Detailed analysis of confirmed cases and their contacts show how household contacts are also responsible for a large fraction of infections (especially during lockdowns) followed by extended family interactions. Children are found to be less susceptible and less at risk of severe outcomes. On the contrary, individuals older than $65$ are found to be more susceptible and at higher risk of death. Crowded cities are affected by higher attack rates and longer epidemic waves.\\
Many articles aimed to measure the effects of NPIs. Across the board, the modeling efforts unanimously show the importance and effectiveness of NPIs in slowing down the spread of COVID-19. Interestingly, the timing of their implementation is a critical variable. Countries that acted early, with respect to the local spread, were most successful in controlling the spread and reported markedly lower death tolls. In several contexts, only a staggered NPI strategy was enough to bring the $R_0$ below the critical value. A range of articles highlights how a patchy and inconsistent geographical NPI strategy is linked to worst health outcomes. It might induce behavioral changes, such as travel further for an activity, which ultimately might accelerate the spreading. The study of the conditions necessary to reach herd immunity, without overflowing the health care system, point out that dynamic mitigation strategies are unrealistic and unpractical. Even more, several articles, some factoring in also economic metrics, point to the fact that suppression is preferable to mitigation in the long run. Many research efforts show the large impact of one of the simpler and less costly NPIs: face masks. 
Analyses of NPIs adopted across many countries as well as data-driven epidemic models highlight how banning public gatherings, limiting social gatherings, school closures, remote working, and lockdowns are the most efficient top-down measures. It is important to notice, how the heterogeneity of infection outcomes as function of age, the difference in socio-demographic stratifications, and intergenerational contacts across different countries suggest that one-fits-all NPIs strategy might be far from optimal. Measures should be custom-tailored to the specific context of each population. \\
Despite their efficacy NPIs came with high societal costs. Several articles in the sample highlight the impact on mental health, loneliness, domestic abuse, sex, food consumption, and substance abuse among others. The way health care facilities operated had to drastically change. The quick shift towards telemedicine and remote care allowed to provide help to a countless number of patients. However, several surveys highlight the challenges that this induced. From lack of enough support and socio-economic disparities to the tragic increase of preventable deaths for several types of cancers due to reduction of screening. Also the education sector had to adapt drastically. The results point to the effectiveness of remote learning as well as the challenges that it entails which go from the lack of social interactions, the key to the development of children, to the impossibility of meeting some learning outcomes linked to practical skills. Even though strict NPIs have increased the time spent at home, a large survey conducted among scientists in the USA and Europe points to the fact that female scientists, those in experimental fields and with young children had to reduce the time spent doing research. \\
When it comes to adoption and adherence to NPIs a range of articles paint a very complex picture pointing out several factors influencing behavioral changes. Gender, education, age, and political leanings are examples. Several surveys suggest that women are more likely to comply with NPIs and to consider COVID-19 as a real risk. Similar patterns are observed in individuals with higher education levels. When it comes to age, the results indicate how young and older adults are less likely to comply with NPIs. Few articles, in the context of the USA, indicate how republican leaning counties were less likely to reduce mobility and were associated with a higher impact of the disease. The research points also to a big urban-rural divide and the importance of socio-economic factors. People living in urban settings were found more informed, and in the context of the developing world, more equipped to adopt protective behaviors. Across the board, thus also in the richest countries, adoption of NPIs is linked to socio-economic indicators: individuals in high-income brackets are able to drastically reduce their mobility with respect to low-income brackets. Heterogeneity in terms of infection rates in disadvantaged socio-economic groups is also reported. These effects, together with the worst access to health care and higher levels of comorbidities, induced disproportionally higher disease burden in low-income groups. There is evidence suggesting that food programs and financial support, implemented as part of containment strategies, help NPIs adoption. Several articles point to the fact that, though many NPIs were imposed top-down, big variations in mobility patterns were observed before official orders. This suggests that bottom-up, spontaneous, behavioral changes took place in several contexts.\\

Despite the incredible effort of many scientific communities, there are still open questions and challenges for future work.\\
\textbf{We don't have a validated theory to describe the feedback loop between behaviors and diseases}. Before COVID-19 the research on NPIs was mainly theoretical. Lack of empirical data describing behavioral changes was the main limitation. During the COVID-19 pandemic unprecedented high-resolution datasets, capturing different aspects of NPIs, have been collected and shared. The large majority of models used them as input. Thus, we transitioned from theoretical approaches in the pre-COVID-19 world to data-driven modeling post-COVID-19. In fact, with few exceptions, the vast theoretical literature on the subject has been neglected during the pandemic. As mentioned in one of the perspectives~\cite{Vespignani_2020_x2}, pandemics are a war period for modelers. Information is sparse, knowledge limited, answers a real commodity. Access to data has allowed the luxury to avoid developing (or adapting) models of behavioral changes and focus on other aspects. As result, despite great progress, we still don't have a validated theory to describe NPIs. The data collected during the COVID-19 pandemic will be of crucial importance to tackle this challenge in the future. \\ 
\textbf{We don't have standardized metrics to compare or judge epidemic models}. Epidemic models are the most represented category in the sample of papers considered here. We find a wide range of approaches and methodologies. While they all seem to do a decent job in reproducing the epidemic patterns, it is far from clear how to compare them or how to judge their efficacy. As noted in one of the perspectives~\cite{Poletto_2020_x2}, going forward the challenge is to define standard metrics to evaluate their performance. Competition is not the goal. The aim is to compare their predictive power, quantify their reliability and sensitivity to assumptions.\\
\textbf{We need to strengthen Data for Good initiatives.} Many companies have provided unprecedented datasets to estimate the effects of NPIs on our behaviors. The data have been shared thanks to several Data for Good programs. Their value and impact in the fight against COVID-19 cannot be overstated. As we move towards the end of the pandemic, I hope these initiatives will become even more popular rather than shrink dramatically. Extensions, definition of standard metrics, and privacy concerns are key challenges for future work.\\
\textbf{We need to improve availability and  accessibility of public health data}. Within few months, we saw incredible progresses on this front. Many health agencies, organizations and initiatives created APIs that allow to download official cases and death counts at different geographical resolutions. However, much more work is needed. Some countries still does not share all the data collected. For example, detailed information about onset of symptoms, and breakdown of data per age-bracket is often not shared or provided by regional initiatives in very unhelpful formats (i.e., pdf). In other cases, we see still long delays before the data is shared. The new SARS-CoV-2 variant spreading in the UK is an example. After several weeks, data is still spread across sources and sparse. One of the major challenges going forward is to improve data access and sharing standards. \\ 
\textbf{We don't know much about the long terms of NPIs}. NPIs have affected virtually any activity and societal process. As noted in many articles, especially those written in the context of health care, going forward key challenges are linked to study their long-term effects and to develop strategies/policies to cope with them. \\
\textbf{Promoting high quality information is key}. The adoption of health promoting behaviors is a complex phenomenon affected by many variables. In the current landscape access to information and misinformation is critical. As we transition towards the largest vaccination campaign ever attempted and as we prepare for future health emergencies, understanding how to promote high-quality information and demote misinformation is a key challenge for future work.


\end{document}